\documentclass[useAMS,usenatbib]{aa}

\usepackage{graphicx}
\usepackage[outdir=./]{epstopdf}
\DeclareGraphicsExtensions{.eps}
\graphicspath{{plots/}}
\usepackage{pdflscape}
\usepackage[varg]{txfonts}
\bibpunct{(}{)}{;}{a}{}{,}
\usepackage{natbib}
\usepackage[breaklinks=true,colorlinks=true,linkcolor = blue,
            urlcolor  = blue,
            citecolor = blue,
            anchorcolor = blue]{hyperref} 
\bibpunct{(}{)}{;}{a}{}{,} 

\newcommand{\high}{H{\sc i}GH}

\newcommand{\hi}{H{\sc i}}
\newcommand{\MHI}{M_{\text{H{\sc i}}}}

\newcommand{\citeb}[1]{\citeauthor{#1} \citeyear{#1}}
\makeatother

\usepackage{color}

\begin{document}

\title{A systematic metallicity study of DustPedia\thanks{DustPedia is a project funded by the EU under the heading `Exploitation of space science and exploration data'. It has the primary goal of exploiting existing data in the \textit{Herschel} Space Observatory and Planck Telescope databases.} galaxies reveals evolution in the dust-to-metal ratios}

\titlerunning{DustPedia dust-to-metal ratios}
\author{P. De Vis$^{1,2}$\thanks{E-mail: pieter.devis1@gmail.com}\and A. Jones$^{1}$\and S. Viaene$^{3}$\and V. Casasola$^{4,5}$\and C.\,J.\,R. Clark$^6$\and M. Baes$^{3}$\and 
S. Bianchi$^4$\and L. P. Cassara$^{7}$ \and J.~I.~Davies$^2$\and I.~De Looze$^{3,8}$\and M. Galametz$^9$\and 
F. Galliano$^9$\and  S. Lianou$^9$\and S. Madden$^9$\and A. Manilla-Robles$^{10}$\and A.~V.~Mosenkov$^{3,11,12}$\and 
A.~Nersesian$^{3,7,13}$\and S. Roychowdhury$^{1}$\and E. M. Xilouris$^{7}$\and N. Ysard$^{1}$}
 \institute{$^{1}$  Institut d’Astrophysique Spatiale, CNRS, Université Paris-Sud, Université Paris-Saclay, Bât. 121, 91405, Orsay Cedex, France \\
$^{2}$ School of Physics \& Astronomy, Cardiff University, Queen's Buildings, The Parade, Cardiff CF24 3AA, United Kingdom \\
$^{3}$ Sterrenkundig Observatorium, Universiteit Gent, Krijgslaan 281, B-9000 Gent, Belgium \\
$^{4}$ INAF, Osservatorio Astrofisico di Arcetri, Largo E. Fermi 5, 50125, Firenze, Italy \\
$^{5}$ INAF - Istituto di Radioastronomia, Via P. Gobetti 101, 4019, Bologna, Italy
$^{6}$ Space Telescope Science Institute, 3700 San Martin Drive, Baltimore, Maryland, 21218, USA \\
$^{7}$ National Observatory of Athens, Institute for Astronomy, Astrophysics, Space Applications and Remote Sensing, \\
Ioannou Metaxa and Vasileos Pavlou, 15236 Athens, Greece\\
$^{8}$ Department of Physics and Astronomy, University College London, Gower Street, London WC1E 6BT, UK\\
$^{9}$ AIM, CEA, CNRS, Universit\'e Paris-Saclay, Universit\'e Paris Diderot, Sorbonne Paris Cit\'e, F-91191 Gif-sur-Yvette, France \\
$^{10}$ Department of Physics \& Astronomy, University of Canterbury, Private Bag 4800, Christchurch, New Zealand\\
$^{11}$ St. Petersburg State University, Universitetskij Pr. 28, 198504 St. Petersburg, Stary Peterhof, Russia \\
$^{12}$ Central Astronomical Observatory of RAS, Pulkovskoye Chaussee 65/1, 196140 St. Petersburg, Russia \\
$^{13}$ Department of Astrophysics, Astronomy \& Mechanics, Faculty of Physics, University of Athens, Panepistimiopolis, 15784 Zografos, Athens, Greece
}

\label{firstpage}
\abstract{Observations of evolution in the dust-to-metal ratio allow us to constrain the dominant dust processing mechanisms. In this work, we present a study of the dust-to-metal and dust-to-gas ratios in a sub-sample of $\sim500$ DustPedia galaxies. Using literature and MUSE emission line fluxes, we derived gas-phase metallicities (oxygen abundances) for over 10000 individual regions and determine characteristic metallicities for each galaxy. We study how the relative dust, gas, and metal contents of galaxies evolve by using metallicity and gas fraction as proxies for evolutionary state. The global oxygen abundance and nitrogen-to-oxygen ratio are found to increase monotonically as galaxies evolve. Additionally, unevolved galaxies (gas fraction $>60\%$, metallicity $\rm 12+log(O/H)<8.2$) have dust-to-metal ratios that are about a factor of 2.1 lower (a factor of six lower for galaxies with gas fraction $>80\%$) than the typical dust-to-metal ratio ($M_d/M_Z\sim0.214$) for more evolved sources. However, for high gas fractions, the scatter is larger due to larger observational uncertainties as well as a potential dependence of the dust grain growth timescale and supernova dust yield on local conditions and star formation histories. We find chemical evolution models with a strong contribution from dust grain growth describe these observations reasonably well. The dust-to-metal ratio is also found to be lower for low stellar masses and high specific star formation rates (with the exception of some sources undergoing a starburst). Finally, the metallicity gradient correlates weakly with the \hi-to-stellar mass ratio, the effective radius and the dust-to-stellar mass ratio, but not with stellar mass.
}


\defcitealias{Clark2015}{C15} 
\defcitealias{DeVis2017b}{DV17b}

\maketitle

\section{Introduction}
Dust is a key component in the interstellar medium (ISM) of galaxies as it acts as a catalyst for the formation of molecules \citep{Gould1963,Draine2003,Galliano2018} and strongly affects the observed emission of galaxies. Dust absorbs and scatters stellar radiation and re-emits the absorbed radiation in the far-infrared (FIR; \citeb{Fixsen1996}; \citeb{Hauser2001}; \citeb{Driver2016}). Interstellar dust forms in a range of environments, such as the winds of evolved low-to-intermediate mass stars \citep[LIMS,][]{Ferrarotti2006,Sargent2010}, core-collapse supernovae ejecta (SNe) \citep[e.g.][]{Dunne2003, Rho2008, Matsuura2011, Gomez2012, Indebetouw2014,DeLooze2017,Bevan2017} and grain growth and accretion in the ISM \citep{Dwek2007, Mattsson2012, Asano2013, Zhukovska2014, Rowlands2014b}. 

Dust depletes metals from the gas-phase ISM \citep{Calzetti1994, Calzetti2000, Spoon2007, Melbourne2012}. If dust and metals are created from stars at constant rates and there are only stellar sources of dust (i.e. no metals are converted into dust through grain growth), then one would expect the dust-to-metal ratio to remain constant as galaxies evolve \citep[e.g.][]{Franco1986}. A constant dust-to-metal ratio is also assumed in early chemical evolution models \citep{Silva1998,Edmunds1998}, to determine the dust mass absorption coefficient \citep{James2002,Clark2016} and in studies combining radiative transfer models with hydrodynamical simulations \citep{Yajima2015,Camps2016} or some semi-analytic models of galaxy formation \citep{Lacey2008,Somerville2012}. On the other hand, dust grain growth would increase the dust-to-metal ratio as galaxies evolve, and dust destruction mechanisms (e.g. shocks or thermal sputtering; see \citeb{Jones2004} for a review) would decrease the dust-to-metal ratio (\citeb{Mattsson2012}). Observations of the dust-to-metal ratio over a wide range of evolutionary stages thus allow us to constrain the dominant dust processing mechanisms, which will significantly alter the ISM composition as galaxies evolve. 

Early work on the dust-to-gas vs. metallicity relation have revealed an increase in the dust-to-gas ratio with metallicity \citep{Viallefond1982,Issa1990,Lisenfeld1998}. Linear relationships were found, corresponding to a surprisingly constant dust-to-metal ratio ($M_d/M_Z \sim 0.5$) obtained for a wide range of galaxies, which was explained using  models without grain growth \citep[e.g.][]{Hirashita1999a, Edmunds2001}. However, more recently it has become clear that local unevolved low-mass galaxies have significantly lower dust-to-gas ratio than would be expected for a constant dust-to-metal ratio of $\sim 0.5$ \citep{Draine2007,Galliano2008, Galametz2011,Fisher2014,Remy-Ruyer2014,DeVis2017b}. \citet{Chiang2018} performed a resolved study and found that the dust-to-metal ratio is not constant in M101, but decreases as a function of radius, which is equivalent to lower dust-to-metal ratios for low metallicity regions. \citet{Jenkins2009} also found variations in the depletion of metals onto dust in the diffuse ISM of the Milky Way. \citet{RomanDuval2017} show variations in the dust-to-gas ratios in the Magellanic clouds scale non-linearly with gas surface density, and are consistent with depletion measurements and simple modelling of grain growth. Recent results based on gamma-ray burst (GRB) afterglows, damped Ly-$\alpha$ absorbers in the foregrounds of QSOs \citep{DeCia2013, Zafar2013,Wiseman2017} and distant lens galaxies \citep[e.g.][]{Dai2009} find mixed results on whether there is evidence for lower dust-to-metal ratios in high-redshift galaxies. A compilation by \citet{DeCia2016} shows a decreasing dust-to-metal ratio towards low metallicities for damped Ly-$\alpha$ absorbers, yet the variation is much smaller than observed dust-to-metal variation in the local universe \citep{Galliano2018}. \citet{Mattsson2014} suggest that selection effects or uncertainties could explain the differing observed trends, and propose that an equilibrium mechanism between dust grain growth and destruction might exist that keeps the dust-to-metals ratio close to constant if certain conditions are fulfilled. 

From a theoretical viewpoint, chemical evolution models tracking the dust, gas and metal content of galaxies and including prescriptions for dust formation, dust grain growth, dust destruction and inflows and outflows are able to model the observed trend of increasing dust-to-metal ratio as galaxies evolve, but different works result in different contributions of dust grain growth \citep{Zhukovska2014,Feldmann2015,McKinnon2016,DeVis2017b}. These differences in dust grain growth contributions are in part due to a lack of strong observational constraints at the low-metallicity end. Grain growth is also essential to understanding the dust budget of the Milky Way \citep{deBennassuti2014}, high redshift normal star forming galaxies \citep{Michalowski2015,Mancini2015,Mancini2016,Knudsen2017} and the rapid dust enrichment of z > 6 quasar host galaxies \citep{Valiante2011,Valiante2014,Calura2014}. Additionally, \citet{Calura2017} and \citet{Popping2017} have shown that grain growth is needed to create models consistent with observations at both low and high redshifts.

In this work, we compile metallicities for the DustPedia sample \citep{Davies2017} to increase the sample size for which the dust-to-metal ratio can be studied. DustPedia is a collaborative focused research project working towards a definitive understanding of dust in the local Universe, by capitalising on the legacy of \textit{Herschel}. The full DustPedia sample consists of 875 nearby (v $< 3000$ km/s), extended ($D25 > 1^\prime$) galaxies that have been observed by \textit{Herschel} and have a near-infrared (NIR) detected stellar component. These galaxies have excellent multi-wavelength photometry available (typically 25 bands; \citeb{Clark2018}) and various derived galaxy properties (such as dust mass, stellar mass, star formation rate; see Section \ref{galaxyprop}). DustPedia uses the physically motivated (based on laboratory data) THEMIS dust model \citep{Jones2016,Ysard2016,Jones2017} as reference dust model. 

In this paper, we obtain a database of metallicity measurements and combine these with the rich DustPedia dataset. We studied the global dust and metal scaling relations and improve the observational constraints (466 galaxies) on the evolution of the dust-to-metal ratio. These measurements will be key for constraining chemical and dust evolution models, and the role of dust grain growth in particular. Section \ref{sec:spoctrophot} is dedicated to our literature compilation of emission line fluxes and extraction of spectrophotometry of archival data from the Multi Unit Spectroscopic Explorer (MUSE; \citeb{Bacon2010}) instrument at the ESO VLT telescope. In Section \ref{metalsec}, we describe how we used these line fluxes to derive metallicities and combine measurements for individual regions into a global metallicity. In Section \ref{galaxysec} we explain how the other galaxy properties were derived and briefly discuss the comparative samples used in this work. Our results are presented in Section \ref{resultssec} and discussed in Section \ref{discussion}. In Section \ref{data}, we describe the data we make available to the community. Finally, Section \ref{sec:conclusions} lists our conclusions. 

\section{Spectrophotometry}
\label{sec:spoctrophot}
\subsection{Compilation of spectrophotometry from literature}
\label{Litsec}
To determine gas-phase metallicities  for the DustPedia galaxies, we used multiple strong-line calibrations (see Section \ref{calibrations}). The emission lines used in this work are given in Table \ref{THEMISobs}. We note that [OII] ${\lambda3727}$ and [OII] ${\lambda3729}$ are blended because of the spectral resolution. We have performed a literature search to compile the emission line fluxes for as many of the DustPedia galaxies as possible. We do not claim that this compilation is exhaustive, yet it does include results from many sources (a total of 6818 regions are compiled). A list of the compiled references and an example of the emission lines for a few sources are given in Appendix \ref{appendix}.

\begin{table}
\caption{Emission lines used in this work and extinction coefficients from the THEMIS dust model \citep{Jones2017}.}
\begin{center}
\begin{tabular}{lcc} \hline\hline
Line \hspace{1cm}& \hspace{1cm} $\lambda$ (\AA) \hspace{1cm} & $k(\lambda)$\\ \hline
{[OII]} & 3727,3729 & 5.252 \\
$\rm{H\beta}$ & 4861 & 3.886 \\
{[OIII]} & 4959 & 3.797\\
{[OIII]} & 5007 & 3.755\\
{[NII]}$^1$ & 6548 & 2.728\\
$\rm{H\alpha}$ & 6563 & 2.720\\
{[NII]} & 6584 & 2.710\\
{[SII]} & 6717 & 2.644\\
{[SII]} & 6731 & 2.637 \\
\hline \\
\end{tabular}
\\
\end{center}
$^1$ [NII]$\lambda{6548}$ is not actually fitted but its flux is set to 1/3 of the [NII]$\lambda{6584}$ flux (theoretically expected ratio; e.g. \citeb{Alam2015}).
\label{THEMISobs}
\end{table}

The compiled emission lines can be split into four categories: integrated, grating, fibre, and integral field unit (IFU) spectroscopy. Integrated spectroscopy provides spectra for the galaxy as a whole and can be obtained either by using a spectrograph where the beam comprises the entire galaxy, or by using techniques such as drift scan spectroscopy (see e.g. \citeb{Boselli2013}). Grating spectroscopy collects light along a slit placed over the galaxy. This light is then diffracted along an additional dimension, which allows the spectra for H{\sc ii} regions along the observed slit to be measured. Fibre spectroscopy is a technique where multiple optical fibres can be pointed at different lines of sights and their spectra collected simultaneously. These are often good resolution pointings targeting a small region within the galaxy. There are often multiple fibre pointings within the same DustPedia galaxy, for which the metallicities will be combined into a global metallicity in Section \ref{global}. IFUs are closely packed bundles of fibres that allow to perform a resolved study of the gas-phase metallicities in galaxies and is the preferred method when available. To extend our sample of sources with resolved metallicities, we supplemented the literature IFU data with MUSE data from the ESO archive, as described in Section 2.2.

The emission lines of galaxies are attenuated both by internal and external dust. To account for this, the emission line intensities are corrected, first for Galactic extinction\footnote{We use the IRSA Galactic Dust Reddening and Extinction Service (https://irsa.ipac.caltech.edu/applications/DUST/) and the prescription of \citep{Schlafly2011}.} and then using the Balmer decrement given by \begin{align} C(\rm{H\beta})= \frac{\log\Big(\frac{\rm{H\alpha}}{\rm{H\beta}}\Big)_{\rm theor} - \  \log\Big(\frac{\rm{H\alpha}}{\rm{H\beta}}\Big)_{\rm obs} }{0.4\times [(k(\lambda_{H\alpha})-k(\lambda_{H\beta})]} 
\end{align} where $k(\lambda)=A_{\lambda}/E(B-V)$ is the reddening curve for the diffuse ISM version of the THEMIS \citep{Koehler2014,Jones2017} dust model, $0.4\times [(k(\lambda_{H\alpha})-k(\lambda_{H\beta})]= -0.466$; $\log(\rm{H\alpha}/\rm{H\beta})_{\rm obs}$ is the observed ratio between $\rm{H\alpha}$ and $\rm{H\beta}$, and $\log(\rm{H\alpha}/\rm{H\beta})_{\rm theor}$ the theoretically expected ratio which depends on the electron density and the gas temperature. We assumed case B recombinations with a density of 100 ${\rm cm^{-3}}$ and a temperature of $10^4$ K, which gives the predicted ratio (unaffected by reddening or absorption) of $\rm{H\alpha}/\rm{H\beta} = 2.86$ \citep{Osterbrock1989}. 

The corrected emission line fluxes are then given by \begin{align}F_{\rm corr}(\lambda)=F_{\rm obs}(\lambda)\  10^{0.4 \ (E(B-V)_{Galactic} + C(\rm{H\beta})) \ k(\lambda) } \end{align} where $k(\lambda) $ for the THEMIS dust model is given in Table \ref{THEMISobs}. When available, we used the uncorrected fluxes from the literature. When only reddening-corrected fluxes are given, we determined the uncorrected fluxes using the listed $C(\rm{H\beta})$ and attenuation curve of each work, and then correct them using the THEMIS attenuation law for consistency. For only a few references \citep[e.g.][]{Bresolin1999,Pilyugin2014}, did we not have the necessary information to implement this correction. In these cases we simply used their listed corrected fluxes. We tested different reddening laws \citep{Cardelli1989,Calzetti2000} and found only small ($\sim 0.01$ dex) variations in the resulting metallicities for each region. None of the qualitative conclusions in our work are affected by these variations. 

\subsection{MUSE spectrophotometry}
With its 0.2 arcsec pixel scale,  spectral range of 4750 - 9350~\AA, spectral sampling of 1.25~\AA\ (R = 1770 - 3590), and field of view of 1 arcmin $ \times $ 1 arcmin, MUSE provides the most high-resolution IFU observations to date.
This enables resolved studies of the dust-to-gas and dust-to-metal ratio (at a resolution set by the dust maps; e.g. this work), as well as resolved attenuation studies through Balmer decrements and in the continuum \citep[e.g.][]{Viaene2017}, and studies of the gas and stellar kinematics \citep[e.g.][]{Guerou2017}. This wealth of data will undoubtedly be of use in future DustPedia papers and we have thus supplemented our literature fluxes by MUSE spectroscopy. By querying the ESO archive\footnote{http://archive.eso.org/}, we found 79 of the DustPedia galaxies have public MUSE data available, often with multiple data cubes for the same galaxy. After downloading the data cubes we inspected the astrometry, which revealed in some cases it was offset by up to 12 arcsec. We fixed the astrometry using GAIA\footnote{Starlink/GAIA  (The Starlink Project was a UK Project supporting astronomical data processing, now maintained by the East
Asian Observatory. GAIA is an extension of the RTD (real time display tool) which has been written at ESO as
part of the VLT project.) } and the NOMAD \citep{Zacharias2004} and 2MASS \citep{Skrutskie2006} point source catalogues. 
 
Once the astrometry was corrected, we masked the stars in the images using an adapted version of PTS\footnote{http://www.skirt.ugent.be/pts/} (Verstocken et al. in prep.). The NOMAD and 2MASS point source catalogues were again used to identify the stars and the radii of the masked regions were determined using a curve-of-growth technique. This method also erroneously masks bright H{\sc ii} regions, which are often included in the point source catalogues, and we thus use an additional criterion to only mask the stars. If the $\rm{H\alpha}$ emission is clearly detected ($5\sigma$) and its peak flux is more than twice as bright as the stellar continuum, we identify this region as an H{\sc ii} region and the pixels are not masked. These masking limits are somewhat arbitrary, but perform well upon visual inspection. Additionally, a small amount of stellar contamination will barely affect the final fluxes, and our method is thus robust to moderate changes to the masking limits.

Given our focus on the dust-metal interplay for which our sub-sample of DustPedia galaxies with MUSE (and \textit{Herschel}) data is uniquely suited, it is of little use to have metallicities to much better resolution than the dust maps. Consequently, we bin all the MUSE pixels to have the same pixel size and positions as the \textit{Herschel} SPIRE 250 $\rm \mu m$ maps. The SPIRE 250 $\rm \mu m$ pixel size is 6 arcsec. We will thus bin 900 MUSE pixels for each \textit{Herschel} pixel and determine the mean flux and standard deviation in each binned pixel. Binning 900 pixels increases the signal-to-noise ratio of the MUSE spectra by a factor 30. We note that we have not smoothed our maps to be consistent with the 18 arcsec beamsize of the SPIRE 250 $\rm \mu m$ maps. The choice of the 250 $\rm \mu m$ pixel size as our resolution is somewhat arbitrary. This pixel size is large enough to drastically reduce the number of MUSE spectra, yet small enough to still reliably convolve our data to any of the SPIRE beams. We have not performed any convolution as this is not important for our determination of the characteristic metallicity (Section \ref{global}), though interested users can still convolve this data if their analysis requires it. 
 
\begin{figure*}
\begin{center}
\includegraphics[width=0.99\textwidth]{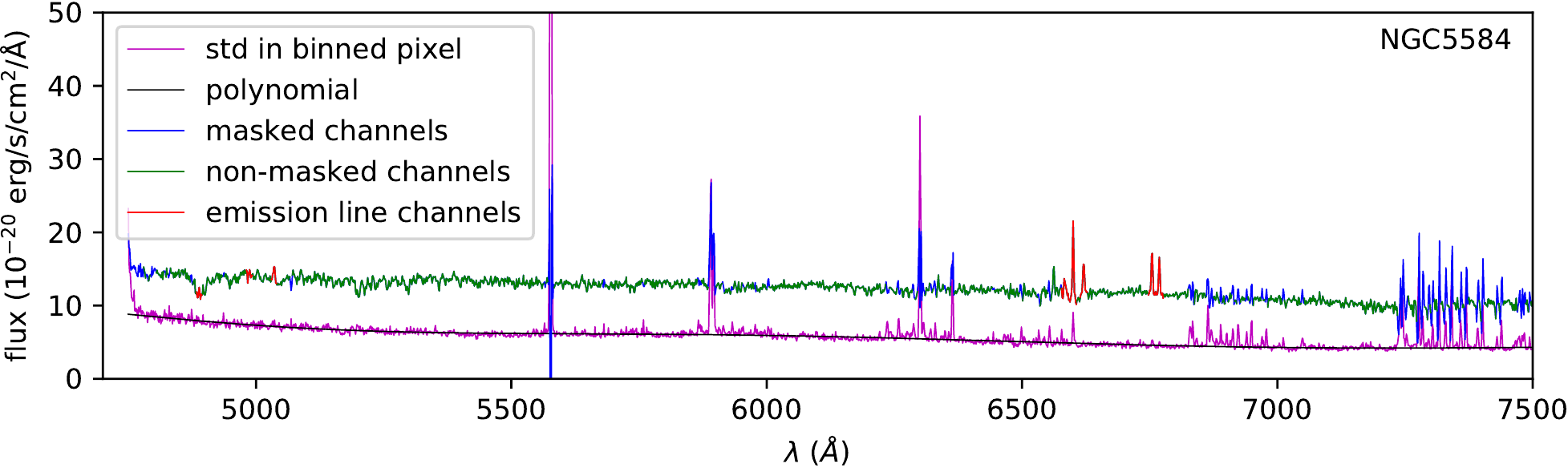}
\includegraphics[width=0.99\textwidth]{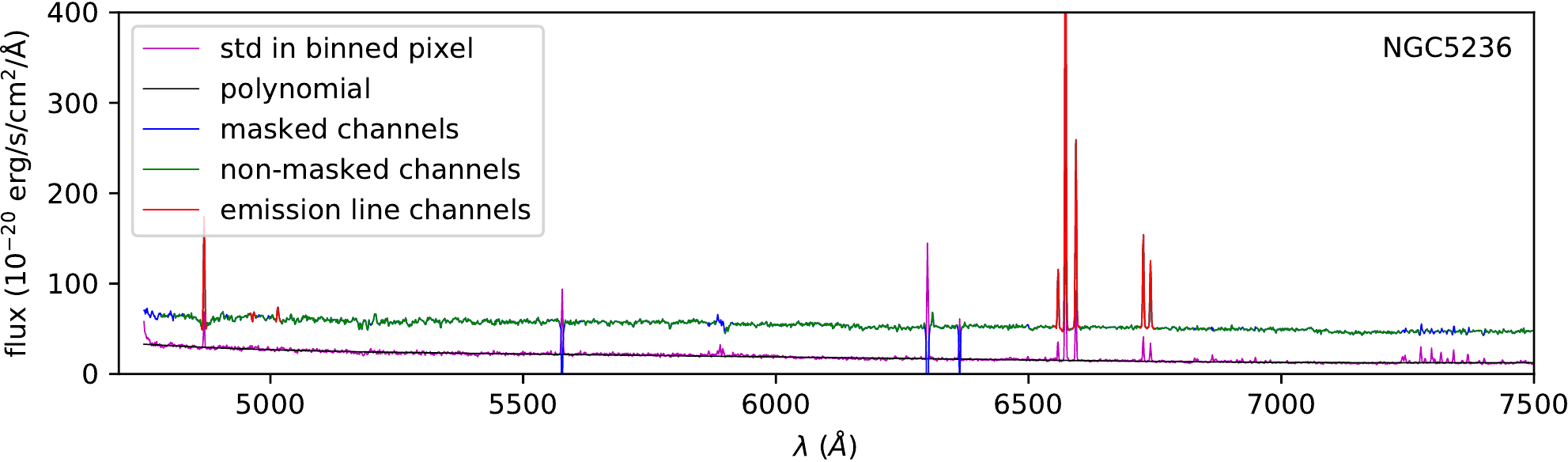}
\includegraphics[width=0.99\textwidth]{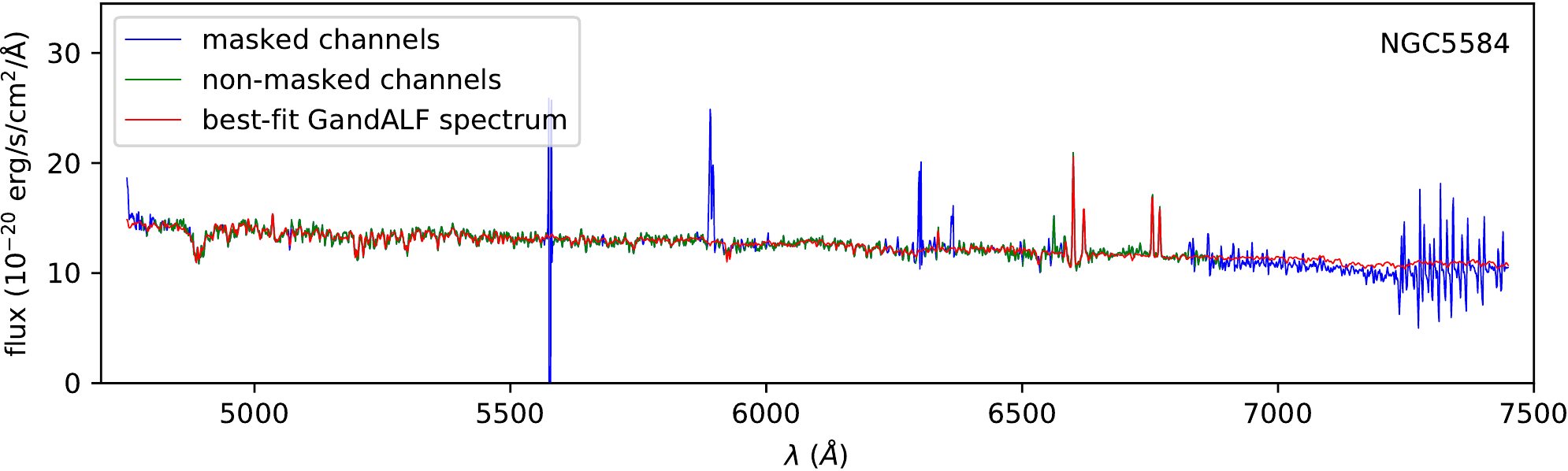}
\includegraphics[width=0.99\textwidth]{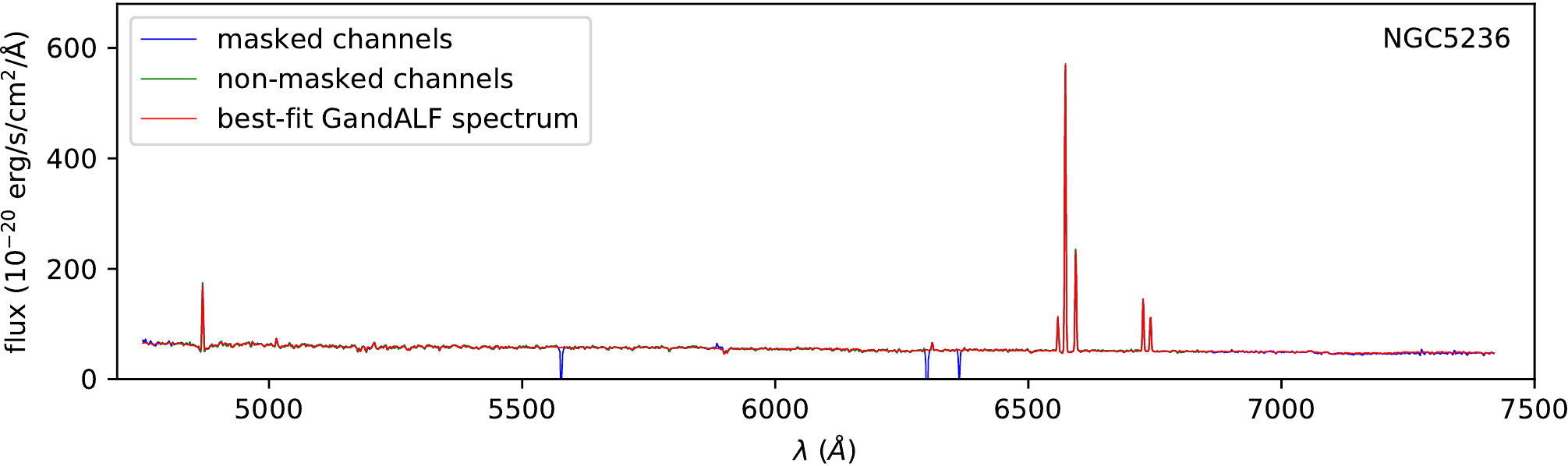}
\end{center}

\caption{Examples of the sky contamination masking (\textit{top two panels}) and GANDALF fit (\textit{bottom two panels}) for a random binned pixel of NGC5584 (heavily contaminated by sky emission; \textit{first and third panel}) and NGC5236 (typical spectra; \textit{second and last panel}). \textit{Top}: These panels show the average flux (green) and standard deviation (magenta) of binned MUSE pixels within one \textit{Herschel} pixel. The best fitting polynomial to the standard deviations is shown in black. Channels where the standard deviation is $3\sigma$ higher than the polynomial are masked (blue). The sky contamination present in the NGC5584 pixel is masked effectively. \textit{Bottom}: Reddening-corrected MUSE spectrum (masked channels are shown in blue, non-masked channels in green) together with the best GANDALF (red) fit.}
\label{skycont}
\end{figure*}

Once the spectra of each binned pixel are extracted and inspected, we find that a few galaxies are significantly contaminated by residual sky emission lines (Figure \ref{skycont}). These emission lines can easily be identified by looking at the standard deviation in the binned pixel. This standard deviation increases if either the mean flux increases (bright emission line) or if a channel is very noisy. In both cases we masked these emission lines, with the exception of the emission lines given in Table \ref{THEMISobs} as for these lines the emission originates in the galaxy rather than our atmosphere.
To identify the channels to mask, we first fitted an eighth-order polynomial\footnote{This high order polynomial was taken to ensure small-scale variation in the baseline did not result in masked channels. Since we do use this polynomial no further, there is no harm in using such a high-order polynomial.} to the standard deviation in each channel (Figure \ref{skycont}; \textit{top}). We then masked the channels in each spectrum where the standard deviation is $3\sigma$ higher than the  polynomial. We then again fitted an eighth-order polynomial, ignoring the masked channels, and repeat the process iteratively until convergence is reached. Finally we unmasked the channels within 300 km/s (6.6~\AA) of the central velocity of each known emission line.

\subsection{GANDALF line fitting}
Emission lines were measured by running each spectrum through a modified (as detailed below) version of the Gas AND Absorption Line Fitting algorithm (GANDALF\footnote{http://star-www.herts.ac.uk/$\sim$sarzi/gandalf\_releases/}; \citeb{Sarzi2006}). GANDALF simultaneously fits the emission and absorption lines and is designed to separate the relative contribution of the stellar continuum and of the nebular emission in the spectra of nearby galaxies, while measuring the gas emission and kinematics. GANDALF uses a combination of stellar templates based on the MILES stellar library \citep{Sanchez2006} to describe the stellar continuum and the pPXF code of \citet{Cappellari2004} to derive the stellar kinematics. For each MUSE data-cube, GANDALF is first run on the full spectrum (i.e. averaging all the binned pixels in the cube). The stellar templates that had a weight of 2\% or higher are then stored and only these templates are used when fitting the spectra of each binned pixel. This significantly speeds up the fitting, as well as ensuring that nearby regions in the galaxy have somewhat similar stellar populations. Using only these templates still results in excellent fits to the data for both pPXF and GANDALF (Figure \ref{skycont}; \textit{bottom}).

The present version of GANDALF (v1.5) includes an uncertainty determination on the fluxes as well as reddening by interstellar dust using a \citet{Calzetti2000} dust model (using $\rm{H\alpha}/\rm{H\beta} = 2.86$ as in Section \ref{Litsec}). However, in order to allow consistent comparison with the dust emission and other dust properties within the DustPedia framework, we instead use the reddening curve from the THEMIS dust model. In the MUSE wavelength range, the THEMIS reddening curve is well described by: 
\begin{align}
k(\lambda)=21850/\lambda  - 0.609
\end{align}
where $\lambda$ is the wavelength in \AA. We note that the values in Table \ref{THEMISobs} are consistent with this curve, and the reddening correction for MUSE is thus consistent with that for the compiled literature data. We use reddening by two dust components \citep[e.g.][]{Charlot2000,Kreckel2013,Battisti2016}, where one component is for reddening the stellar continuum (attenuation by diffuse dust) and the other is for reddening the emission lines using the Balmer decrement (attenuation by dust in star-forming regions). Both these components are free parameters in our fit. The stellar continuum thus has a different E(B-V) from the H{\sc ii} regions, for which both reddening components are added. 

We have tested the effects of using a better resolution than binning to the \textit{Herschel} pixels. When 2-arcsec pixels are used instead of our 6-arcsec pixels, we find good agreement for the vast majority of pixels. There are a few fainter pixels towards the outskirts of some galaxies where the 6 arcsec pixels have somewhat higher metallicities. This is likely due to remaining diffuse ionised gas (see also Section \ref{specclas}) enhancing the [S II]/$\rm{H\alpha}$ and  [N II]/$\rm{H\alpha}$ ratios \citep{Zhang2017} in these larger pixels. However, this happens for few enough pixels that the global metallicities derived in Section \ref{global} are barely affected.

Finally, we rejected some binned pixels where the GANDALF fit does not describe the measured spectra well. To this end, we measured the standard deviation between the best fit GANDALF spectrum and the measured spectrum (excluding masked channels) over three wavelength ranges: over the whole spectrum ($\sigma_{\text{full}}$), around $\rm{H\alpha}$ ($\sigma_{\text{H$\alpha$}}$) and around [NII] ${\lambda6584}$ ($\sigma_{\text{NII}}$).  Pixels are rejected if $\sigma_{\text{full}}$ is larger than the mean of the spectrum divided by 3, $\sigma_{\text{H$\alpha$}}$ is larger than the peak flux of the $\rm{H\alpha}$ line divided by 5, or $\sigma_{\text{NII}}$ is larger than the peak flux of the [NII] ${\lambda6584}$ line divided by 2. This effectively removes all poorly fitted spectra. We have tested that changing the rejection criteria only has very minor effects on the conclusions of this work. Finally we obtain a sample of 8272 MUSE regions with reliable spectrophotometry.

\section{Gas-phase oxygen abundances}
\label{metalsec}
\subsection{Spectral classification}
\label{specclas}
Various methods can be used to determine gas-phase metallicities from the emission line fluxes of galaxies. However, one complication is that active galactic nuclei (AGN) also affect the emission line fluxes and thus bias the metallicity estimates. Therefore, we need to discard the sources which are affected by AGN. AGN have a very energetic radiation field, which causes high intensities of collisionally excited lines (e.g. [OIII]$\lambda5007$, [NII]$\lambda6584$) relative to recombination lines (such as $\rm H\alpha$ and $\rm H\beta$). In normal star-forming galaxies, the emission lines are powered by massive stars, and there is an upper limit on the intensity ratios of collisionally excited lines relative to recombination lines. Diffuse ionised gas can also affect the line ratios of the strong emission lines at fixed metallicity \citep[e.g.][]{Zhang2017}, which could result in a bias in our measurements. The ionisation of this diffuse ionised gas is a subject of active research. It is thought the radiation from hot evolved stars may have an important contribution to the ionisation \citep{Oey1997,Hoopes2003,Zhang2017}. Since our metallicity calibrations are based on H{\sc ii} regions, it is important to exclude these low ionisation emission line regions (LIERs). 

\begin{figure}
  \center
    \includegraphics[width=\columnwidth]{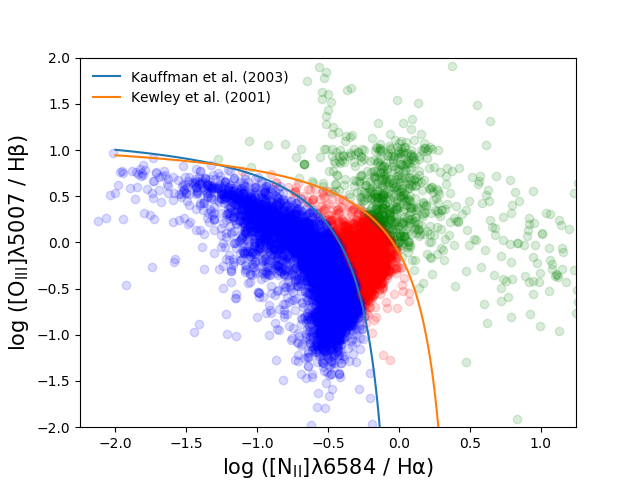}
  \caption{BPT \citep{Baldwin1981} diagram used to classify our spectra of DustPedia galaxies based on classification curves \citep{Kewley2001,Kauffmann2003}. Star-forming H{\sc ii} regions are shown in blue, composite regions in red, and H{\sc ii} regions containing an AGN in green. In this study, we computed metallicities only for the regions which are classified as star forming from this diagram, and discard AGN or composite regions.}
  \label{BPT}
\end{figure}

For our sample, we selected star-forming (H{\sc ii}) regions using the criteria in \citet{Kauffmann2003} by placing sources on the Baldwin, Phillips \& Terlevich (BPT; \citeb{Baldwin1981}) diagram. Similarly, we separated AGN and composite spectra using the curve from \citet{Kewley2001}. We discarded all AGN and composite regions. For galaxies with AGN or composite regions as well as star-forming regions, we still included the galaxy in our sample, yet only used the available starforming regions. A density plot of all the literature regions and MUSE binned pixels on the BPT diagram is given in Figure \ref{BPT}. It has also been shown that LIERs have low equivalent widths of $\rm{H\alpha}$ \citep[$\rm EW_{H\alpha}$][]{CidFernandes2010,Sanchez2015}. Therefore, we discarded all binned MUSE pixels where $\rm EW_{H\alpha}<6\ \AA$ (2\% of the sample). Of the 15090 regions with $\rm EW_{H\alpha}>6\ \AA$, 886 (5.9\%) are classified as AGN, 3216 (21.3\%) are composite regions and 10988 (72.8\%) are star-forming regions. Out of the 683 DustPedia galaxies for which we have spectroscopy, there are 412 galaxies that contain a region that can be classified with $3\sigma$ confidence (the three sigma errorbars on the BPT diagram do not cross the \citet{Kauffmann2003} or \citet{Kewley2001} curves). There are 124 DustPedia galaxies that have at least one confidently classified AGN region. 

\subsection{Strong line calibrations}
\label{calibrations}
To derive metallicities from the emission line spectra, we compared the results from different empirical and theoretical methods to understand any systematic differences that may result from our methods. Direct metallicity estimates require detections of the faint [OIII]$\lambda4363$ line, which is often lacking in our observations. However, numerous empirical calibrations have been determined in the literature that use some of the strong lines (i.e. much brighter than [OIII]$\lambda4363$) listed in Table \ref{THEMISobs}. Empirical calibrations are only valid for the same range of excitation and metallicity as the H{\sc ii} regions that were used to build the calibration. Since they are determined assuming an electron temperature, these methods may systematically underestimate the true metallicity if there are temperature inhomogeneities in a galaxy. This is thought to be more severe in metal-rich H{\sc ii} regions because the higher efficiency of metal-line cooling leads to stronger temperature gradients \citep{Garnett1992,Stasinska2005,Moustakas2010}.  On the other hand, theoretical calibrations require inputs including stellar population synthesis and photoionisation models; often the theoretical metallicities are higher than those found with the empirical calibrations.

We therefore chose to compare four common empirical methods:
\begin{itemize}
\item O3N2 from \citet{Pettini2004}, which uses the [OIII]$\lambda5007$, [NII]$\lambda6584$, $\rm{H\beta}$ and $\rm{H\alpha}$ lines. Their derived relation is only valid for metallicities $12+\mathrm{log(O/H)} > 8.09$ and therefore limited for describing some of the low-metallicity sources in our sample.
\item N2 from \citet{Pettini2004}, using the third order polynomial to determine the metallicity from the ratio of [NII]$\lambda6584$ and $\rm{H\alpha}$ fluxes. The N2 method also runs into difficulties at the lowest metallicities due to the large scatter observed in $\mathrm{N/O}$ ratios \citep{Morales-Luis2014} and instead provides upper limits to the true metallicity for galaxies when $12+\mathrm{log(O/H)_{N2}} < 8$.
\item $R$ calibration from \citet[][hereafter PG16R]{Pilyugin2016}, which uses all lines in Table \ref{THEMISobs} except for the [SII] lines. This calibration performs well, but many of the regions in our sample do not have the necessary [OII]$\lambda3727,3729$ measurements.
\item $S$ calibration from \citet[][hereafter PG16S]{Pilyugin2016}, which uses all lines in Table \ref{THEMISobs} except for [OII]$\lambda3727,3729$.  \citet[][hereafter \citetalias{DeVis2017b}]{DeVis2017b} found PG16S is the most reliable calibration for the low-metallicity sources (and performs significantly better than the \citet{Pilyugin2005} calibration which is often used for low-metallicity sources). 
\end{itemize}
And three theoretical calibrations:
\begin{itemize}
\item \citet[][hereafter KK04]{Kobulnicky2004}. This calibration uses the $\rm R_{23}$ diagnostic ($\rm R_{23} \equiv$ ([OII]$\lambda3727,3729$ + [OIII]$\lambda5007$ + [OIII]$\lambda5007$) / $\rm{H\beta}$). This diagnostic is sensitive to temperature and ionisation, and as a result the $\rm R_{23}$ diagnostic can be degenerate with both a high and low-metallicity solution. The [OII]$\lambda3727,3729$, [NII]$\lambda6584$ and $\rm{H\alpha}$ lines are used to break this degeneracy.
\item \citet[][hereafter T04]{Tremonti2004}. As we do not have access to their code we used the scaling relation between O3N2-T04 from \citet{Kewley2008}, calibrated against 27,730 star-forming galaxies from the Sloan Digital Sky Survey (SDSS). We note that this conversion is only valid for $8.05<12+\mathrm{log(O/H)_{O3N2}}<8.9$. 
\item The Bayesian-based IZI tool \citep{Blanc2015}, which provides a theoretical calibration based on photo-ionisation models and uses all the available lines.
\end{itemize}

Throughout this work we use PG16S as the reference calibration, yet for completion we also include the other calibrations. In addition, we have also derived $\rm log(N/O)$ using the calibration from \citet{Pilyugin2016}. For each of the calibrations, we only computed a metallicity if for each of the used lines the measured flux is larger than its uncertainty. All our calibrations used at least [NII] ${\lambda6584}$ and $\rm H\alpha$.

\subsection{Uncertainties}
\label{errors}
Errors on the line measurements were provided by GANDALF or obtained directly from the literature. We then bootstrapped the measurements by generating 1000 new emission line fluxes assuming a normal distribution with the extinction-corrected emission line fluxes as mean and the measured error as the standard deviation of the distribution. For each set of emission lines we then determined the oxygen abundances for each metallicity calibration. The measurement uncertainties on the metallicity are then given as the 16th and 84th percentiles of the distribution. For sources for which the measured errors on the fluxes were not provided in the literature, we assigned a large artificial uncertainty of 0.2 dex.  

Apart from these measurement uncertainties, there are also uncertainties associated with the extraction method of the spectra. To quantify these, we carried out some consistency checks when we had multiple measurements available for the same region or galaxy. For a number of sources (especially when there are multiple MUSE cubes over the same galaxy) there are multiple metallicity measurements over the same region. We first compared about 700 overlapping MUSE measurements (which have all been consistently measured with GANDALF). We find that the measurements compare very well, with no significant outliers for any of the calibrations. However, the remaining scatter is larger than the typical uncertainties ($\sigma$). Therefore, we added an additional uncertainty $\sigma_{add}$ so that the average $\chi^2 = (Z_1-Z_2)^2/(\sigma_1^2+\sigma_2^2+\sigma_{add}^2)$ is equal to one. For this additional uncertainty, we find values of $\sigma_{\text{MUSE}}=$ 0.036, 0.025, 0.033 and 0.018  for the N2, O3N2, T04 and PG16S respectively (due to the lack of the OII line in the MUSE spectra, PG16R and KK04 cannot be measured). The IZI metallicity did not need any further uncertainty\footnote{IZI has an average $\chi^2\sim 0.5$. The average IZI uncertainties could thus be somewhat overestimated.}. 

There are also fibre spectra from the literature that overlap with the MUSE coverage. There is again good agreement between the measurements and we again add uncertainty so that the average $\chi^2 =1$ for the 114 overlapping pointings. We found that we needed an additional uncertainty of $\sigma_{\text{Lit}}=$ 0.062, 0.060, 0.089 and 0.024 for the N2, O3N2, T04 and PG16S respectively and again no need for additional uncertainty for the IZI metallicities. These sources of uncertainty ($\sigma_{\text{MUSE}}$ or $\sigma_{\text{Lit}}$, not both) were added to the bootstrapped measurement uncertainties for each of the appropriate regions before fitting the radial profiles and determining the global metallicities. This additional uncertainty barely affects the final uncertainty of the global metallicity, yet it does change the weighting of the individual regions.    

For completeness, we note that there is a calibration uncertainty between the different calibration methods, with discrepancies between the different calibration methods as high as 0.6 dex (see the dashed lines in Figure \ref{MZ} in Section \ref{DPoxabund}).
In addition, the empirical calibrations used in this work are derived using the electron temperature method. The uncertainty in the absolute metallicity determination by this method is $\sim 0.1$ dex \citep{Kewley2008}. The above calibration errors\footnote{We note that this is regarding the uncertainty between different calibration methods (e.g. O3N2 or PG16S). The calibration errors on the optical spectra have been included in our error estimates.} were not included in the error-budget in this work, as they should not affect relative differences between galaxies or regions. They should, however, be kept in mind when comparing our measurements with models.

\subsection{Global oxygen abundances}
\label{global}

\begin{figure*}
  \center
\includegraphics[height=0.3\textwidth]{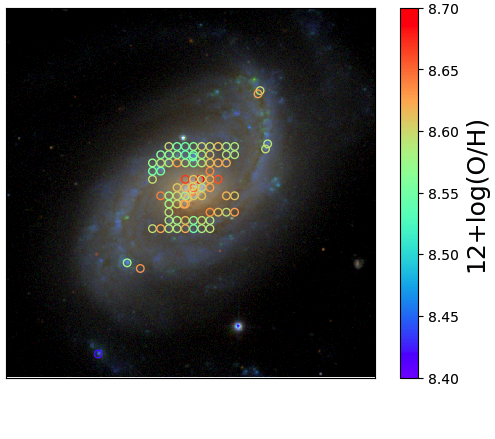}
\includegraphics[height=0.31\textwidth]{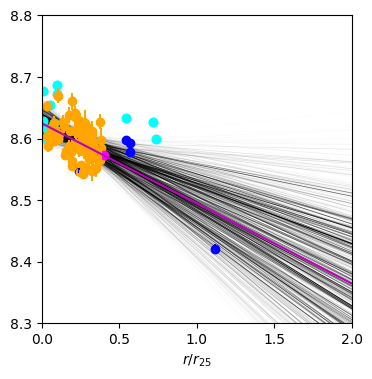}  
\includegraphics[height=0.3\textwidth]{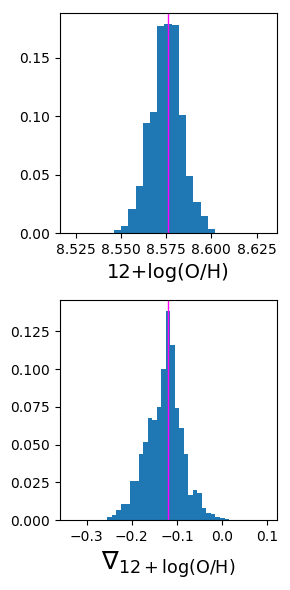}
\includegraphics[height=0.3\textwidth]{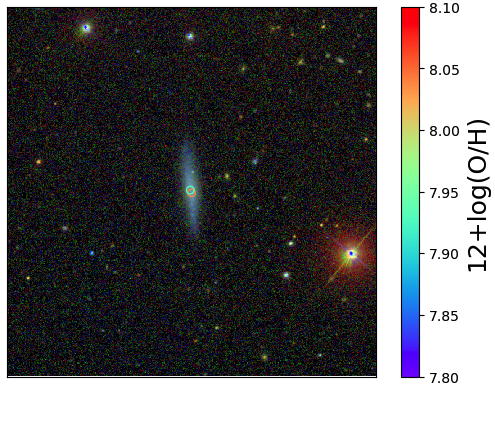}
\includegraphics[height=0.31\textwidth]{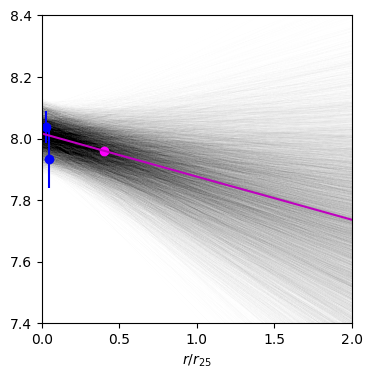}  
\includegraphics[height=0.3\textwidth]{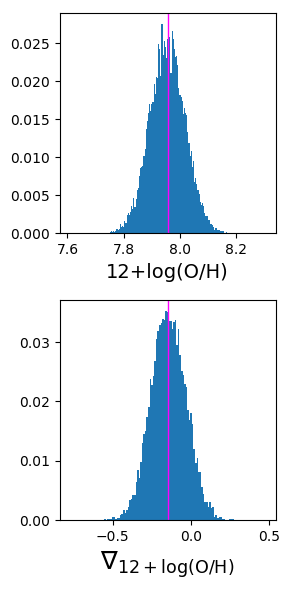}
  \caption{\textit{Left}: $4.5\times4.5$ arcmin$^2$ SDSS gri colour image together with the distribution of metals (coloured circles) in NGC5248 (well sampled; \textit{top}) and UGC00931 (sparsely sampled; \textit{bottom}). \textit{Middle:} Radial metallicity profile for both galaxies. MUSE data points are shown in orange, literature data with measured uncertainties in blue and literature data without measured uncertainty (uncertainty of 0.2 dex was assigned) in cyan. The magenta line gives the best radial fit and the black lines give the 30000 individual fits attempted in our Bayesian approach, weighted by their probability from Eq. \ref{probdatapoint}. \textit{Right:} The resulting PDFs for the gradient $\nabla_{\rm12+log(O/H)}$ and characteristic metallicity. The median is shown in magenta (see also middle panel) and is used for the remainder of this work. }
  \label{gradfig}
\end{figure*}

Many of the galaxies in our sample have multiple spectra or even IFU data available (e.g. Figure \ref{gradfig}; \textit{left}), yet do not have integrated emission lines. It is thus not trivial to determine the best global oxygen abundance. Given the importance of the global metallicity in scaling relations and chemical evolution modelling, we aim to derive the most reliable integrated oxygen abundance possible. One very useful relationship that has been identified in the literature \citep{Kobulnicky1999,Pilyugin2004,Moustakas2006,Moustakas2010} is that the luminosity-weighted integrated metallicity is statistically consistent with the characteristic abundance, which is defined as the oxygen abundance at a radius of $r=0.4\times r_{25}$, where $r_{25}$ is the radius at which the B-band surface brightness reaches 25 mag/arcsec$^2$. When integrated emission lines are not available, we use the oxygen abundance at $0.4\times r_{25}$ as the global metallicities. The best estimate of the oxygen abundance at $0.4\times r_{25}$ is given by performing a linear fit to the radial oxygen abundance ($12+\log(O/H)$) profile. In this work, we define the metallicity gradient as
\begin{equation}
\nabla_{\rm12+log(O/H)} \equiv \frac{{\rm d}\log({\rm O/H})}{{\rm d} r}
\label{gradient}
.\end{equation}
As many of the galaxies in this work do not have enough metallicity measurements to derive a reliable gradient, we use a Bayesian approach (see also Clark et al. in prep. for more details) to determine the most likely gradient for each galaxy. 

Within our Bayesian framework we used a Gaussian prior for the radial gradient, intercept and intrinsic scatter. These priors will be different for each galaxy, and we used slightly different approaches for poorly sampled and well sampled galaxies. We refer the reader to Appendix \ref{priors} for further detail. In short, we use individual priors based on the available data for well sampled galaxies, yet for poorly sampled galaxies we use a prior based on the average gradients for the well sampled galaxies (Table \ref{gradtable} in Section \ref{radialgrad}). In order to determine the best gradient and characteristic metallicity for each galaxy, we generate 30000 samplings from the combined priors. Each of these samplings corresponds to a radial metallicity profile (Figure \ref{gradfig}; \textit{middle}) and are compared to the observed metallicities by computing $\chi^2$. We then built probability density functions (PDF) by assigning a probability to each sampling as
\begin{align}
\label{probdatapoint}
P&=P\ (\nabla_{\rm12+log(O/H)} ,intercept,scatter_{int}) \nonumber \\
P &\propto \exp\left( \chi^2 \right)   \\ 
                                              &\propto \exp\left( \sum_{i} \left[-\frac{(y_i-[\nabla_{\rm12+log(O/H)}  \times r_i-intercept])^2}{2(\sigma_{y_i}^2+scatter_{int}^2)}\right] \right), \nonumber                                         
\end{align}
where the sum is over all the regions in the galaxy, $\nabla_{\rm12+log(O/H)} $ is the gradient, $intercept$ is the intercept, $y_i$ is the $12+\log(O/H)$ oxygen abundance for each region, $\sigma_{y_i}$ is the corresponding uncertainty (see Section \ref{errors}), and $scatter_{int}$ is the intrinsic scatter in metallicity between the different regions. For the gradients, we then make a histogram from -0.8 to 0.55 with steps 0.01 and assign the probability to the appropriate bin. In this way, we obtained a PDF (Figure \ref{gradfig}; \textit{right}). At the same time a PDF for the characteristic metallicity is made between $7<12+log(O/H)<10$ with steps of 0.001.

The best gradient and characteristic metallicity for each galaxy are then taken as the 50$\rm ^{th}$ percentile (i.e. the median, where the cumulative probability reaches 50\%) of their respective PDFs and the uncertainties as the 16th and 84th percentiles. We note that this method even works for sources with only very low signal-to-noise ratio (S/N) regions and even for sources with only one pointing. This is important as 176 of our galaxies have only a central pointing, which is on average overestimated compared to the global metallicity. By using the prior on the gradients, sources with one pointing just end up using the most likely gradient for the overall sample, and the uncertainty on the gradient is propagated to an uncertainty on the characteristic abundance.  

\begin{figure}
  \center
    \includegraphics[width=\columnwidth]{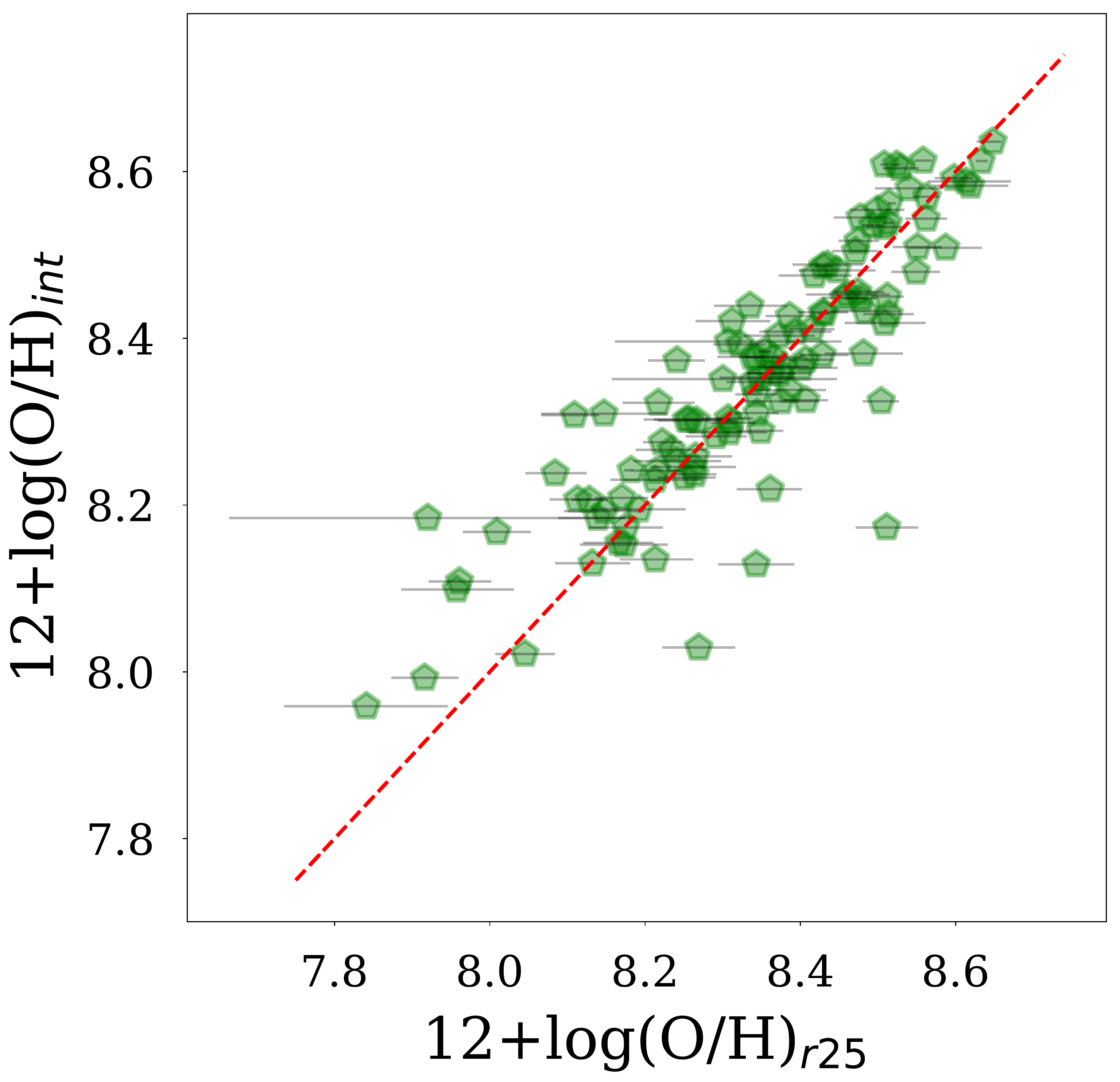}
    \includegraphics[width=\columnwidth]{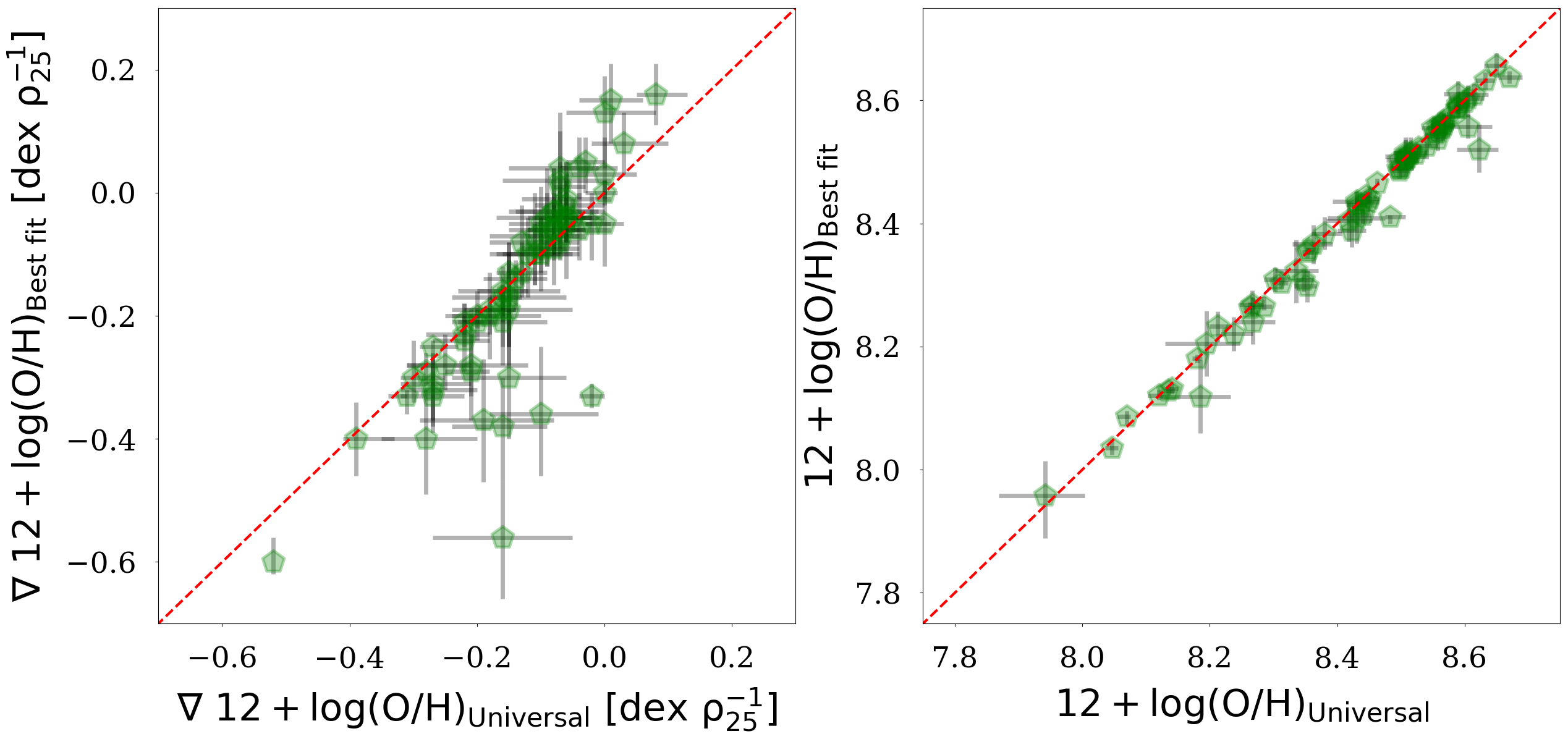}
  \caption{\textit{Top:} Comparison of the characteristic metallicity 12+log(O/H)$_{r25}$ (Bayesian estimate of the metallicity at a radius of $0.4\ r_{25}$) and the metallicity from integrated spectroscopy 12+log(O/H)$_{int}$ for all the galaxies where both estimates are available. The red dashed line gives the one-to-one relation. \textit{Bottom:} Comparison of the gradients (\textit{left}) and characteristic metallicities (\textit{right}) for the well constrained sample, comparing the best fit method that is used in this work, to the `universal prior' method that is used for the unconstrained sample (though here applied to the well constrained sample). The red dashed line gives the one-to-one relation. The characteristic metallicities are very independent on the method, though the `universal prior' gradients are somewhat biased towards the mean.}
  \label{compGlobalZ}
\end{figure}

We compare our global metallicities estimated from IFU and fibre metallicities to measured global metallicities for the 110 sources which have both estimates available. The resulting comparison is shown in Figure \ref{compGlobalZ}. There is a good match between the two estimates, indicating that our method performs well. Finally, we also tested using the `universal prior' for the gradients for all galaxies, rather than just for the unconstrained sub-sample. When the results are compared, we find the effect on the global metallicities is negligible, though there are some significant offsets on the resulting gradients, where the `universal prior' results are biased towards the mean of the sample\footnote{When the `universal prior' is used, the comparison with other samples is less good (Section \ref{Raddiscussion}).}. Therefore, we consider our global metallicities to be reliable, even for the unconstrained sub-sample. However, the measured gradients for the unconstrained sub-sample are likely to be unreliable and we used them no further. 

\section{Other galaxy properties}
\label{galaxysec}
\subsection{DustPedia}
\label{galaxyprop}
Dust masses, stellar masses and star formation rates (SFR) for the DustPedia galaxies will be presented in Nersesian et al. (in prep.). These results are derived from the aperture-matched DustPedia Photometry presented in \citet{Clark2018} using the SED fitting package CIGALE \citep{Noll2009}. CIGALE creates a library of SED templates for which the energy balance between the energy absorbed by dust in the UV-optical and the energy re-emitted in the infrared is maintained. This library is build assuming a delayed and truncated SF history \citep{Ciesla2016} along with stellar emission from the stellar population models of \citet{Bruzual2003} and a Salpeter initial mass function. For the dust emission, the THEMIS \citep{Jones2017} dust grain model is used, which is the reference DustPedia dust model. Using Bayesian statistics the best values and uncertainties for the $M_d$, $M_*$ and SFR are determined. 

To obtain \hi\ masses for the DustPedia sample we have performed a literature compilation (see also Casasola et al. in prep.). Integrated \hi\ fluxes were found for 764 out of the 875 DustPedia galaxies. The various references used for this compilation are listed in Table \ref{HIrefs}. 569 sources were found with \hi-detections and listed uncertainties, 96 galaxies have \hi-detections but no listing for the uncertainty and 99 galaxies have upper limits available only. When multiple references were available, preference was given to the measurements with the smallest uncertainty. There are 67 ALFALFA sources for which the DustPedia aperture from \citet{Clark2018} is larger than $\rm 5^{\prime}$ and there might thus be some \hi\ flux outside the ALFALFA aperture. The \hi\ flux for these sources was therefore corrected to account for \hi\ outside of the ALFALFA aperture using the \hi-profile from \citet{Wang2014}, who found the \hi\ discs of galaxies exhibit a homogeneous radial distribution in their outer regions, and provide scaling relations between the scale length and $\MHI$.

This method of correcting the ALFALFA \hi\ flux seemed to give more reliable results than using alternative observations or methods available in the literature (which often differed by a factor of 1.5 or more). We note that the upper limits are not calculated consistently between all the different references. The most realistic $5\sigma$ upper limit should be given by $f_{uplim}=5\ rms_{ch}\ \sqrt{W_{gal}\ W_{ch}}$, where $rms_{ch}$ and $W_{ch}$ are the rms (in Jy) and width (in km/s) for a given channel, and $W_{gal}$ is the estimated width of the \hi-line. Some of the references in our compilation do not take into account the channel width. We have not used upper limits in this work, but include them (as published) in our compilation for completion. \hi\ fluxes were converted to $\MHI$ using:
\begin{equation}
\MHI=2.36\times 10^5\   f_{\text{H{\sc i}}} \ D^2 
,\end{equation}  
where $f_{\text{H{\sc i}}}$ is the compiled \hi\ flux in Jy km s$^{-1}$ and $D$ is the best distance measure from \citet{Clark2018} in Mpc.
To obtain gas masses, we first added $\rm H_2$ to the \hi. We refer the reader to Casasola et al. (in prep.) for a study of the $\rm H_2$ content of DustPedia galaxies. Unfortunately, we do not have global $\rm H_2$ masses available for all our galaxies, so instead we use a scaling relation between the $\rm H_2$-to-\hi\ ratio and the \hi-to-stellar mass ratio to estimate global $\rm H_2$ masses. Casasola  et al. (in prep.) find the following relation. 
\begin{equation}
\rm log(M_{H_2}/\MHI) = -0.72 \ log(\MHI/M_*) - 0.78 
\label{H2HI}
\end{equation}  
The gas mass is then determined as
\begin{equation}
M_{g} = \xi\ \MHI\  (1 + M_{H_2}/\MHI) 
\label{gastot}
\end{equation}  
where $M_{H_2}/\MHI$ is taken from Casasola  et al. (in prep.) when detections are available and estimated using Eq. \ref{H2HI} if not. $\xi$ is a correction factor to account for the fraction of the gas that is made up of elements heavier
than hydrogen. We follow \citet{Clark2016} and define $\xi$ as
\begin{equation}
\xi = \frac{1}{1-(f_{\rm He_p}+f_Z[\frac{\Delta f_{\rm He_p}}{\Delta _Z}])-f_Z}
\end{equation}

where $f_{\rm He_p}$ is the primordial Helium mass fraction of 0.2485 \citep{Aver2011}, $f_Z = Z \times f_{Z_\odot}$ is the fraction of metals by 
mass and $\frac{\Delta f_{\rm He_p}}{\Delta _Z} = 1.41 $ \citep{Balser2006} is the evolution of the helium mass fraction with metallicity.
The correction factor varies from $\xi=1.33$ for zero metallicity to $\xi=1.39$ for solar metallicity. The measurement uncertainties on $M_{H_2}/\MHI$ and the uncertainties on the estimated $M_{H_2}/\MHI$ (about 0.5 dex) are propagated into the uncertainty on $M_d/M_g$. This uncertainty on $M_{H_2}/\MHI$ often dominates the total uncertainty.  

In this work, we followed \citetalias{DeVis2017b} in using gas fraction ($M_{g}/(M_{g}+M_*)$) as a rough proxy for evolutionary stage. Due to inflows and outflows of gas, there is not necessarily a monotonic relation between the gas fraction and the evolutionary stage of a galaxy. Even so, gas fraction remains a good tracer of the evolutionary stage as it is a measure of how much future star formation can currently be sustained, relative to the past star formation. 

\subsection{Comparative samples}
\begin{table*}
\caption{Summary of the parameters used for the three chemical evolution models used in this work. See \citetalias{DeVis2017b} for further details.}
\begin{center}
\begin{tabular}{lccccccc} \hline\hline
Name & SFH & Reduced SN dust &  Destruction &  Grain growth &Inflow &Outflow \\ \hline
Model I & Milky Way & N & N &  N  & N  & N \\
Model V & Delayed & $\times 12$ & $m_{\rm ISM}=1500$ &  $\epsilon=5000$ &  $2.5\times$ SFR &  $2.5\times$ SFR \\ 
Model VI & Delayed/3 & $\times 100$ &  $m_{\rm ISM} = 150$ &  $\epsilon = 8000$ &  $2.5\times$ SFR &  $2.5\times$  SFR \\ \hline
\end{tabular}
\end{center}
\label{chemevmodels}
\end{table*}

Due to its 5$\sigma$ detection in the WISE W1 band and diameter ($D25>1'$) selection criteria, DustPedia is somewhat biased against dwarf galaxies. Yet it is exactly these dwarf galaxies that have the lowest metallicities, which makes them a key stage to observe the evolution in the dust-to-metal ratios. Therefore, we increase our statistics at the low-metallicity end by adding the Dwarf Galaxy Survey (DGS; \citeb{Madden2013}), the dust-selected HAPLESS \citep{Clark2015} and \hi-selected \high\ \citep{DeVis2017a} sample to our study. Only sources that are not in the DustPedia sample are added. Galaxy properties for each of these samples were compiled in \citetalias{DeVis2017b}. We used their metallicities, SFR, stellar masses, and \hi\ masses as published, though their dust masses were derived with MAGPHYS \citep{daCunha2008} rather than CIGALE, and thus do not use the THEMIS dust model. MAGPHYS uses a dust mass absorption coefficient of $\kappa_{850}= 0.077 {\rm m^2 kg^{-1}}$, and the THEMIS dust mass absorption coefficient is well described by $\kappa_{\lambda} = 0.64 \times  (250/ \lambda)^{1.79}  {\rm m^2 kg^{-1}}$ for $\lambda$ in $\mu m$ \citep{Galliano2018}. We therefore scaled the \citetalias{DeVis2017b} dust masses down by a factor of 1.075. Total gas masses for these samples were calculated using the \citetalias{DeVis2017b} \hi\ masses and Equation \ref{gastot}.    

In addition, we compared our observations to some of the chemical evolution models presented in \citetalias{DeVis2017b}. In summary, these models track the global gas, stellar, metal, and dust content of a galaxy as gas is converted into stars (using a Chabrier IMF) as a result of a given star formation history. Dust and metals are expelled into the ISM when stars reach the end of their life at an age appropriate for their mass. The models separately track the oxygen (used for $12+\log(O/H)$) and total metal content (total mass of metals). The models also include prescriptions for inflows and outflows (proportional to the SFR), dust destruction and dust grain growth. The dust mass evolution is described by
\begin{align}
\frac{d(M_d)}{dt}&=\left(\frac{d(M_d)}{dt}\right)_{stellar}-\left(\frac{d(M_d)}{dt}\right)_{astr}-\left(\frac{d(M_d)}{dt}\right)_{destr}\nonumber \\
     &+ \left(\frac{d(M_d)}{dt}\right)_{gg} +\left(\frac{d(M_d)}{dt}\right)_{inf}-\left(\frac{d(M_d)}{dt}\right)_{outf}.
\label{eq:dustmasst}
\end{align}
The first term accounts for dust formed in stars and supernovae. The second term describes the removal of dust due to astration (destruction of dust mixed with the gas that is consumed during star formation) and the dust destruction and grain growth are given in terms three and four. The fifth and sixth term are simple parameterisations of dust contributed or removed via inflows and outflows. In more detail, the stellar term integrates over all stellar masses using a Chabrier IMF, the lifetimes of the stars at each mass, and LIMS and SN dust yields for each mass. 
The dust destruction term is due to SN-driven shocks and is described by
\begin{equation}
\left(\frac{d(M_d)}{dt}\right)_{destr}=\left(1-f_c\right) \frac{M_d}{\tau_{\rm dest}}
.\end{equation}
Here $f_c$ gives the fraction of the dust that is in the dense phase, which we have set to 50\%. The timescale for dust destruction ($\tau_{\rm dest}$, following \citealt{Dwek2007}) is described as a function of the rate of SN ($R_{\rm SN}$):
\begin{equation}
\tau_{\rm dest} = \frac{M_g}{m_{\rm ISM}R_{\rm SN}(t)},
\label{destr}
\end{equation}
where $m_{\rm ISM}$ is the mass of ISM that is swept up by each individual SN event. 
The following prescription is used for the dust grain growth:
\begin{equation}
\left(\frac{d(M_d)}{dt}\right)_{gg}=f_c \frac{M_d}{\tau_{\rm grow}}
.\end{equation}
The dust grain growth timescale uses the prescription of \citet{Mattsson2012} and is given by 
\begin{equation}
\tau_{\rm grow} = \frac{M_g}{\epsilon \, Z \, SFR}\  \left(1 - \frac{M_d}{M_Z}\right)^{-1}
\label{growth}
\end{equation}
where Z is the fraction of heavy elements, $M_Z$ is the mass of metals, and $\epsilon$ is a dimensionless free parameter which is varied to set the appropriate rate of dust grain growth. The inflows and outflows of gas are simple parametrisations proportional to the SFR. The inflows in this work are pristine gas and the inflow term in Equation \ref{eq:dustmasst} thus becomes zero. The outflows drive out dust at a rate which is the product of the current dust-to-gas ratio and the gas outflow rate. In this work we compare the DustPedia observations to three models from \citetalias{DeVis2017b}. The parameters used for these three models are given in Table \ref{chemevmodels}. The aim here is not to find the best model or to constrain the dust evolution parameters, but rather to put our DustPedia observations into context and to provide a basic interpretation for the observed trends. The DustPedia data will be used in future work to constrain a next generation of chemical evolution models.

Model I is the simplest model, with only stellar source of dust and no inflows or outflows. It uses a Milky Way-type exponentially declining SFH. The dust yield per SN is mass and metallicity dependent and is taken from \citet{Todini2001}. This is not the most realistic model, but is included here to illustrate the failures of such a simple model. Models V and VI are more realistic and include prescriptions for dust grain growth, dust destruction, delayed SFH and inflows and outflows. Both Model V and VI have outflows with constant mass loading factors (outflow $\propto 2.5\times$ SFR) and inflows at the same rate as the outflows. Model V has stronger dust destruction ($m_{\rm ISM} = 1500\,\rm M_{\odot}$; \citealp{Dwek2007}), consistent with dust destruction in the diffuse ISM. For Model VI $m_{\rm ISM} = 150\rm \,M_{\odot}$, indicative of SN shocks ploughing into typical interstellar densities of $10^3\,\rm cm^{-3}$ (\citealt{Gall2011,Dwek2011}). Additionally, Model VI has faster grain growth (higher $\epsilon$) than Model V. Finally, the \citet{Todini2001} dust yield per SN is also reduced in both models. This reduction in dust yield is likely necessary to account for the reverse shock in the remnants of SN \citep[e.g.][]{Bocchio2016}. For Model V the dust yield is reduced less than for Model VI. We note that there is a degeneracy between the dust destruction (Eqn. \ref{destr} and reduced dust yield) and dust grain growth, and as a result both Model V and VI provide reasonably well fitted models (though further work on improving the models is underway). 

\section{Results}
\label{resultssec}
\subsection{Radial gradients}
\label{radialgrad}
\begin{figure*}
  \center
\includegraphics[height=0.4\textwidth]{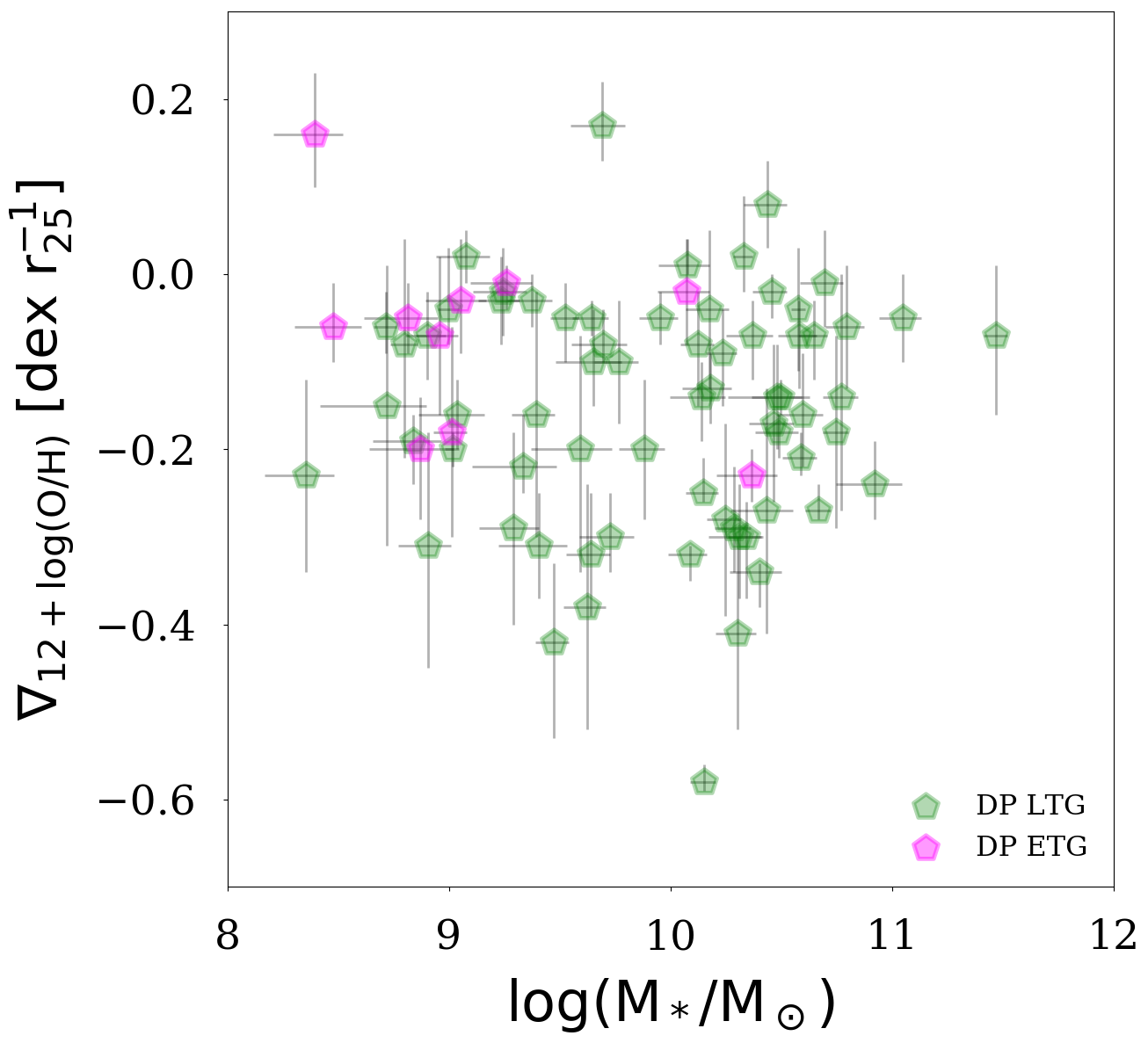}
\includegraphics[height=0.4\textwidth]{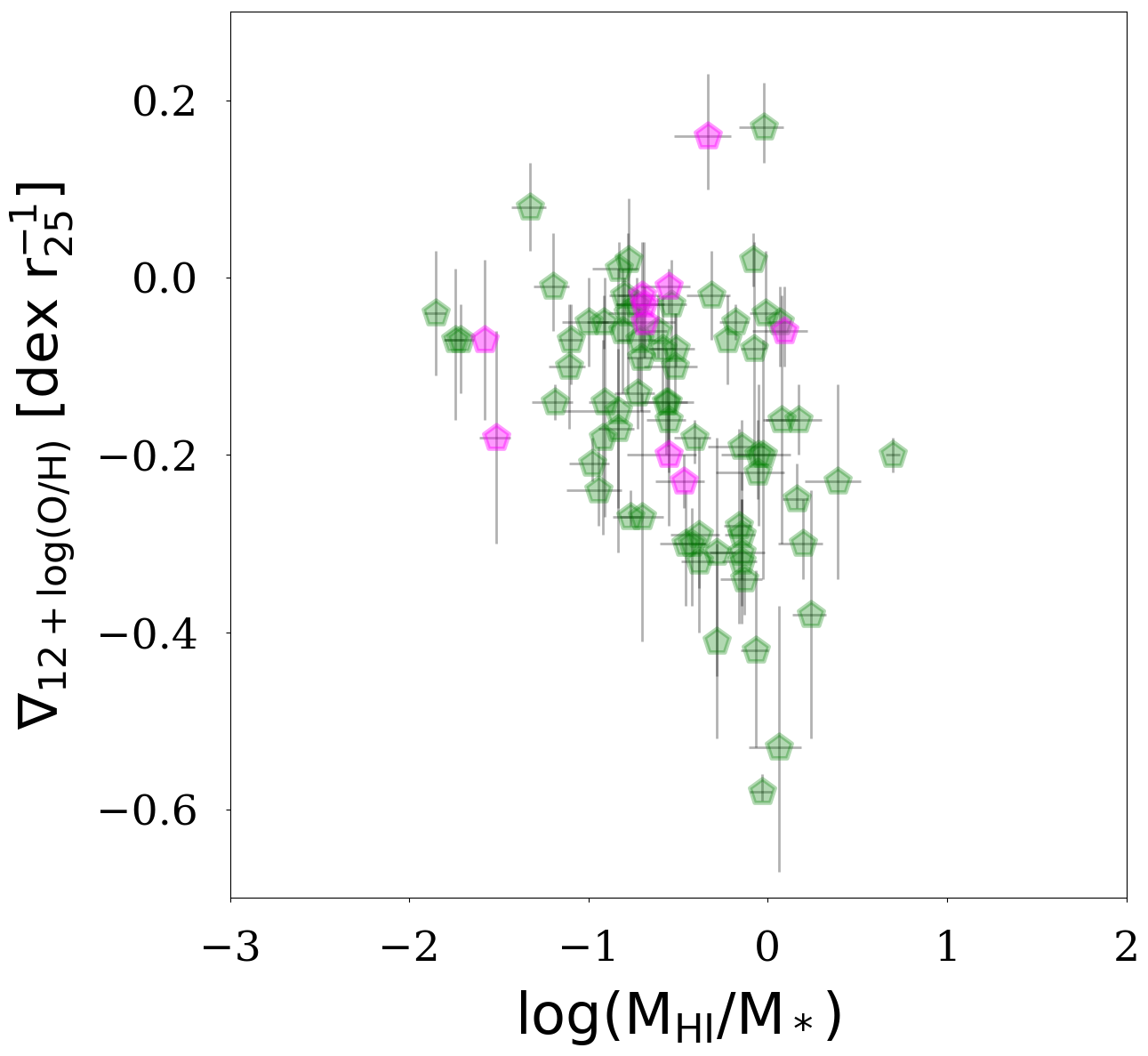}
\includegraphics[height=0.4\textwidth]{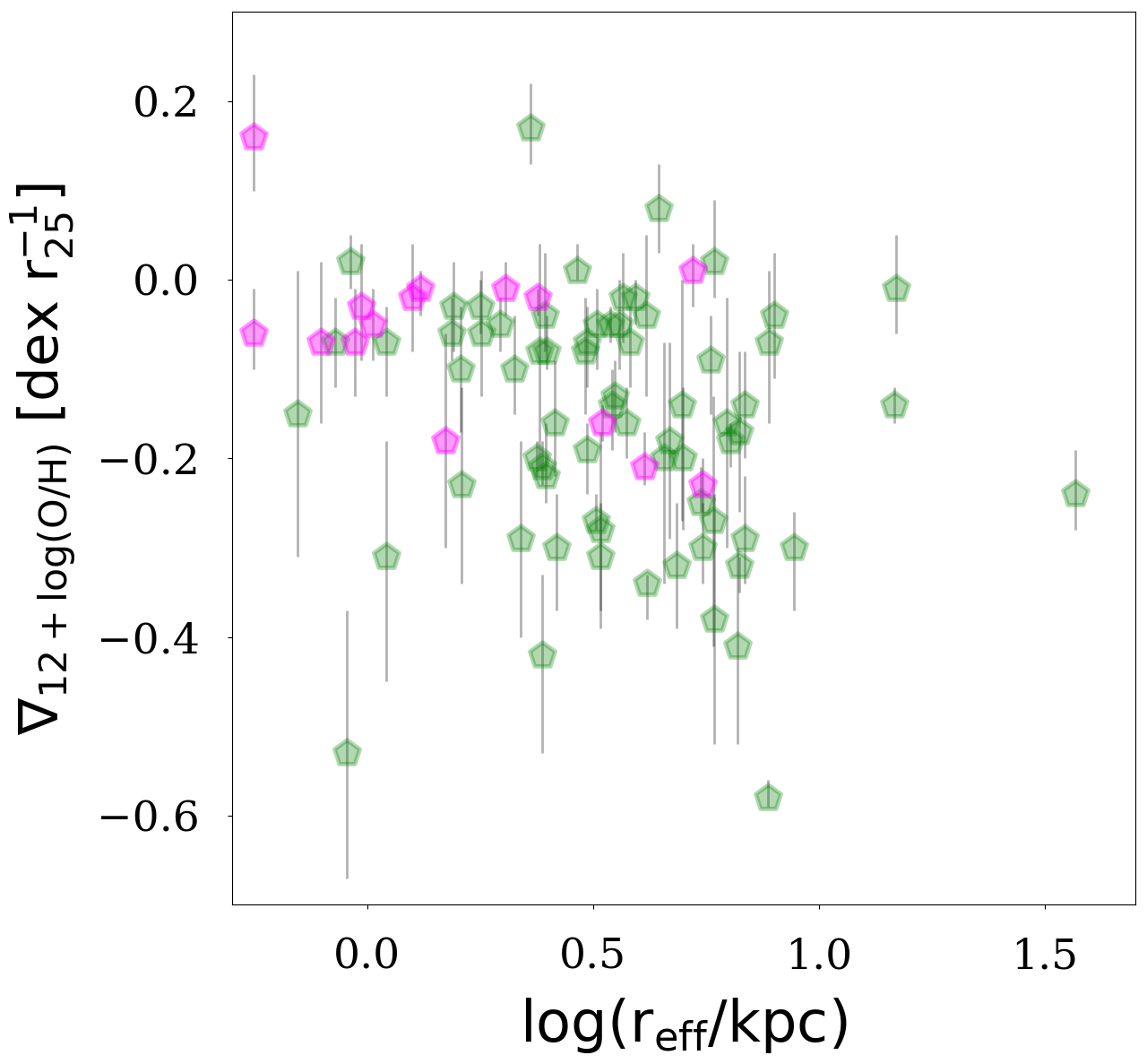}  
\includegraphics[height=0.4\textwidth]{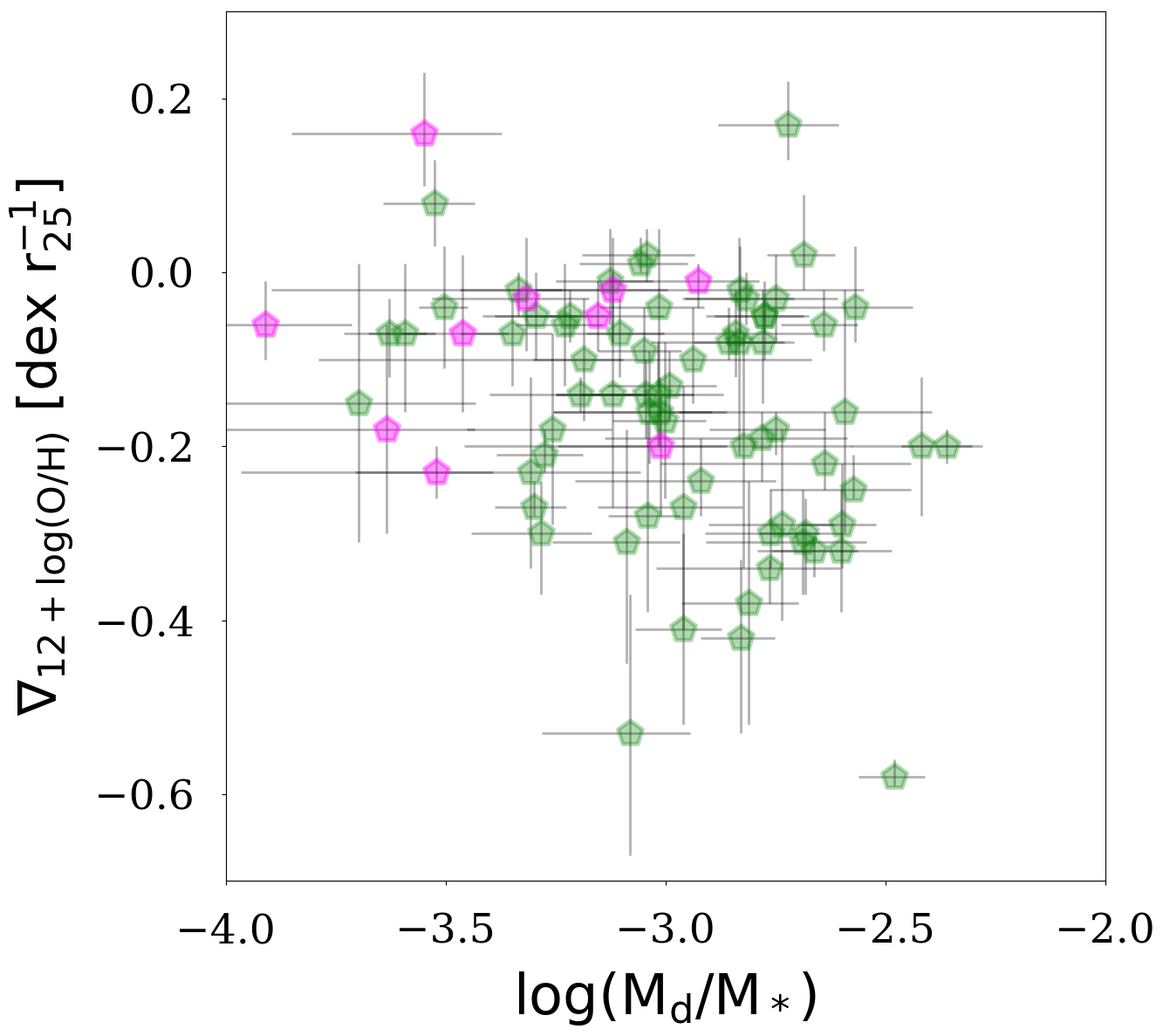}
  \caption{Radial metallicity gradients plotted against stellar mass (\textit{top left}), the \hi-to-stellar mass ratio (\textit{top right}), the effective radius from a single S\'ersic model fit \citep{Mosenkov2018} in kpc (\textit{bottom left}) and the dust-to-stellar mass ratio (\textit{bottom right}). The DustPedia sample is divided in LTG (green) and ETG (magenta).  We find weak correlations.}
  \label{gradfig2}
\end{figure*}

\begin{table}
\caption{Average gradients and standard deviations between the gradients for each of the calibrations used in this work.}
\begin{center}
\begin{tabular}{lcc} \hline\hline
Calibration & $\langle \nabla_{\rm12+log(O/H)} \rangle$ & $\sigma_{\nabla_{\rm12+log(O/H)}}$ \\ \hline
PG16S & -0.15 & 0.16 \\
PG16R &  -0.20 & 0.21 \\
N2 & -0.09 & 0.21 \\
O3N2 & -0.16 & 0.25 \\
IZI &  -0.17  & 0.26 \\
KK04 & -0.24 & 0.25 \\
T04 & -0.18 & 0.29 \\
$\rm log(N/O)$  &  -0.38 & 0.34 \\ \hline 
\end{tabular}
\end{center}
\label{gradtable}
\end{table}

Before studying the global DustPedia metallicities, we investigated in this section whether the gradients within our well-sampled galaxies correlate with any other galaxy properties. The average gradients and standard deviations for each calibration are given in Table \ref{gradtable}. There are significant differences between the different calibrations (see Section \ref{Raddiscussion}). In Figure \ref{gradfig2} (\textit{left}), we have plotted how the gradients depend on stellar mass for all the sources in the well constrained sub-sample (at least five data points covering a range of radii at least $0.5\ r_{25}$ wide). No significant correlation is found (Spearman's rank correlation coefficient $\rho=-0.007$ ). This is consistent with the findings of \citet{Sanchez2012,Sanchez2014,Ho2015,Sanchez-Menguiano2016}. However, we refer the reader to Section \ref{Raddiscussion} where we show that this result depends on the calibration used. 

When the parameter-space is further explored, we find a number of galaxy properties which yield a stronger (though still relatively weak) correlation with the metallicity gradient than stellar mass. When the gradient is plotted against $\MHI/M_*$, Spearman's rank correlation coefficient increases to $\rho=-0.415$. With respect to the dust properties, the strongest correlation with the metallicity gradient is found with $M_d/M_*$ (Spearman $\rho=-0.271$). We also plot the gradient against the effective radius $r_{\text{eff}}$ (Spearman $\rho=-0.274$) from S\'ersic \citep{Sersic1963, Sersic1968} fits\footnote{These S\'ersic fits are based on WISE 3.4 $\mu$m imagery (i.e. they trace the old stellar population).} from \citet{Mosenkov2018}, we find that galaxies with the largest physical extent, have the strongest gradients\footnote{Using the effective radius instead of $r_{25}$ in the calculation of the gradients (i.e. using $r/r_{\text{eff}}$ instead of $r/r_{25}$) yields a correlation that is not very different.}. \citet{Tortora2010} have found a similar correlation between the optical colour gradients and $r_{\text{eff}}$ in a sample of 50000 nearby SDSS galaxies. The metallicity gradients and colour gradients are likely an effect of the same mechanism. Hydrodynamical simulations suggest galaxies with a larger extent have larger rotational velocities relative to the velocity dispersion. This could result in less mixing of the gas and dust at different radii and thus steeper gradients.

\subsection{DustPedia oxygen abundances}
\label{DPoxabund}
\begin{figure}
  \center
    \includegraphics[width=\columnwidth]{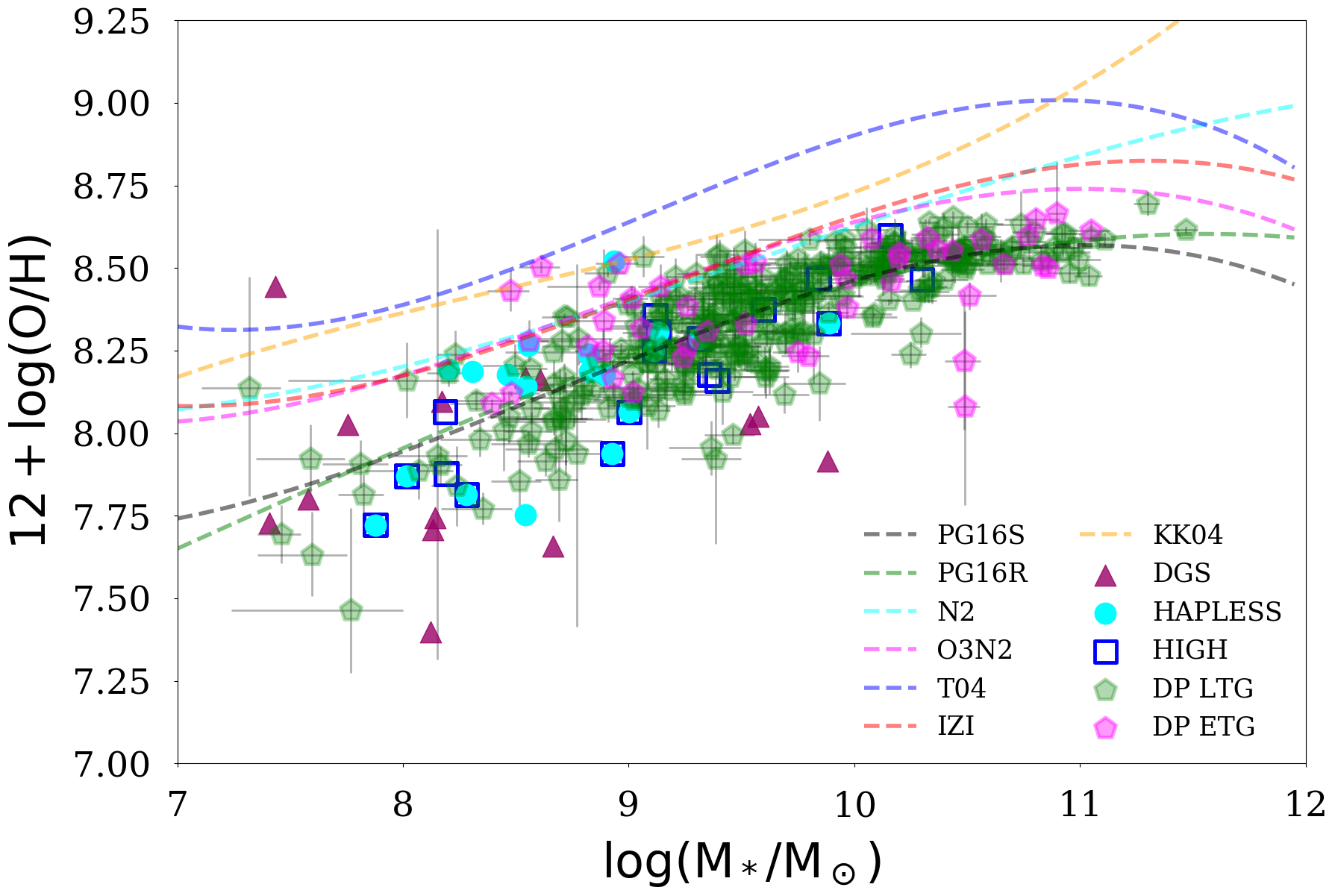}
  \caption{M-Z relation for the PG16S calibration used throughout this work. The DustPedia sample is divided in LTG (green) and ETG (magenta). The \high (blue), HAPLESS (cyan), and DGS (purple) samples are added to improve statistics at the low-metallicity end. Additionally, we show the best-fit M-Z relations (third order polynomials) for the different metallicity calibrations listed in Table \ref{evolved}. The y-axis offset for the various calibrations differs substantially, though the overall shape is similar.}
  \label{MZ}
\end{figure}

To highlight the differences between the various calibrations used in this work, we plot the M-Z relation for each calibration in Figure \ref{MZ}. We plot the global DustPedia metallicities using PG16S (our reference calibration; see section \ref{calibrations}), and add the \high, HAPLESS and DGS samples to obtain better statistics at the low-metallicity end. We also have global metallicities available for each of the different calibrations (not plotted as this would overcrowd the plot), and have fitted a third order polynomial to the metallicities for each of the different calibrations (plotted in dashed lines). The best-fitting third order polynomials are similar in shape, though have a different y-axis intercept. This discrepancy towards higher metallicities for theoretical calibrations (based on photo-ionisation models) is well known in the literature \citep[e.g. ][]{Kewley2008,Moustakas2010}. We provide metallicity calibration conversions between PG16S and the other calibrations in this work in Appendix \ref{convPG16}. 

\begin{figure}
  \center
\includegraphics[width=\columnwidth]{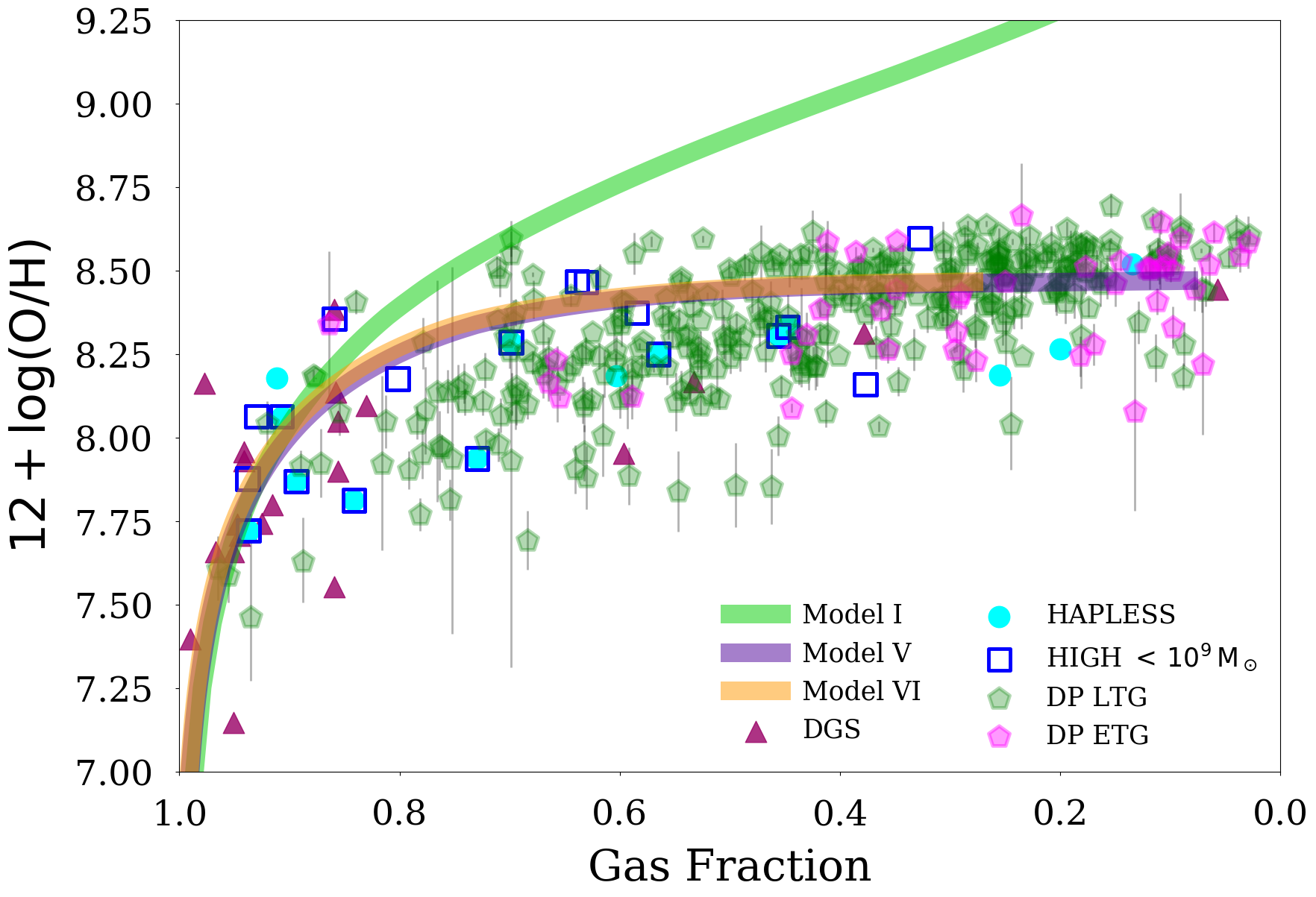}
  \caption{Evolution of metallicity with gas fraction for the various samples and chemical evolution models used in this work. The monotonic increase in metallicity is relatively well described by models including inflows and outflows, though stronger outflows are likely necessary at early evolutionary stages.}
  \label{Z_fg}
\end{figure}

Next we show how the metallicity increases as galaxies evolve\footnote{Gas fraction is used as a proxy for evolution as it is a good measure of how much star formation can be sustained from the current gas reservoir, compared to the star formation that has already happened. Though inflows and outflows of gas and mergers will also affect the gas fraction.} in Figure \ref{Z_fg}. The metallicity is found to increase monotonically with decreasing gas fraction. When the data are compared to chemical evolution models, we find that significant inflows and outflows are necessary to avoid significantly overestimating the model metallicity at low gas fractions (Models V and VI; \citetalias{DeVis2017b}). Using our larger DustPedia sample, it now becomes clear that Models V and VI still overestimate the metallicity for the earlier stages of evolution (gas fraction $>0.5$). This is likely because for these models, the strength of the inflows and outflows shows the same proportionality with SFR for all galaxies, whereas in reality low mass galaxies (which have higher gas fractions) will be affected by outflows more strongly than high mass galaxies because they have a weaker gravitational potential to counteract the outflows. The remaining mismatches between models and observations might be alleviated by including models with higher mass loading factors for low mass galaxies, which would decrease the model metallicity at early evolutionary stages. We are working on a new set of chemical evolution models which will include this dependency (De Vis et al. in prep.).

\begin{figure}
  \center
\includegraphics[width=\columnwidth]{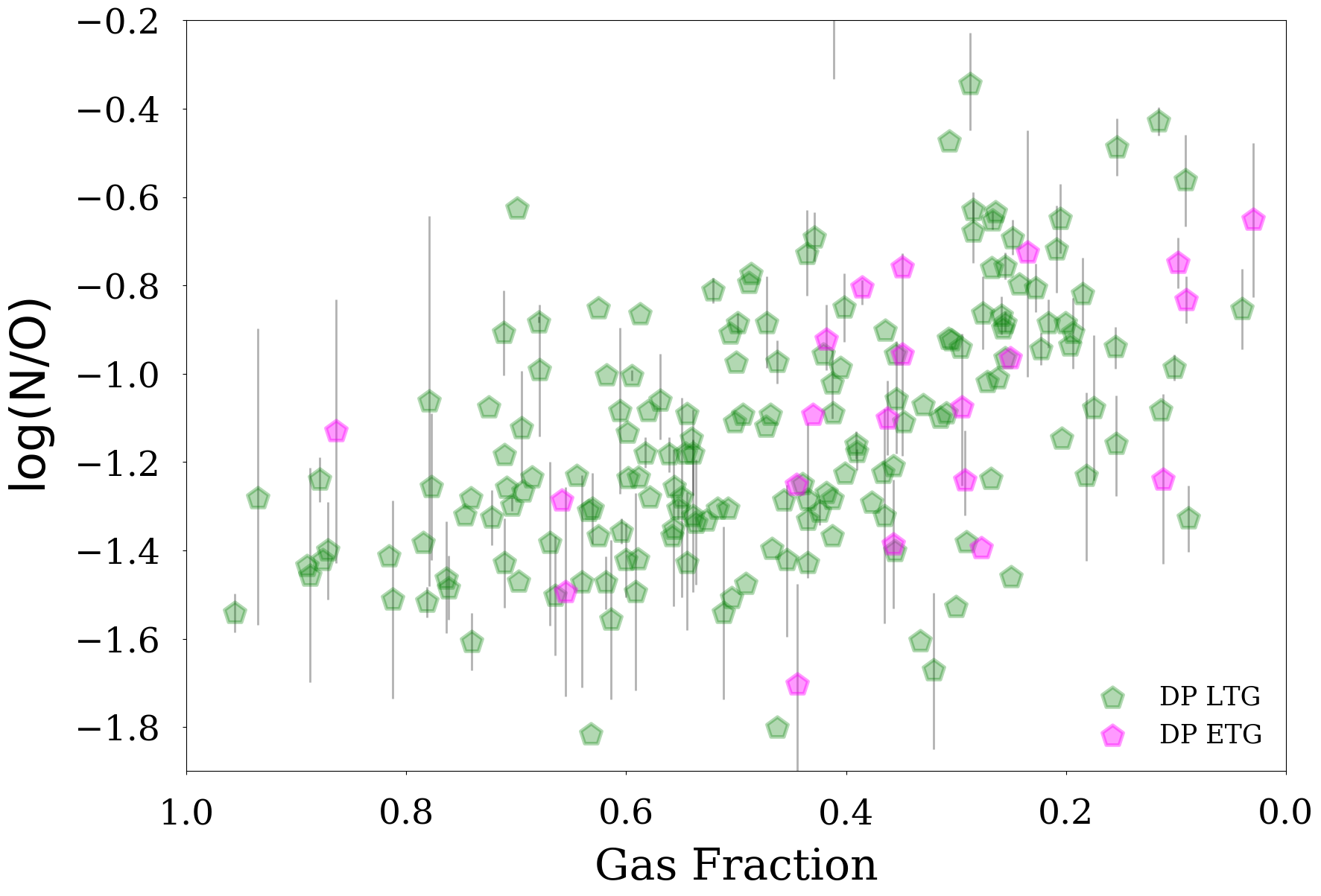}
  \caption{Variation in the nitrogen-to-oxygen ratio as a function of gas fraction for the DustPedia galaxies which have the necessary [OII]$\lambda3727,3729$ measurements available.}
  \label{NO_fg}
\end{figure}

In Figure \ref{NO_fg}, we show how the nitrogen to oxygen ratio (N/O) evolves with gas fraction. There is a significant correlation (Spearman's rank correlation coefficient $\rho=-0.590$). Here only DustPedia galaxies are shown as \citetalias{DeVis2017b} did not compile N/O ratios. The N/O  is found to be more or less steady around $\rm log(N/O) \sim -1.4$ until a gas fraction of 0.5 and then increases towards lower gas fractions. We attribute this to the fact that for high gas fractions, only primary N and O are available, yet for higher metallicities secondary (i.e. the yield depends on the previous amount of carbon and oxygen in the stars) N and O become available \citep{Perez-Montero2013}. The increasing N/O can be explained by the faster production rate of secondary N than for O \citep{Henry2000,Thuan2010}.
We note that many galaxies in this plot have no uncertainties available. Many of these estimates come from references which present integrated spectroscopy without uncertainties.  The lack of resolved data is due to the lack of [OII]$\lambda3727,3729$ emission line measurements (e.g. MUSE only provides spectra down to $\rm 4750\ \AA$) necessary for our N/O determination.

\subsection{Evolution in the dust-to-metal ratio}
\begin{figure}
  \center
\includegraphics[width=\columnwidth]{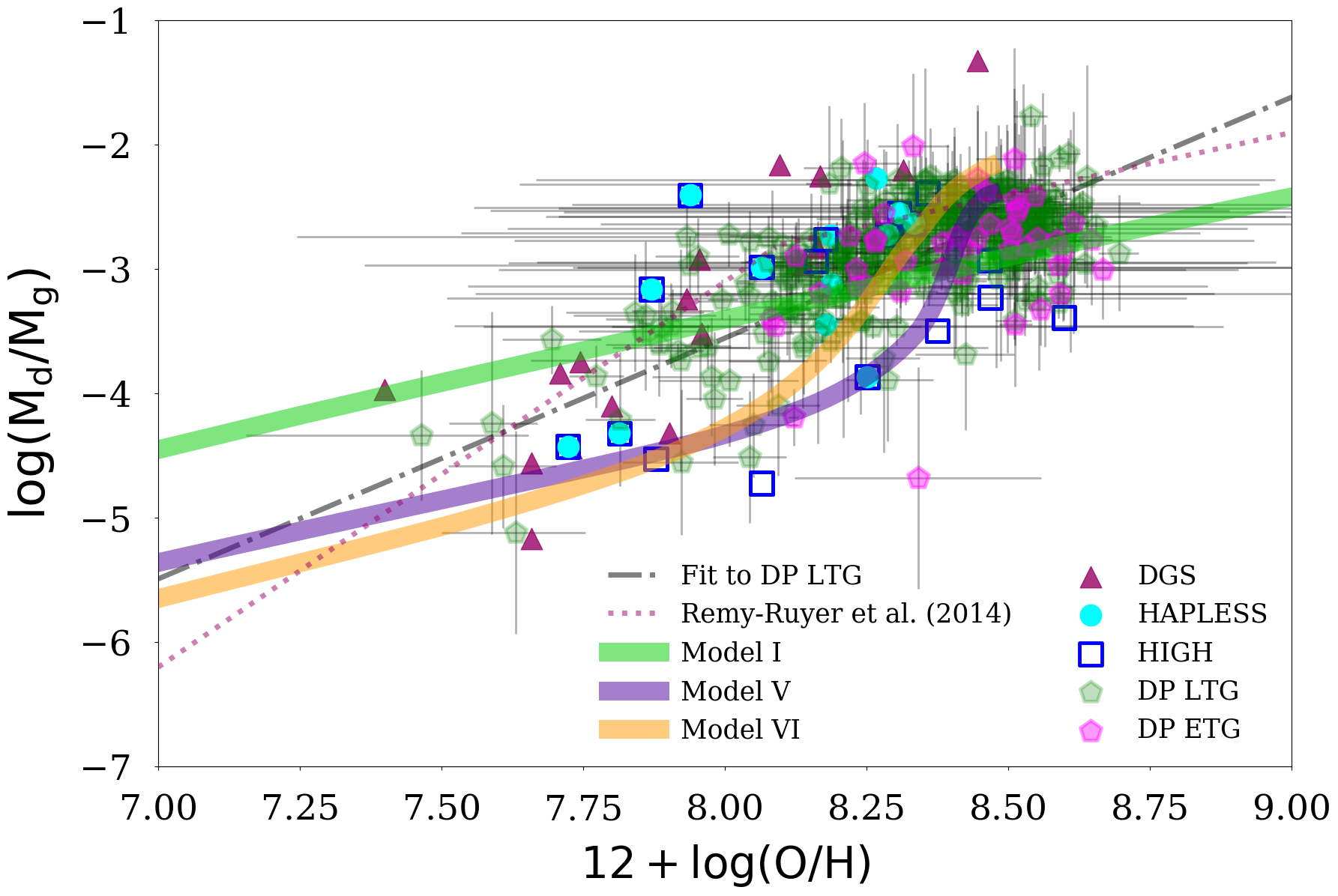}
  \caption{Dust-to-gas ratio is plotted against $\rm 12+log(O/H)$ (proxy for metal-to-gas ratio). The observations show a steeper increase than would be expected from stellar dust sources alone (Model I).}
  \label{mdmg}
\end{figure}

In this section, the dust and metal content are compared to better understand the processes driving the dust evolution. We start by plotting the dust-to-gas ratio against $\rm 12+log(O/H)$ (which by definition is a measure of the metal-to-gas ratio) for the 466 DustPedia galaxies which have all three of these measures available. In Figure \ref{mdmg}, we find an increasing dust-to-gas ratio with increasing metallicity. The dust-to-gas ratio increases more steeply than Model I of the chemical evolution models. Model I only includes stellar sources of dust, and thus has a constant dust-to-metal ratio. The only way to obtain a steep enough increase of the dust-to-gas ratio, is to include dust grain growth in the models (Model V and VI). When \citet{Remy-Ruyer2014} studied this relationship, they found a broken power law provides a good empirical fit to the data. We find this relation results in slightly higher dust-to-gas ratios than the average for our sample. Some of this offset might be due to the use of a different metallicity calibration\footnote{The \citet{Pilyugin2005} metallicity calibration used in this work cannot be reliably scaled into the different calibration methods used in this work \citep{Kewley2008}.}. The \citet{Remy-Ruyer2014} trend traces the DGS data at low metallicity, yet the observed DustPedia dust-to-gas ratios are lower in general. Our more complete sample of DustPedia galaxies is thus key to properly constrain models of the dust and chemical evolution of galaxies. 

We have fitted a power law and broken power law to the DustPedia\footnote{We have not included the DGS, HIGH and HAPLESS samples in our fit to avoid any potential bias from the different selection criteria or methods used.} LTGs for each of the metallicity calibrations using orthogonal distance regression (taking into account uncertainties on both the dust-to-gas ratios and metallicities). ETGs have not been included as dust destruction by hot gas sputtering might lower their dust-to-gas ratio (see also Figure \ref{selection} in Section \ref{dustometaldisc}). The results for single power laws are listed in Table \ref{powerlaws}. Contrary to \citet{Remy-Ruyer2014}, we find broken power laws do not provide a better description to the data as a single power law. When we followed \citet{Remy-Ruyer2014} in fixing the slope for high metallicities to one \citep[see also e.g.][]{James2002,Galliano2008}, the fits to the data are formally worse than the single power law for all calibrations except N2 and K04. If this slope is left free, the fits are only marginally better than the single power law (in spite of two additional free parameters), and the slope for high metallicities is consistent within the errors with the single power laws listed in Table \ref{powerlaws}. The single power laws thus provide the best description of the dust build up with increasing metallicity. For each of the calibrations, the relation is super-linear, indicating that stellar dust sources alone (which would result in a slope of one) cannot explain these relations. 

\begin{table}
\caption{Power law fits to the DustPedia LTGs for each of the metallicity calibrations used in this work. The slope $a$ and intercept $b$ are given together with their uncertainties ($\log(M_d/M_g)= a\ \times\ 12+\log(O/H)\ +\ b$).}
\begin{center}
\small
\begin{tabular}{lcccc} \hline\hline
calib. & $a$ & $\sigma_a$ &  $b$ & $\sigma_b$ \\ \hline
PG16S & 1.94 & 0.11 & -19.04 & 0.91 \\ 
PG16R &  1.83 & 0.15 & -18.13 & 1.22 \\ 
N2 & 1.62 & 0.10& -16.71 & 0.85 \\ 
O3N2&  1.77 & 0.10 & -17.96 & 0.86 \\
IZI&  1.45 & 0.10 & -15.30 & 0.82 \\ 
KK04 &1.22 & 0.14 & -13.50 & 1.26 \\
T04& 1.50 & 0.10 & -16.01 & 0.84\\ 
\hline
\end{tabular}
\end{center}
\label{powerlaws}
\end{table}

\citetalias{DeVis2017b} found a steep initial increase of the dust-to-gas ratio, followed by a more gradual increase (with constant dust-to-metal ratio). Using our larger DustPedia sample, there is no such clear break between these two regimes. The steep increase in these previous works could be explained by galaxies reaching the critical metallicity at which dust grain growth becomes effective. This steep increase is also seen in Models V and VI (at slightly different critical metallicities). However, in reality, the metallicity in a galaxy is not uniform (as is assumed in these models). The critical metallicity will be reached at different points in time for different regions in the galaxy (inside-out evolution, which also results in the observed metallicity gradients), and as a result the increase in the dust-to-gas ratio will be more gradual. Further resolved chemical evolution modelling \citep[such as][]{Aoyama2016,McKinnon2016,McKinnon2018} is necessary to study this behaviour in detail.

In what follows, we have studied the evolution of the dust-to-metal ratio with other galaxy properties. The mass of metals is calculated as $M_Z=f_Z \times M_g + M_d$ where $f_Z$ is the fraction of metals by mass calculated using $\rm f_Z=27.36\times 10^{(12+log(O/H)\  -\ 12)}$. The factor of 27.36 is found from assuming $\rm 12+log(O/H)_\odot = 8.69$ and a Solar metal mass fraction $Z_\odot = 0.0134$ following \citet{Asplund2009}. Throughout this work we track the total mass of metals in the ISM (i.e. including metals locked up in dust). 

In reality, the fraction of oxygen to the total mass of metals will not be constant throughout the evolution (see e.g. varying N/O in previous section). From studying the \citetalias{DeVis2017b} chemical evolution models where the mass of oxygen and total mass of metals are tracked separately, we find that these differences are of the order of $\sim 25\%$ (0.1 dex). It should be kept in mind that in what follows, we are essentially tracking changes in the dust-to-oxygen mass ratio (re-scaled to dust-to-metal ratio using a fixed oxygen-to-metal ratio) rather than in the intrinsic dust-to-metal mass ratio. 

\begin{figure}
  \center
\includegraphics[width=\columnwidth]{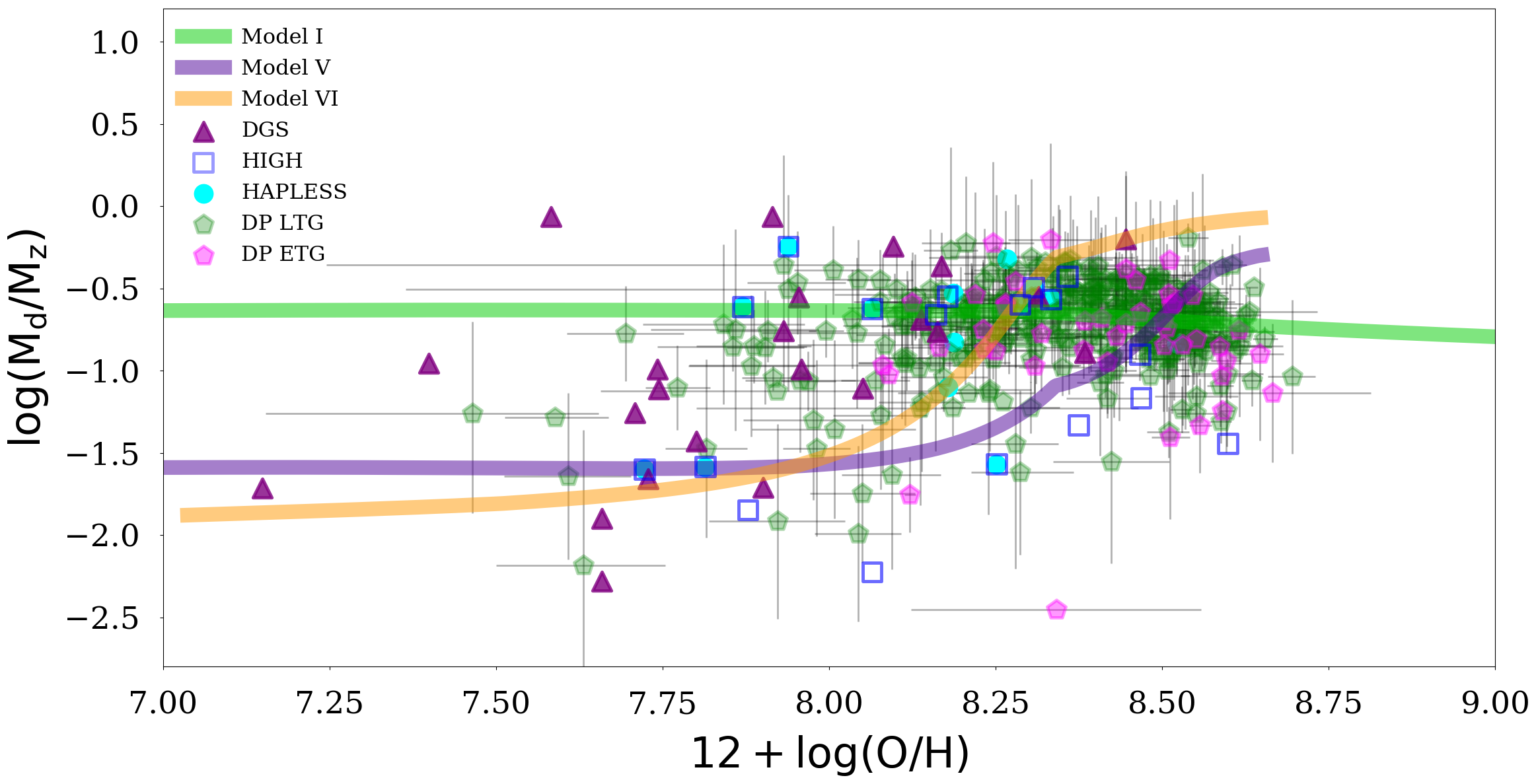}
\includegraphics[width=\columnwidth]{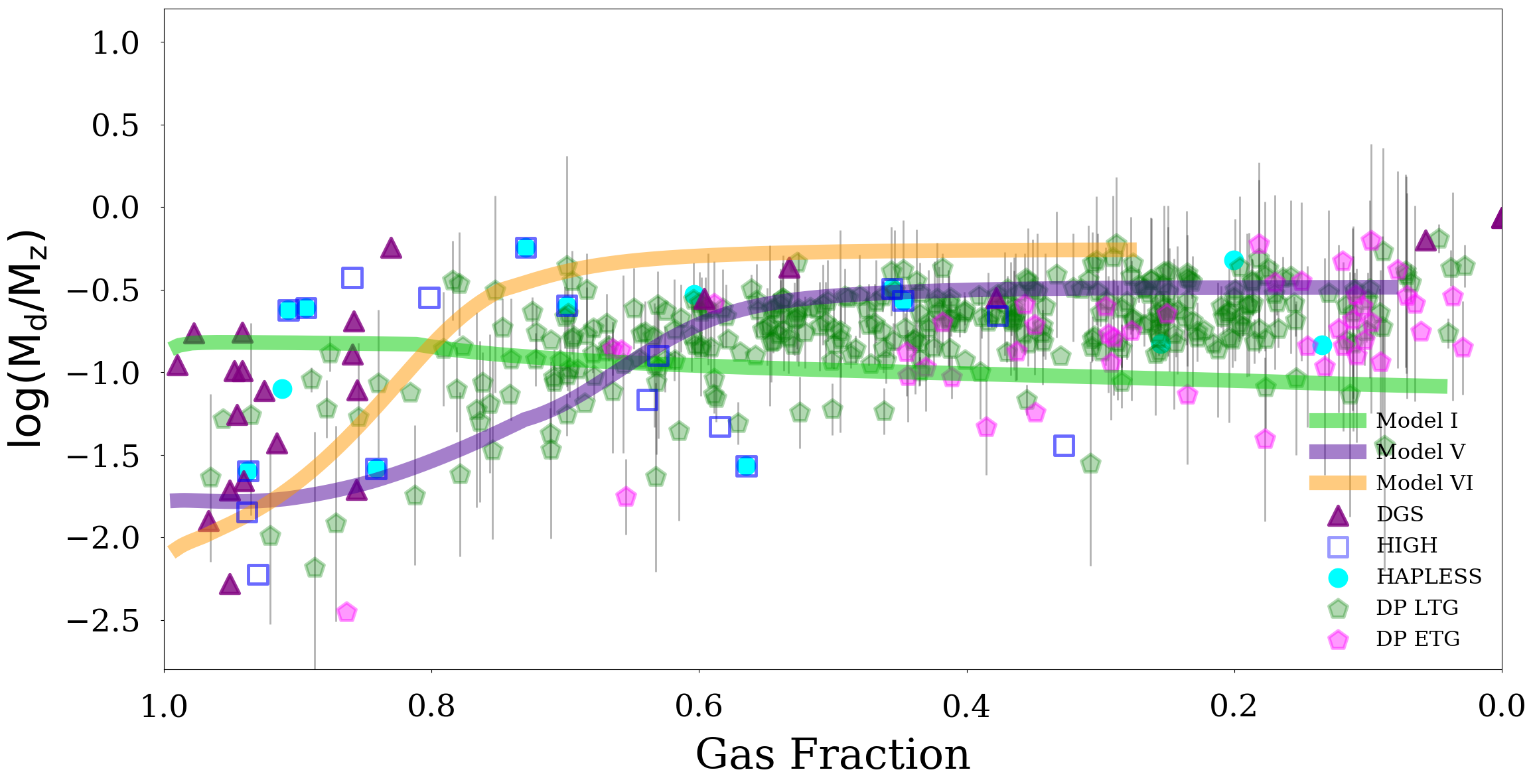}
  \caption{Dust-to-metal ratio is plotted against metallicity (\textit{top}) and gas fraction (\textit{bottom}). The dust-to-metal ratio at early evolutionary stages (low metallicity, high gas fraction) is significantly lower than for more evolved sources. Models including dust grain growth are necessary to match the observations.}
  \label{mdmz}
\end{figure}

Plotting the dust-to-metal ratio against metallicity and gas fraction (both tracers of the evolutionary stage) in Figure \ref{mdmz}, a significantly lower $\rm M_d/M_Z$ is found for the earliest stages of evolution. This behaviour is also followed by Models V and VI (see Section \ref{discussion} for further discussion), where for some galaxies Model V provides the best fit, yet for others Model VI is better. The overall correlations are quite weak (Spearman $\rho=-0.121$ and $\rho=-0.441$ for metallicity and gas fraction respectively). We find that DustPedia galaxies with gas fractions below $60\%$ (or above $\rm 12+log(O/H) \sim 8.2$) have a more or less constant dust-to-metal ratio of $\rm M_d/M_Z\sim 0.214$. For the other calibrations in this work the dust-to-metal ratio is also constant over the same range of gas fraction (or metallicity), though due to the discrepancy towards higher metallicities, the dust-to-metal ratio is lower for the other calibrations. Table \ref{evolved} shows the average $\rm M_d/M_Z$ for evolved galaxies for each calibration. 

\begin{table}
\caption{Average dust-to-metal ratios and standard deviation using different metallicity calibrations for DustPedia galaxies with gas fractions $<60\%$, which is the regime where the dust-to-metal ratio remains relatively constant. Galaxies with higher gas fractions have dust-to-metal ratios significantly below these values.}
\begin{center}
\small
\begin{tabular}{lccccccc} \hline\hline
calib. & $\langle\rm M_d/M_Z\rangle$ & $\langle\rm log(M_d/M_Z)\rangle$ &  $\sigma_{\rm log(M_d/M_Z)}$) \\ \hline
PG16S & 0.214 & -0.67 & 0.21 \\ 
PG16R &  0.206 & -0.69 & 0.21  \\
N2 & 0.162 & -0.79 & 0.23 \\
O3N2& 0.151 & -0.82 & 0.23 \\
IZI& 0.141 & -0.85 & 0.26 \\
KK04 & 0.116 &-0.93 &  0.30\\
T04& 0.092 & -1.04 & 0.25\\ 
\hline
\end{tabular}
\end{center}
\label{evolved}
\end{table}

For gas fractions greater than $60\%$ and metallicities below $\rm 12+log(O/H) \sim 8.2$, the average dust-to-metal ratio starts to differ significantly. DustPedia galaxies with gas fractions greater than $60\%$ have an average $\rm M_d/M_Z\sim 0.101$, and galaxies with gas fractions greater than $80\%$ have an average $\rm M_d/M_Z\sim 0.037$. However, even at these early evolutionary stages, some galaxies still have high $\rm M_d/M_Z$, and the scatter in $\rm M_d/M_Z$ is thus quite high (standard deviation of $\sim 0.4$ dex). Part of this scatter can be attributed to the increased uncertainties as a result of these galaxies being fainter. We do indeed see larger error bars for the unevolved galaxies in Figure \ref{mdmz}. Yet even for the well constrained sources, the scatter remains large. The remaining differences in $\rm M_d/M_Z$ at these evolutionary stages can be explained by differences in the local conditions and SFH (see also Section \ref{dustometaldisc}). In particular, \citet{Schneider2016} have shown differences in the density of the cold ISM can result in large differences in the dust mass. Galaxies with high $\rm M_d/M_Z$ in spite of high gas fractions, probably have an unusually dense ISM and a corresponding fast dust grain growth timescale.
         
\begin{figure}
  \center
\includegraphics[width=\columnwidth]{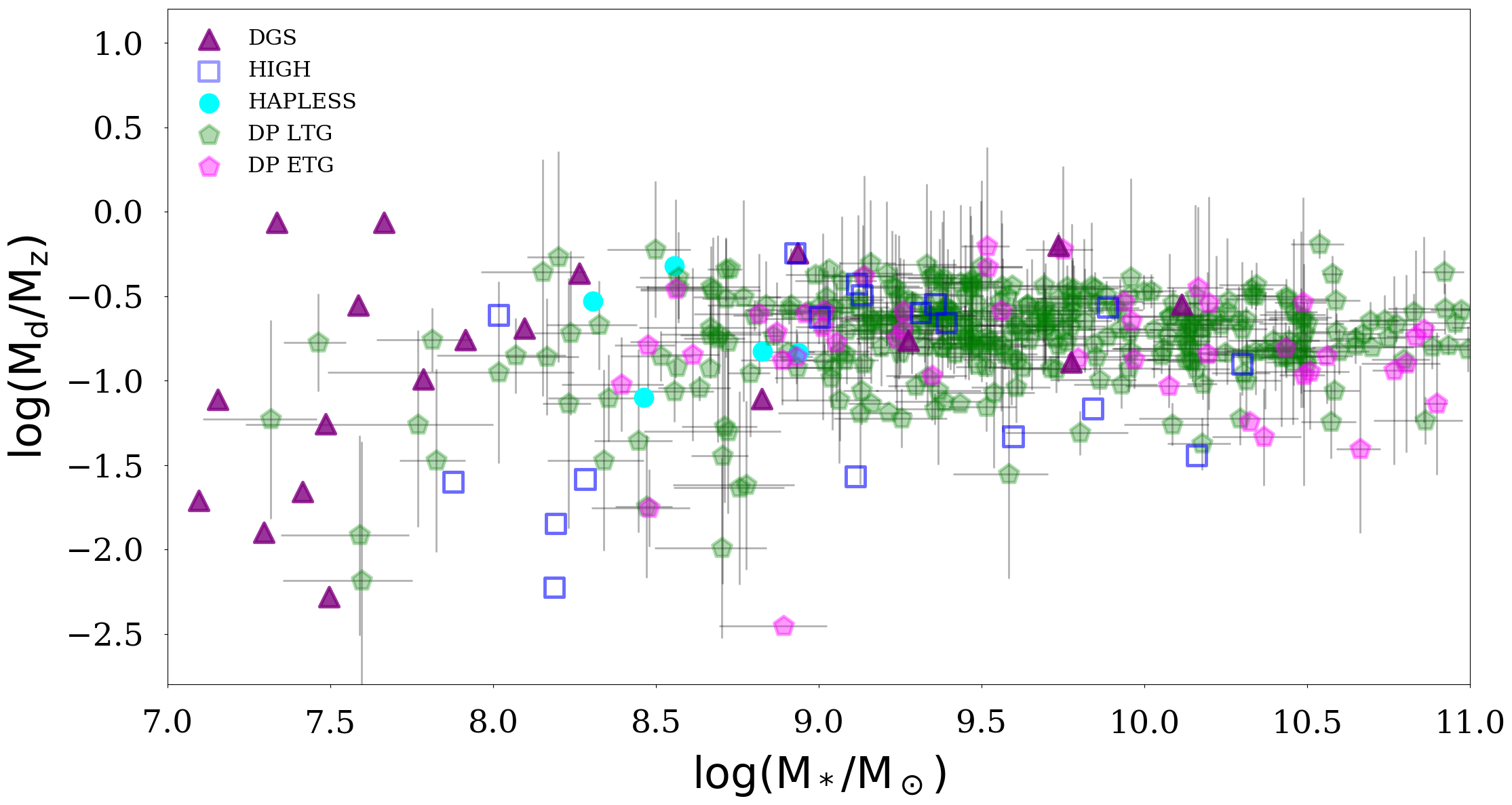}
\includegraphics[width=\columnwidth]{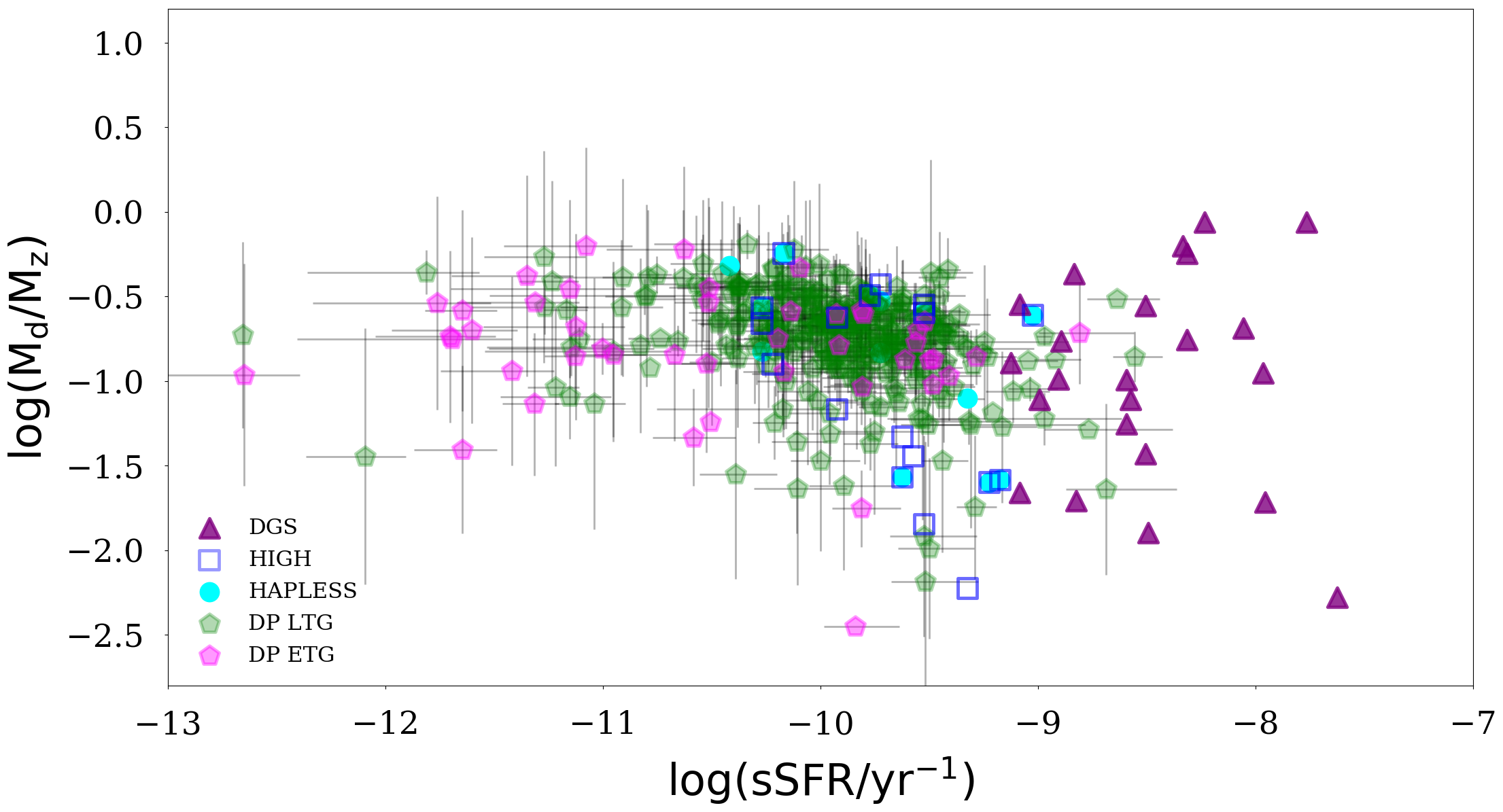}
  \caption{\textit{Top:} Variation of the dust-to-metal ratio with stellar mass. Low mass galaxies have low dust-to-metal ratios. \textit{Bottom:} Variation of the dust-to-metal ratio with sSFR. The outlying DGS sources are bursty galaxies which have high sSFR in spite of having already gone through some enrichment.}
  \label{mdmz_extra}
\end{figure}

In Figure \ref{mdmz_extra} we explore how the dust-to-metal ratio scales with other galaxy properties such as stellar mass and the specific SFR (sSFR). For low stellar masses, we find a very weak correlation (Spearman rank correlation coefficient $\rho=0.109$) and again find a lower dust-to-metal ratio and large scatter, as was expected since low stellar mass sources typically have high gas fractions. Sources with stellar masses larger than $\rm M_*=10^9\,M_\odot$ have a fairly constant dust-to-metal ratio of $\rm log(M_d/M_Z) \sim -0.60$ with a standard deviation of $0.24$ dex. This is slightly lower, yet consistent with the average value for low gas fraction ($<60\%$) sources. 
When the dust-to-metal ratio is plotted against sSFR, we find a weak but significant correlation (Spearman $\rho=-0.330$) over the whole range of sSFR, though with significant outliers. There are a number of DGS sources which have very high sSFR, in spite of having already gone through some evolution in the past (they have moderate gas fractions, metallicities and stellar masses). These galaxies are currently undergoing a starburst (\citetalias{DeVis2017b}), which results in an increased sSFR that is not matched by an equivalent decrease in $M_d/M_Z$. 

\section{Discussion}
\label{discussion}
\subsection{Radial gradients}
\label{Raddiscussion}

\begin{figure}
  \center
\includegraphics[height=0.4\textwidth]{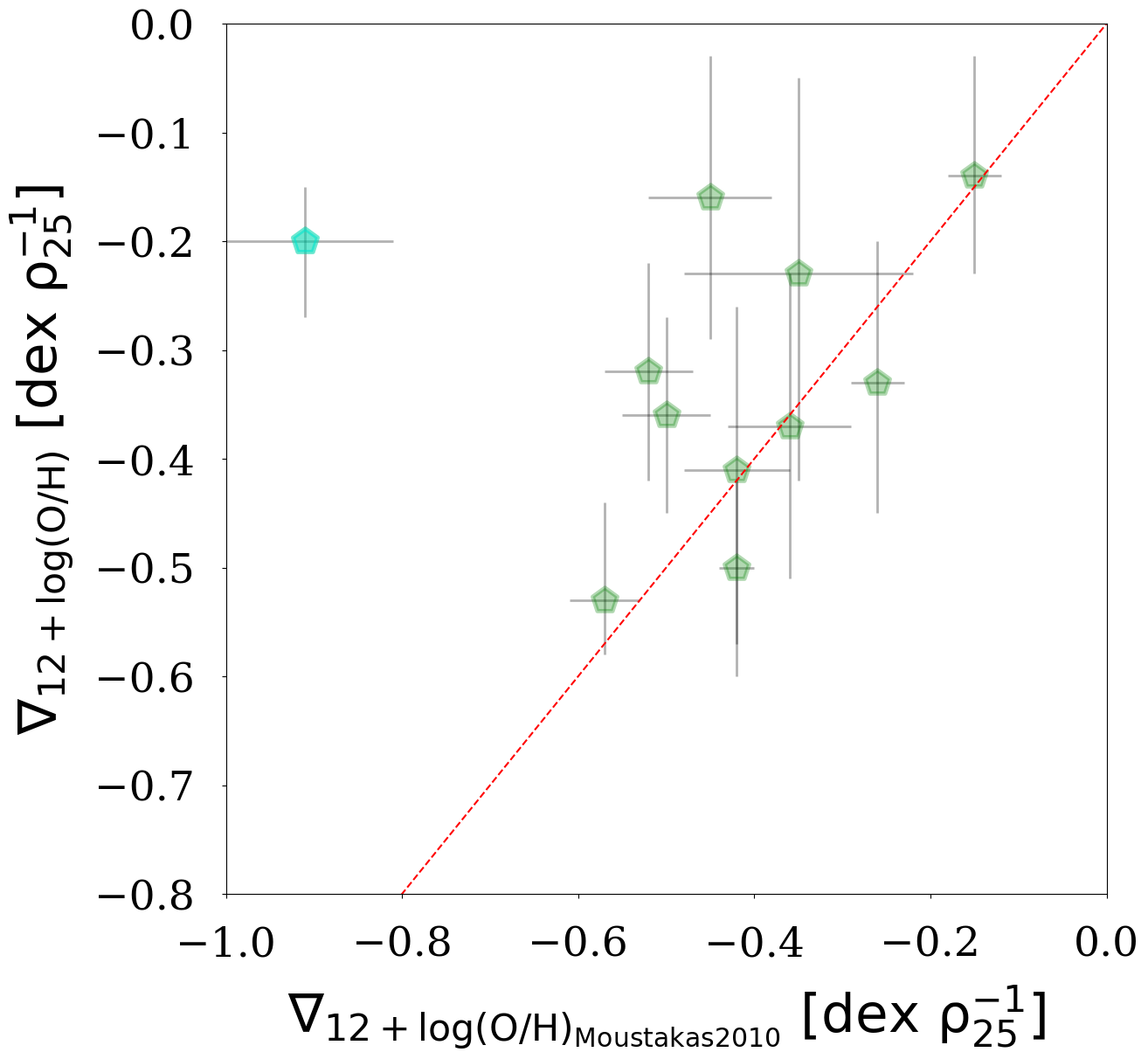}
  \caption{Comparison of the KK04 metallicity from our work, compared to the overlapping results from \citet{Moustakas2010}. Both gradients use the same calibration and the same normalisation of radii by the $r_{25}$, though different data points and different methods were used to derive the gradients. Most of the sample compares well (\textit{green}), though NGC3621 is a strong outlier (\textit{cyan}).} 
  \label{Moustcomp}
\end{figure}

There are numerous studies that have looked at metallicity gradients. Some of the galaxies in these studies are also in our sample. For these common galaxies, we can thus compare our gradient estimates to validate our method. Figure \ref{Moustcomp} shows the common galaxies between our sample and the sample from \citet{Moustakas2010}. We used the KK04 calibration for both sets of results. We find relatively consistent results, with only one strong outlier (NGC3621). The average $\chi^2\sim 1.7$, which drops to $\chi^2\sim 1.0$ if the outlier is discarded. Our gradient for NGC3621 is likely different since we have data out to larger radii (though there is no evidence for a change in the gradient between the central and outer regions). Our gradient for NGC3621 is also more similar to the gradient for other calibrations (\citet{Moustakas2010} find a big difference between their KK04 and PT05 calibration for NGC3621).

There are significant differences between different metallicity calibrations and the resulting radial gradients (Section \ref{resultssec}). Using KK04, \citet{Moustakas2010} find an average gradient of $\rm -0.42 \pm 0.19 \ dex \ r_{25}^{-1}$. This is significantly steeper than our average KK04 gradient in Table \ref{gradtable}. However, if we only consider our sources in common with \citet{Moustakas2010}, we find an average KK04 gradient of  $\rm -0.34 \pm 0.12 \ dex \ r_{25}^{-1}$, which is consistent with \citet{Moustakas2010} within the uncertainties. 
The average O3N2 gradient from CALIFA is $\rm -0.16 \pm 0.12 \ dex \ r_{25}^{-1}$ \citep{Sanchez2014}, consistent with our O3N2 results.   The average O3N2 gradient of \citet{Ho2015} is $\rm -0.25 \pm 0.18 \ dex \ r_{25}^{-1}$, slightly steeper than our average, which is likely due to \citet{Ho2015} only selecting star-forming field galaxies. 

\begin{figure}
  \center
\includegraphics[height=0.4\textwidth]{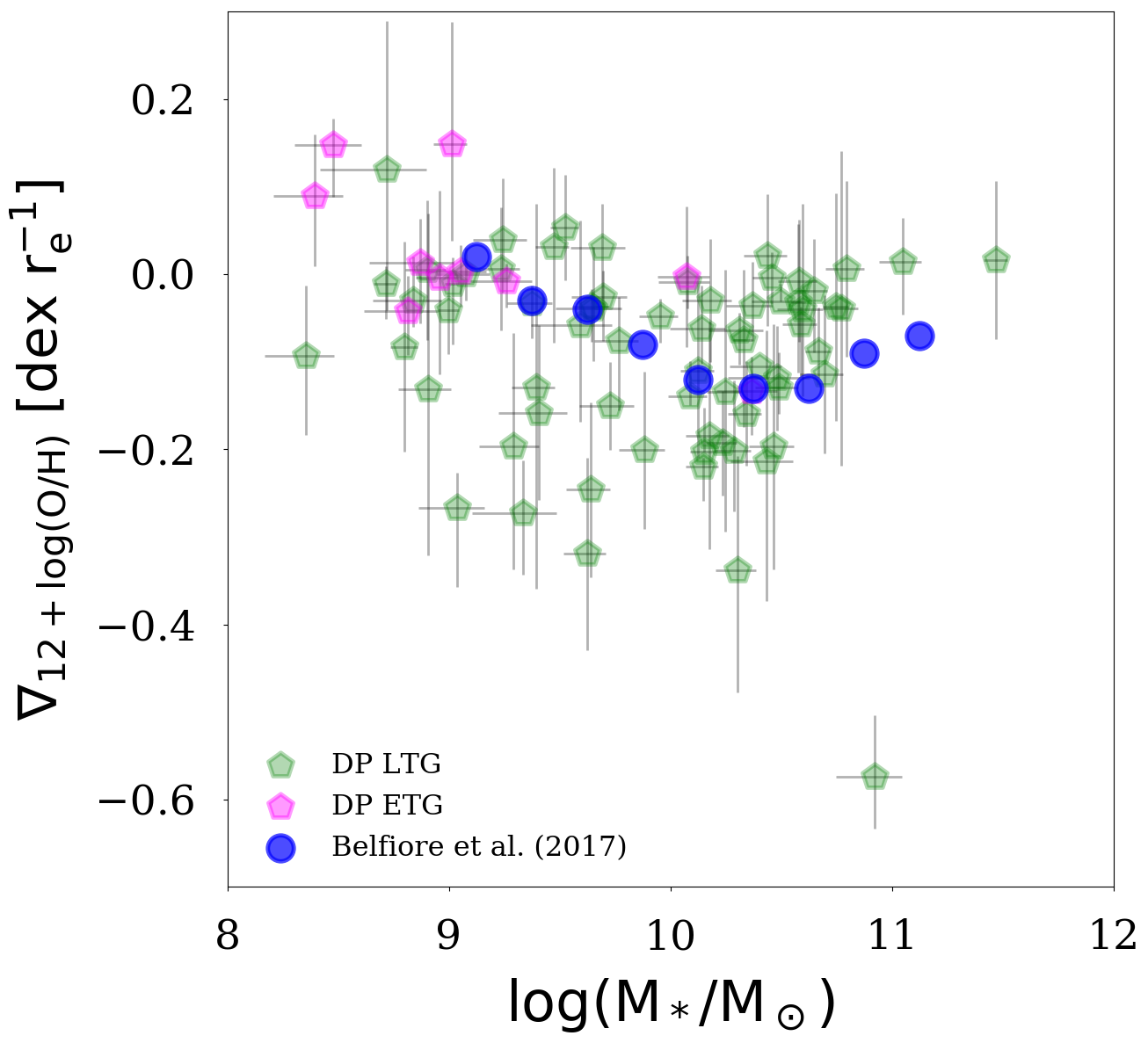}
\includegraphics[height=0.4\textwidth]{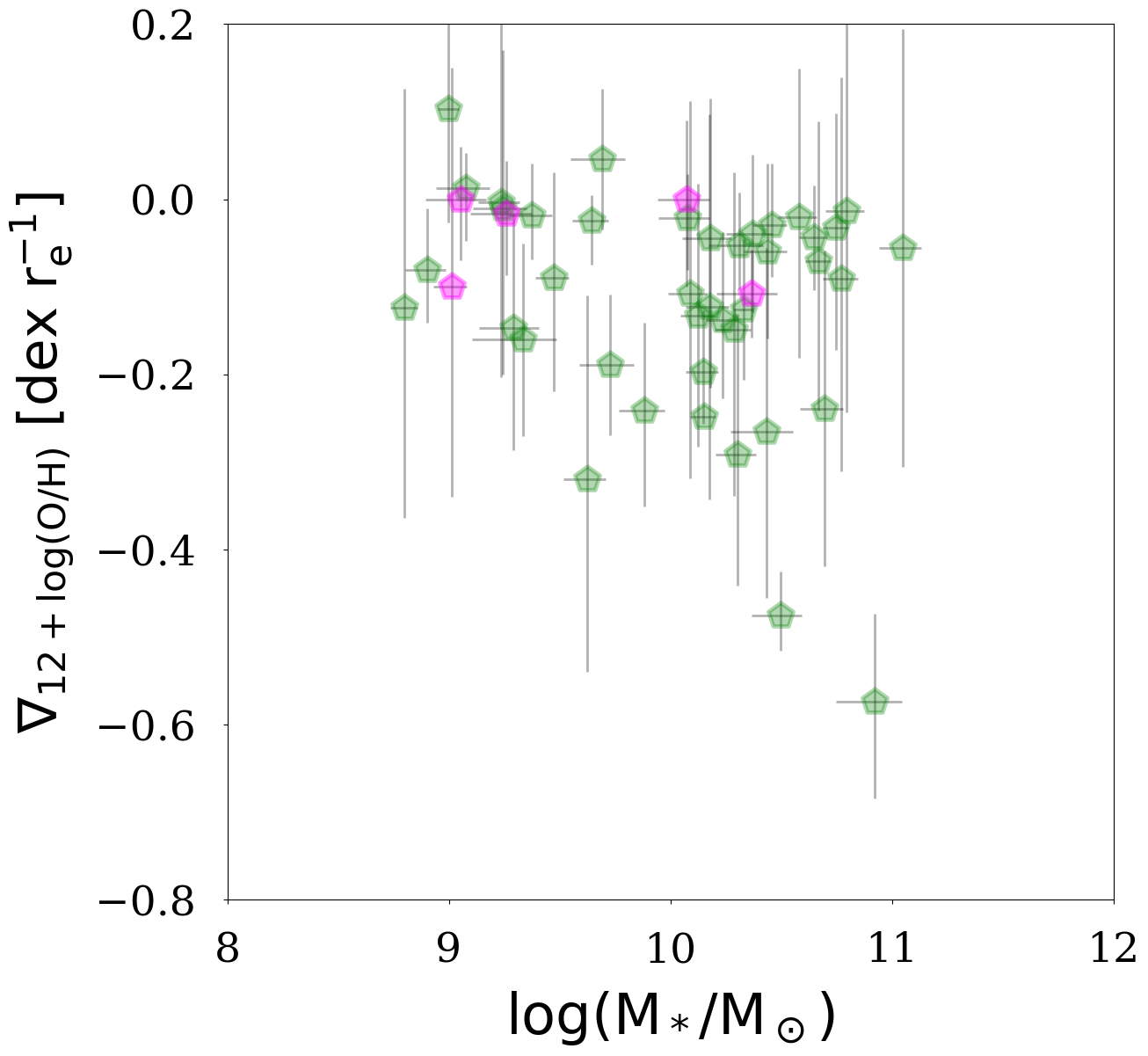}
  \caption{Radial metallicity gradients plotted against stellar mass for the O3N2 calibration (\textit{top}) and the KK04 calibration (\textit{bottom}) for DustPedia LTG (green) and ETG (magenta). The radii are relative to $r_{\text{eff}}$ instead of $r_{25}$. The weak correlations found for these calibrations are consistent with \citet{Belfiore2017} (shown in blue) and \citet{Poetrodjojo2018}.} 
  \label{gradO3N2}
\end{figure}

In addition to different calibrations, some works in the literature \citep[e.g.][]{Belfiore2017,Poetrodjojo2018} normalise the radii by $r_{\text{eff}}$ instead of $r_{25}$. Therefore, caution is necessary when comparing our results to the results in the literature. In contrast with our results in Section \ref{radialgrad} for PG16S, \citet{Belfiore2017} and \citet{Poetrodjojo2018} find a weak but significant trend between the metallicity gradient and stellar mass. To compare consistently, in Figure \ref{gradO3N2}  we plot the relation between metallicity gradient and stellar mass using the O3N2 and KK04 calibrations instead of PG16S. In addition, we normalise our radii by $r_{\text{eff}}$ instead of $r_{25}$ for consistency. 
We find very weak correlations (Spearman $\rho=-0.163$ and $\rho=-0.226$ for O3N2 and KK04 respectively), though now our results are more similar to the results from \citet{Belfiore2017} and \citet{Poetrodjojo2018}.
The binned results from \citet{Belfiore2017} do indeed look sensible when compared to our data and the difference could simply be due to the limited size of our sample with well constrained gradients. These differences show that the choice of calibration and normalisation of the radii can cause important differences in the results.

Hydrodynamical simulations suggest that the underlying process that is responsible for the average negative metallicity profiles is the inside-out formation of discs with specific angular momentum conservation \citep{Tissera2016}. However, dynamical processes such as mergers perturb the metallicity distributions, and typically result in flatter or even positive gradients. However, if this merger is followed by a central starburst, the newly produced heavy elements steepen the gradients again, yet they often remain above the values (i.e. flatter) that would otherwise be expected \citep{Tissera2016,Perez2011}. In addition, outflows of enriched material from the central regions could be re-accreted at larger radii, and thus contribute to flattening the metallicity profiles \citep{Perez2011}. The common slope for PG16S can be explained if all disk galaxies went through very similar chemical evolution when building up their disks. 

Even though our metallicity determinations have been performed consistently across our whole compilation, the sampling of regions in different galaxies is quite heterogeneous. Some galaxies have very well sampled IFU data, whereas others have a few or even only one fibre spectrum. Therefore, we assessed whether our gradient determination is biased towards a particular parameter-space. We studied the dependency of the gradients on any of the following parameters:
\begin{itemize}
\item The number of data points we have available for a specific galaxy.
\item The radial distance to the most distant available data point.
\item The total radial distance covered by our data points.
\item The selection bias of IFU data towards high metallicity galaxies.
\item The number density of the local environment (from Davies et al. in prep.).
\end{itemize} 
We find no dependence of the average gradient, nor the scatter, on any of these parameters. We note that we are only considering the well constrained sub-sample and galaxies with very few data points have thus been discarded (these would be biased towards the mean gradient and show less scatter). There is a small interdependency between the number of data points (selection bias of IFU data) and other galaxy properties such as stellar mass and SFR (massive actively star-forming galaxies often have more metallicity data points available on average), though this does not carry through to any bias in the metallicity gradients.    

\subsection{Dust-to-metal ratio}
\label{dustometaldisc}
Our sample is the largest sample of galaxies for which the dust and metal content are measured consistently over all galaxies. When we combine DustPedia with the 73 galaxies in our comparison samples, we arrive at a sample of 539 sources (compared to 382 for the combined sample in \citetalias{DeVis2017b}). In addition to the better statistics and better consistency across samples, our sample has made use of the resolved information in these galaxies to arrive at a better estimate of the global estimate. Our data has revealed a more gradually and continuously evolving dust-to-metal ratio, rather than two separate regimes for high and low metallicity galaxies \citep[as in][and DV17]{Remy-Ruyer2014}. The more gradual increase is attributed to different regions in the galaxy reaching the critical metallicity at different times.

In the literature, the dust-to-metal ratio of galaxies is often assumed to have a constant value of around $\rm M_d/M_Z\sim 0.3$ (e.g. \citeb{James2002}; \citeb{Clark2016}; \citeb{Camps2016}). This is slightly higher than what we find for evolved DustPedia galaxies ($\rm M_d/M_Z\sim 0.214$), though for unevolved galaxies significantly lower values are found. We note that our estimate of the dust-to-metal ratio should not be used at face value for high redshift galaxies. At high redshifts, high stellar mass galaxies are still in early evolutionary stages (and have high gas fraction and low metallicity), and thus likely have low dust-to-metal ratios.

A consistent picture emerges from our results in the previous sections: Galaxies start out with low dust-to-metal ratios, which then increase as galaxies evolve, and finally remain constant in the later evolutionary stages. Dust grain growth provides the most likely explanation for this behaviour. For unevolved galaxies the metallicities are very low and conversely their dust grain growth timescales are very long (Eq. \ref{growth}). As the metal content increases, the efficiency of the grain growth increases as there are more and more metals available to accrete onto the dust grains. For DustPedia galaxies the critical metallicity is around $\rm 12+log(O/H) \sim 8.1$, consistent with other work \citep{Galliano2018}. The increase in the dust-to-metal ratios levels off towards high metallicities or low gas fractions as most of the available metals are depleted onto the dust grains. At the same time global dust destruction (Eq. \ref{destr}; i.e. not the reverse shock in SN) processes reduce the dust-to-metal ratio. The observed $\rm M_d/M_Z$ increase shows these destruction processes are not dominant, yet they do contribute to some extent. At the start of the evolution, the dust-to-metal ratio is set by a balance between dust destruction and dust production by stars, yet later it is set by a balance between dust destruction and dust grain growth \citep{Mattsson2014}.

Our chemical evolution model without dust grain growth (Model I) clearly fails to explain the observations. Variations of Model I (e.g. including inflows and outflows, different SFH.) also cannot obtain a good match to the observations unless dust grain growth is included and the SN dust yield is reduced. Models V \& VI do include dust grain growth and reduced SN dust and consequently provide a good match to our data. We note that the scatter in our observations at high gas fractions is large, and as a result the observations cannot be described by a single model. In De Vis et al. (in prep.) we will explore a grid of models where each of the free parameters will be varied. In particular we expect the following parameters to cause significant variations between the different galaxies at a fixed gas fraction.

The first important parameter is the SFH. Galaxies typically go through one or more burst of star formation throughout their evolution. The start time and duration of these burst can (temporarily) affect the dust-to-metal ratio \citep[e.g.][]{Zhukovska2014}. During the burst, a lot of metals are expelled, which have not yet had time to accrete onto the grains, and $\rm M_d/M_Z$ decreases. After the burst, dust grain growth continues until all the newly available metals are accreted. Variations in the grain growth timescale will also strongly affect the dust-to-metal ratio. Local conditions (such as ISM density; \citeb{Schneider2016}) could significantly influence the dust grain growth timescales. Very fast ($\sim 5$ Myr) timescales conform with \citet{Feldmann2015} result in high $\rm M_d/M_Z$, even for unevolved galaxies. Additionally, variations in the strength of the reverse shocks could result in scatter in the SNe dust yields and thus scatter in $\rm M_d/M_Z$. 

Each of these three parameters could easily be affected by the individual conditions within a galaxy. A galaxy merger could, for example, result in a burst of star formation, as well as change the density of the ISM, which could in turn result in different grain growth timescales, and a different strength of the reverse SN shock. Additionally, uncertainty remains over the dust and metal yield of SN and AGB stars. AGB dust yields could be metallicity dependent \citep[e.g.][]{Ferrarotti2006}, and more recent SN dust yield prescriptions \citep{Bianchi2007,Marassi2018} could be used. However, \citet{Ginolfi2018} show that, independent of the adopted (metallicity-dependent) AGB and SN dust yields, models without grain growth cannot explain the observed trends.

Instead of dust grain growth as the driver of the $\rm M_d/M_Z$ evolution, varying the SN dust yields could provide an alternative explanation. If the SN condensation efficiency (dust yield per metal yield) increases as galaxies evolve, there is no need for dust grain growth. However, the SN condensation efficiency is not expected to vary by more than a factor of three \citep{Dwek1998,Todini2001} and cannot explain the observed variation of an order of magnitude. However, these yields do not take into account the reverse shock. One option is that dust destruction by the reverse shock is much stronger in unevolved galaxies than at later evolutionary stages. Dust destruction by the reverse shock is expected to be stronger if the local ISM density around the SN is significantly higher in unevolved galaxies \citep{Bianchi2007,Bocchio2016}. There is little information in the literature about how the strength of the reverse shock and the local ISM density change in various environments, so we cannot discount this possibility. However, we note that without any dust grain growth, SN would have to produce more than five solar masses of dust to explain the galaxies with the highest $\rm M_d/M_Z$. This is inconsistent with the dust content of SN remnants (which have not yet been processed by the reverse shock).  

\begin{figure}
  \center
\includegraphics[width=\columnwidth]{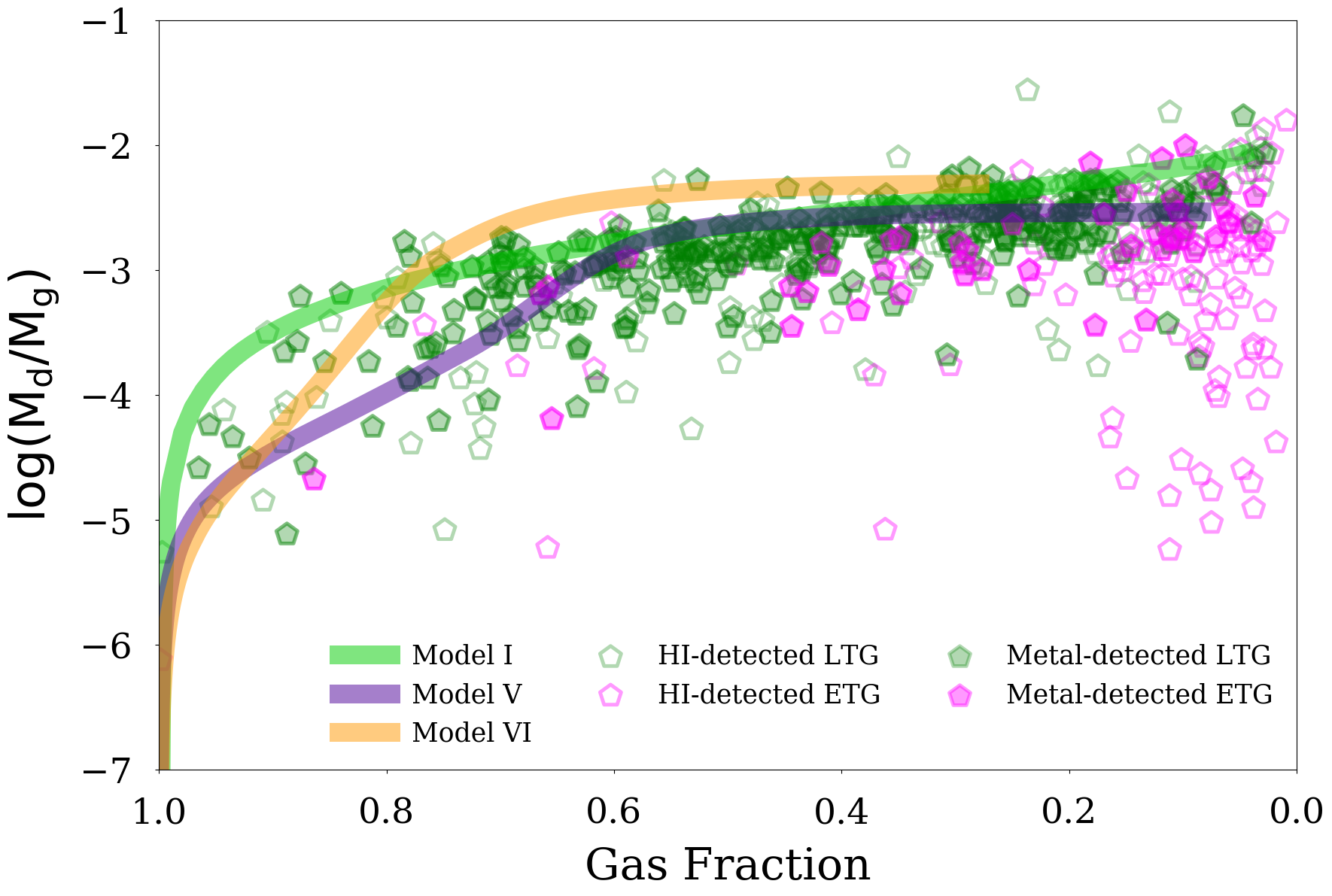}
  \caption{Evolution in the dust-to-gas ratio with gas fraction for \hi-detected sources (open symbols) and the metal-detected sources used throughout this work (filled symbols).  There is a population of low dust-to-gas ratio ETGs that is present in the \hi-detected sample, but not in the metal-detected sample.} 
  \label{selection}
\end{figure}

Another important point to discuss is how the selection of our sample affects the relations found. In particular, by including only sources with detected metallicities, we have excluded a part of the parameter-space. In Figure \ref{selection}, we show the dust-to-gas ratio against gas fraction for the metal-detected sources used in this work, together with all DustPedia sources for which we have a detected \hi\ mass (i.e. including the ones without metallicities). For the \hi-detected sources, there is a population of low gas fraction ETGs with dust-to-gas ratios that are significantly lower (by more than an order of magnitude) than the rest of the sample. Presumably, sputtering by hot gas has destroyed a lot of the dust in these galaxies. When we focussed on the metal-detected sources, however, we find that this population of sources has entirely been missed. Upon inspection we find that these galaxies only have AGN-classified optical spectra available. If we assign these sources a metallicity that is typical for their gas fraction, they have significantly lower dust-to-metal ratios than the rest of the sample. 

Finally, we would like to note two potential caveats to our method. The first is that we assumed, as in almost all studies concerning dust, that the dust mass absorption coefficient $\kappa$ is constant. However, our changes in dust-to-metal ratio could also be interpreted as changes in the dust mass absorption coefficient at a constant dust-to-metal ratio. This alternative interpretation has actually been used to constrain $\kappa$ for evolved galaxies \citep{James2002,Clark2016}. 
Our second caveat is that we used the total gas mass from single dish observations, whereas some of the \hi\ might actually be extended beyond our aperture from \citet{Clark2018}. This \hi\ gas might be extra-planar gas that is not actually associated with the galaxy \citep[e.g.][]{Vargas2017}, and thus not affecting the dust properties. An overestimation of the galaxy gas mass would also lead to an overestimation of the metal mass and thus to lower dust-to-metal ratios. Dwarf galaxies (which typically have higher gas fractions) could be more strongly affected by this issue, thereby introducing a bias in our results. Roychowdhury et al. (in prep.) are investigating DustPedia dwarf galaxies using resolved \hi\ data to study this issue in more detail.

\section{Data products}
\label{data}
This work presents a large, homogeneous set of resolved and global metallicity measurements of 516 nearby galaxies (10143 individual regions). In combination with the available multi-wavelength imagery and photometry \citep{Clark2018}, as well as galaxy properties from SED fitting (Nersesian et al. in prep.) and the \hi\ measurements, this forms an unprecedented database which will be invaluable for future investigations of galaxy (chemical) evolution processes. The data in this work is available as four separate tables, which can be accessed on the DustPedia archive\footnote{http://dustpedia.astro.noa.gr/AncillaryData}, and from the VizieR catalogue service\footnote{Will be uploaded after publication of the  of the paper}. The references used for our compilation and examples of each of the four tables are given in Appendix \ref{appendix}. The first table list the \hi\ fluxes, uncertainties, and masses (Section \ref{galaxyprop}). The next table lists the reddening-corrected emission line fluxes for the compiled emission lines from the literature, combined with the extracted MUSE spectrophotometry (see Section \ref{sec:spoctrophot}), for all the regions that are classified as star-forming. The third table lists the oxygen abundances and corresponding bootstrapped uncertainties using multiple calibration methods for each of these star-forming regions. The reddening coefficients $C(\rm{H\beta})$ are listed in both the emission line and metallicity table for each individual region (for MUSE both reddening components have been added). Our final table presents the available global metallicities for each of the galaxies in our sample, together with their uncertainties as detailed in Section \ref{global}.

\section{Conclusions}
\label{sec:conclusions}
We have studied the relative dust, gas, and metal content of DustPedia galaxies. By performing a literature search, combined with archival MUSE data, we were able to compile metallicities for 10143 individual regions within DustPedia galaxies. Radial profiles were fitted to these regions and global metallicities determined for 516 DustPedia galaxies by taking the metallicity at a radius of $r=0.4\ r_{25}$ . All the metallicity measurements are made available to the community. A total of 466 DustPedia galaxies have all the necessary observations to constrain dust, gas and metal content. 76 of these have metallicities below $\rm 12+log(O/H)=8.2$ (22 sources below $\rm 12+log(O/H)=8.0$), providing key constraints on the evolution of the dust-to-metal ratios in early evolutionary stages. The DustPedia observations were also compared to additional samples, as well as simple chemical evolution models from \citet{DeVis2017b}. Our main conclusions are:
\begin{itemize}
\item The gradient from a linear fit to the radial metallicity profile shows no significant dependence on stellar mass. Yet, the gradient was found to correlate weakly with the \hi-to-stellar ratio, the extent of the galaxy and with the dust-to-stellar mass ratio. The metallicity gradients depend on which calibration is used and whether the radii are normalised by $r_{\text{eff}}$ or $r_{25}$.
\item The oxygen abundance is found to increase monotonically as galaxies evolve from high to low gas fractions. Chemical evolution models including inflows and outflows are necessary to explain the results. The nitrogen-to-oxygen ratio ($\rm log(N/O)$) is also found to increase monotonically with decreasing gas fraction, though it shows more scatter.
\item The relation between the dust-to-gas ratio and metallicity is best described by a single powerlaw ($\log(M_d/M_g)= {[1.94\pm0.11]} \times12+\log(O/H)-{[19.04\pm0.91]}$), rather than a broken power law.
\item For more evolved galaxies, we find a more or less constant dust-to-metal ratio. For galaxies with gas fraction below $60\%$, metallicities above $\rm 12+log(O/H) \sim 8.2$ or stellar masses above $M_*\sim10^9\,M_\odot$, the dust-to-metal ratio is about $M_d/M_Z\sim 0.214$. 
\item At early evolutionary stages (gas fraction above $60\%$, metallicities below $\rm 12+log(O/H) \sim 8.2$), the average DustPedia galaxy has 2.1 times lower dust-to-gas and dust-to-metal ratios (6 times lower for galaxies with gas fraction $>80\%$). However, at these early evolutionary stages there is quite a lot of scatter. Some galaxies have dust-to-metal ratios as high as evolved galaxies ($M_d/M_Z\sim 0.214$), yet other galaxies have dust-to-metal ratio as low as $M_d/M_Z\sim 0.01$. This increased scatter is likely to be due to variations in the local conditions and SFH resulting in different dust grain growth timescales and SN dust yields, combined with increased uncertainties due to these galaxies being fainter.
\item We also find a decrease in the dust-to-metal ratio for low stellar masses and high sSFR. Though the larger scatter indicates these are likely indirect correlations. Bursty sources such as some of the DGS galaxies, have high sSFR yet have a high dust-to-metal ratio that is consistent with their gas content rather than with their star formation activity.
\item By including only sources with detected metallicities (spectra classified as star-forming), we have missed a population of low gas fraction, low dust-to-gas ratio ETGs. These sources presumably have lower dust-to-metal ratios than the rest of the sample. Most likely, sputtering by hot gas has destroyed a lot of dust in these galaxies.
\item We have compared these results to simple chemical evolution models. We find models with a dominant contribution to the dust budget from grain growth do a decent job at describing the observations. Resolved chemical evolution models would result in a less steep increase in the dust-to-metal ratio (thereby providing a better match to the observations), as well as allowing the study of metallicity and dust-to-metal gradients, which is the logical next step.

\end{itemize}
\begin{acknowledgements}
DustPedia is a collaborative focused research project supported by the European Union under the Seventh Framework Programme (2007-2013) call (proposal no. 606847). The participating institutions are: Cardiff University, UK; National Observatory of Athens, Greece; Ghent University, Belgium; Université Paris Sud, France; National Institute for Astrophysics, Italy and CEA, France. We thank the referee, Raffaella Schneider, for helpful comments. We gratefully acknowledge Sebastian Sanchez and Laura Sanchez-Menguiano for their help with the CALIFA data. We also thank Leonid Pilyugin for sharing the data from \citet{Pilyugin2014} and useful comments. This work is based on publicly available MUSE data obtained from the ESO Science Archive Facility under request number pdevis291121. This research made use of the NASA/IPAC Extragalactic Database (NED),
which is operated by the Jet Propulsion Laboratory, California Institute of Technology,
under contract with the National Aeronautics and Space Administration.
We acknowledge the usage of the HyperLeda database (\citeb{Makarov2014}; http://leda.univlyon1.fr). This research made use of the VizieR catalogue access tool \citep{Ochsenbein2000}, CDS, Strasbourg, France.
\end{acknowledgements}
\bibliographystyle{aa}
\bibliography{Library}

\begin{appendix}
\section{Fitting radial profiles to heterogeneously sampled data}
\label{priors}
In Section \ref{global}, we describe our use of a Bayesian approach to fit radial profiles to the available metallicity data. The literature compilation of emission lines we have performed results in some galaxies having many metallicity regions, yet other have only a few detected regions available. There are also a few literature sources that did not publish the positions of their regions, and we can thus not determine the radii of these regions. In this appendix we will address how we have dealt with these issues. 

In order to deal with the difference in how well certain galaxies are sampled, we determined the prior for our Bayesian approach differently for well-sampled and poorly sampled galaxies. Ideally, we wanted to use a Gaussian prior with our best estimate (for the gradient, intercept, or intrinsic scatter) as the mean and its uncertainty as the standard deviation. However, since our sample is quite heterogeneous in terms of radial coverage, we cannot make a reliable guess for the gradient for many of the galaxies in our sample. Therefore, we divide our sample in two sub-samples. One sub-sample consists of all galaxies with at least five data points covering a range of radii at least $0.5\ r_{25}$ wide. The other sub-sample consists of all other galaxies. 
For the well-sampled galaxies, we simply used the best fitting gradients (using a weighted least squares fit) as the mean of the Gaussian for the radial gradients. 
For the unconstrained sub-sample, we used a `universal prior', that is, each galaxy has the same value for the mean of the prior on the metallicity gradient. This value is given by the average gradient for the galaxies in the good quality sub-sample. The average gradients and standard deviations for the well constrained sub-sample for each calibration are found in Table \ref{gradtable} in Section \ref{radialgrad}. 

For the intercept, the mean of the Gaussian is set to the value that results in the best fit to the observed data for each individual galaxy, and the standard deviation is given by the quadratic sum of the standard deviation of the regions and the mean uncertainty on this data. 
For the intrinsic scatter we used a mean of 0.0 and standard deviation of 0.094 (determined iteratively from the results of the bootstrapping). However, for the intrinsic scatter all values below zero or above the scatter in the residuals (data values after subtracting the best fitting line) are rejected. Using these priors, the best values and uncertainties on the gradient and global metallicity are then determined as detailed in Section \ref{global}

A few literature sources did not publish the positions of their regions, and we can thus not determine the radii of these regions. In total there are 65 regions (0.6\% of the total sample) for which we have no positions. Therefore, we gave these sources an additional source of uncertainty and perform a weighted average with the weights being $w=1/(\sigma_{y_i}^2+scatter_{int}^2)$ for each region with uncertain position and $w=\sum 1/(\sigma_{y_i}^2+scatter_{int}^2)$ for the best fitting metallicity from the gradient. The additional source of uncertainty is determined from generating 10000 random gradients from the Gaussian prior, and storing the metallicity at a random radius uniformly between 0 and $2\, r_{25}$ (the intercept is determined from keeping the metallicity at $0.4\ r_{25}$ constant). The extra source of uncertainty is then the standard deviation of the newly generated metallicities. This uncertainty is typically much larger than the uncertainty on the best fitting metallicity from the gradient. These data will thus barely affect the final metallicity, unless there are no high S/N sources with positions available to determine the best fitting metallicity from the gradient.

\section{Metallicity calibration conversions }
\label{convPG16}

\begin{figure}
\centering
    \includegraphics[width=0.49\columnwidth]{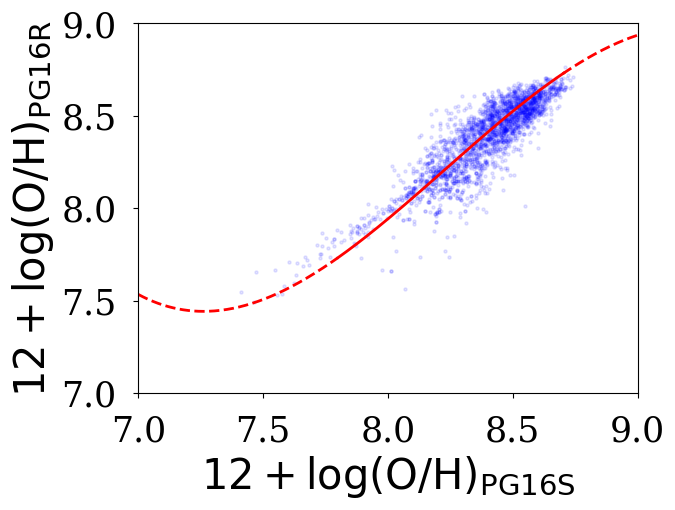}
    \includegraphics[width=0.49\columnwidth]{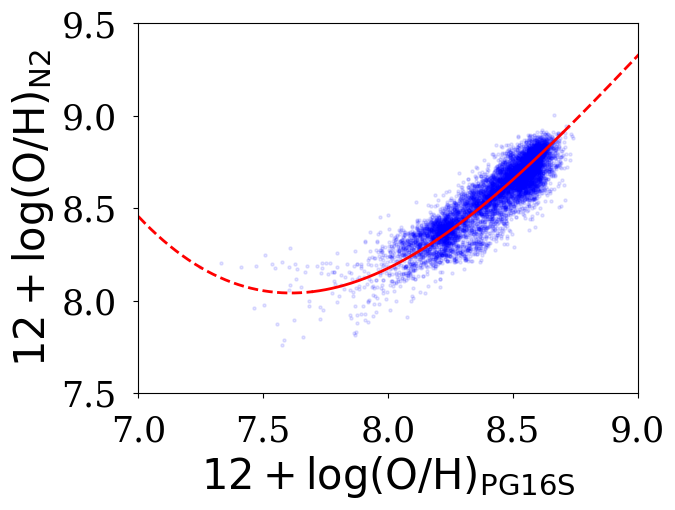}
    \includegraphics[width=0.49\columnwidth]{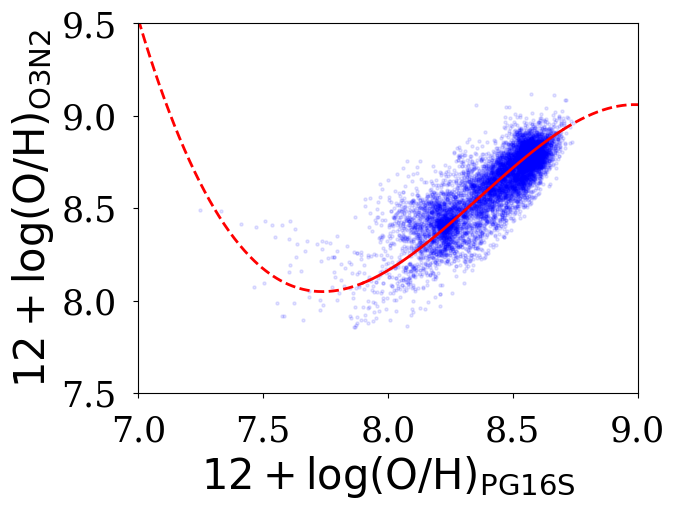}
    \includegraphics[width=0.49\columnwidth]{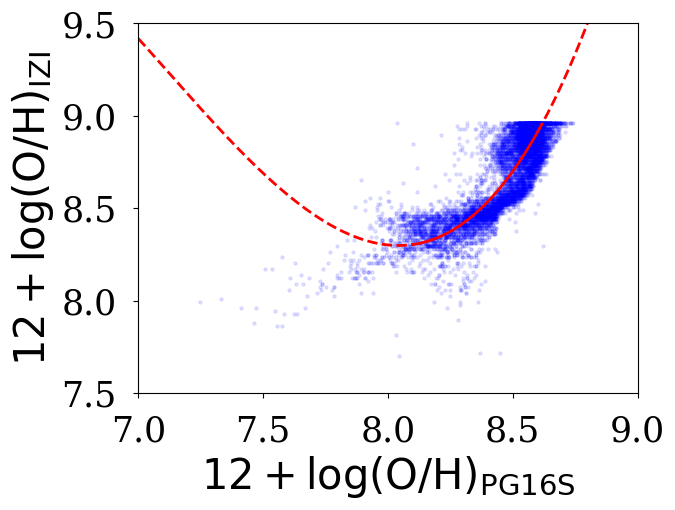}
    \includegraphics[width=0.49\columnwidth]{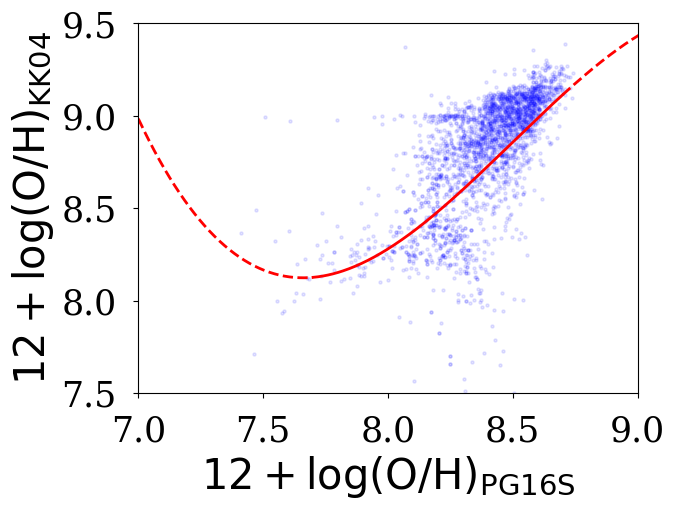}
  \caption{Relationships between PG16S and the other metallicity calibrations used in this work. The red line gives the best fitting third order polynomial, with the dashed part indicating ranges for which we do not trust the conversion (see Table \ref{convtable}). }
    \label{convfig}
\end{figure}

\begin{table*}
\caption{Conversion relations between the PG16S metallicity calibration ($x$) and the other calibrations ($y$) used in this work. The range columns indicate the range over which we trust the conversions. $n_{reg}$ lists the number of regions that were used to determine the relation.}
\begin{center}
\scriptsize
\begin{tabular}{llclcc} \hline\hline
Calibration  & Relation & x-range & Inverse relation & y-range & $n_{reg}$ \\ \hline
PG16R& $y = -0.42249 x^3 + 10.4323 x^2 - 84.675 x + 233.99$ & 7.8-8.7 & $x = 2.77861 y^3 - 68.8204 y^2 + 568.714 y - 1559.76$ & 7.8-8.7 & 2511 \\
N2& $y = -0.22112 x^3 + 6.0219 x^2 - 53.239 x + 161.90$ & 7.7-8.7 & $x =-0.87614 y^3 + 21.8528 y^2 - 180.559 y + 502.31 $ & 8.1-8.9 & 8862 \\
O3N2& $y = -1.03517 x^3 + 25.9719 x^2 - 215.993 x + 603.91$ & 7.9-8.7 & $x = -0.97935 y^3 + 24.8169 y^2 - 208.612 y + 589.94$ & 8.1-8.9 & 8862 \\
IZI & $y = 0.59523 x^3 - 12.7114 x^2 + 88.959 x - 194.60$ & 8.1-8.6 & $x = -2.57677 y^3 + 64.5087 y^2 - 536.678 y + 1491.83$ & 7.9-8.6 & 8051 \\
KK04 & $y = -0.62342 x^3 + 15.8860 x^2 - 133.653 x + 380.08$ & 8.1-8.6 & $x = -0.10625 y^3 + 2.9875 y^2 - 27.400 y +  90.54$ & 7.9-8.6 & 2641 \\
\hline \\
\end{tabular}
\\
\end{center}
\label{convtable}

\setcounter{section}{3}
\setcounter{table}{0}
\caption{Example table showing the \hi\ data for 10 random galaxies. $D_{\text{best}}$ and $v_\text{helio}$ give the best distance measurement and heliocentric velocity from \citet{Clark2018}. The \hi\ flux and its error are given by $\rm F_{\text{H{\sc i}}}$ and $\rm E_{\text{H{\sc i}}}$ and the \hi\ mass and uncertainty by $\rm \MHI$ and $\rm E_{\MHI}$. The Ref. column lists which reference has been used (see Table \ref{HIrefs}). The flag column identifies whether the source has a well measured flux (0), an upper limit (1), the \hi\ emission is confused (2) or that the \hi\ measurements have been corrected for flux outside the ALFALFA beam (3; see Section \ref{galaxyprop}). Sources with flags (1) and (2) are discarded in this study.}
\begin{center}
\begin{tabular}{lrrrrrrlc} \hline\hline
\scriptsize
Name & $D_{\text{best}}$ & $v_\text{helio}$ & $\rm F_{\text{H{\sc i}}}$ & $\rm E_{\text{H{\sc i}}}$ & $\rm \MHI$ & $\rm E_{\MHI}$ & Ref. & Flag  \\
 & Mpc & $km\ s^{-1} $ & $Jy\ km\ s^{-1} $ & $Jy\ km\ s^{-1} $ & $10^9 M_{\odot}$ & $10^9 M_{\odot}$ &  &  \\ \hline
ESO097-013 & 4.20 & 434 & 654.0 & 103.0 & 2.73 & 0.43  & Huchtmeier1989 & 0  \\
NGC0628 & 10.14 & 657 & 843.5 & 0.4& 20.46 & 0.01 &  Haynes2018 & 3  \\
NGC3381 & 28.74 & 1626 & 14.9 & 0.1 & 2.91 & 0.02 &  Haynes2018 & 0  \\
NGC4038 & 24.54 & 1630 & 37.1 & \ldots & 5.27 & \ldots  & Casasola2004 & 0  \\
NGC4651 & 23.23 & 799 & 62.9 & 0.2 & 8.00 & 0.02 &  Haynes2018 & 3  \\
NGC5236 & 4.90 & 507 & 361.0 & 18.1 & 2.04 & 0.10 &  Wong2006 & 0  \\
NGC5457 & 7.11 & 237 & 1100.0 & 55.1 & 13.13 & 0.66 &  Wong2006 & 0  \\
NGC5480 & 28.63 & 1904 & 9.1 & 1.0 & 1.76 & 0.19 &  Springob2005 & 0  \\
NGC7320 & 14.43 & 776 & 7.9 & 0.1 & 0.387 & 0.004 &  Haynes2018 & 0  \\
UGC09299 & 29.24 & 1548 & 45.5 & 0.1 & 9.18 & 0.02 &  Haynes2018 & 0  \\ \hline
\end{tabular}
\end{center}
\label{HIex}
\end{table*}
\setcounter{section}{2}
\setcounter{table}{1}
Here we provide metallicity calibration conversions between PG16S and the other calibrations in this work (excluding T04, which is already a conversion from O3N2) following \citet{Kewley2008}. A third-order polynomial is fitted to all individual regions within the DustPedia galaxies with measured metallicities from both calibrations. A least-squared minimization is used including uncertainties for both metallicity calibrations. The resulting fits are given in Table \ref{convtable} and Figure \ref{convfig}. 

\section{Example tables and list of references}
\setcounter{table}{1}
\label{appendix}
The data in this work is published in four separate data tables on the DustPedia archive\footnote{http://dustpedia.astro.noa.gr/AncillaryData}, and from the VizieR catalogue service\footnote{To be uploaded after acceptance of the paper}. Here we present examples of each of these tables and the references used in our compilation. Table \ref{HIex} gives the \hi\ fluxes, uncertainties, and masses for ten random sources. The references used to compile the \hi\ data are listed in Table \ref{HIrefs}. Table \ref{metalrefs} lists the references used in the emission line flux compilation. Table \ref{app:lines} list ten randomly selected entries in our emission line fluxes table. We also give examples of the resolved and global metallicity files in Table \ref{app:reg} and Table \ref{app:glob} respectively. We selected ten random entries from the emission line fluxes table and used the same galaxies in our other example tables. 

\begin{table}
\caption{References used in the compilation of \hi\ fluxes.}
\begin{center}
\begin{tabular}{ll}\hline\hline
\hi\ ref ID & Reference \\ \hline
Bettoni2003 & \citet{Bettoni2003} \\ 
Boselli2014 & \citet{Boselli2014} \\ 
Bottinelli1980 & \citet{Bottinelli1980} \\ 
Bottinelli1982 & \citet{Bottinelli1982} \\ 
Bouchard2003 & \citet{Bouchard2003} \\ 
Bureau1996 & \citet{Bureau1996} \\ 
Casasola2004 & \citet{Casasola2004} \\ 
Chamaraux1999 & \citet{Chamaraux1999} \\ 
Clark2015 & \citet{Clark2015} \\ 
Courtois2009 & \citet{Courtois2009} \\ 
Davoust2004 & \citet{Davoust2004} \\ 
Garcia1992 & \citet{Garcia1992} \\ 
Haynes2018 & \citet{Haynes2018}\\ 
Huchtmeier1989 & \citet{Huchtmeier1989} \\ 
Huchtmeier1995 & \citet{Huchtmeier1995} \\ 
Huchtmeier2000 & \citet{Huchtmeier2000} \\ 
Huchtmeier2003 & \citet{Huchtmeier2003} \\ 
Huchtmeier2005 & \citet{Huchtmeier2005} \\ 
Kilborn2002 & \citet{Kilborn2002} \\ 
Koribalski2004 & \citet{Koribalski2004} \\ 
Lang2003 & \citet{Lang2003} \\ 
Martin1991 & \citet{Martin1991} \\ 
Masters2014 & \citet{Masters2014} \\ 
Mathewson1992 & \citet{Mathewson1992} \\ 
Meyer2004 & \citet{Meyer2004} \\ 
Nordgren1998 & \citet{Nordgren1998} \\ 
Ott2012 & \citet{Ott2012} \\ 
Paturel2003 & \citet{Paturel2003} \\ 
Schneider1992 & \citet{Schneider1992} \\ 
Smoker2000 & \citet{Smoker2000} \\ 
Springob2005 & \citet{Springob2005} \\ 
Staveley-Smith2016 & \citet{Staveley-Smith2016} \\ 
Theureau1998 & \citet{Theureau1998} \\ 
Theureau2005 & \citet{Theureau2005} \\ 
Theureau2007 & \citet{Theureau2007} \\ 
Tifft1988 & \citet{Tifft1988} \\ 
vanDriel2016 & \citet{vanDriel2016} \\
Walter2008 & \citet{Walter2008} \\
Wong2006 & \citet{Wong2006} \\  \hline
\end{tabular}
\end{center}
\label{HIrefs}
\end{table}

\begin{table}
\caption{References used in the emission line flux compilation for DustPedia galaxies. Every listing in the sample column in Tables \ref{app:lines} and \ref{app:reg} here have a corresponding reference.}
\begin{tabular}{ll}\hline\hline
Metal sample ID & Reference \\ \hline
2df$ ^e$ & \citet{Colless2001} \\ 
6df$ ^e$ & \citet{Jones2009} \\ 
Anni10 & \citet{Annibali2010} \\ 
Bres99$ ^d$ & \citet{Bresolin1999} \\ 
Bres02 & \citet{Bresolin2002} \\ 
Bres05 & \citet{Bresolin2005} \\ 
Bres09 & \citet{Bresolin2009} \\ 
Bres12 & \citet{Bresolin2012} \\ 
CALIFA$ ^c$ & \citet{Sanchez2012,Sanchez2016b} \\ 
CHAOS & \citet{Berg2015,Croxall2015,Croxall2016}\\
Crox09 & \citet{Croxall2009} \\ 
DGS & \citet{DeVis2017b} \\ 
Disney77 & \citet{Disney1977} \\ 
DV17 & \citet{DeVis2017b} \\ 
GAMA & \citet{Liske2015} \\ 
Gavazzi04$ ^a$ & \citet{Gavazzi2004} \\ 
Gavazzi13$ ^b$ & \citet{Gavazzi2013} \\ 
Gon95 & \citet{Gonzalez1995} \\ 
Gus11 & \citet{Guseva2011} \\ 
Ho95$ ^b$ & \citet{Ho1995} \\ 
HRS$ ^a$ & \citet{Boselli2013} \\ 
Jansen00I$ ^a$ & \citet{Jansen2000} \\ 
Jansen00N$ ^b$ & \citet{Jansen2000} \\ 
Kim95 & \citet{Kim1995} \\ 
Kniazev2004 & \citet{Kniazev2004} \\ 
Lee03 & \citet{Lee2003} \\ 
Lira07 & \citet{Lira2007} \\ 
MUSE$ ^c$ &  This work (ESO archival data) \\ 
Moust06I$ ^a$ & \citet{Moustakas2006} \\ 
Moust06N$ ^b$ & \citet{Moustakas2006} \\ 
Moust10circ & \citet{Moustakas2010} \\ 
Moust10nuc$ ^b$ & \citet{Moustakas2010} \\ 
Moust10rad & \citet{Moustakas2010} \\ 
Pilyu14$ ^d$ & \citet{Pilyugin2014} \\ 
Rodr14$ ^c$ & \citet{Rodriguez2014} \\ 
Ros11$ ^c$ & \citet{Rosales2011} \\ 
SAMI$ ^c$ & \citet{Green2018} \\ 
SDSS & \citet{Alam2015} \\ 
Sanchez12$ ^c$ & \citet{Sanchez2012b} \\ 
UZW$ ^e$ & \citet{Falco1999} \\ 
vanZee97 & \citet{vanZee1997} \\ 
vanZee98 & \citet{vanZee1998} \\ 
vanZee06 & \citet{vanZee2006} \\ \hline
\end{tabular}
\vspace{0.1cm}
 \\ $ ^a$ These references provide integrated spectroscopy.\\
  $ ^b$ These references provide nuclear spectroscopy.\\
  $ ^c$ These references provide IFU spectroscopy.\\
  $ ^d$ We were not able to make the reddening correction consistent with THEMIS since these references did not provide $C(\rm{H\beta})$.\\
  $ ^e$ These spectra were not properly flux calibrated. Only calibrations using lines that are very close in wavelength (i.e. N2 and O3N2) can be used reliably.\\
  \label{metalrefs}
\end{table}

\begin{landscape}
\begin{table}
\caption{Example of emission line flux table for ten random entries from the full table. The catID column provides an identifier for each region used (when possible the identifier from the reference table was used). The fluxes in this table are relative to $H\beta$ (i.e. they are dimensionless) and have been corrected for extinction using the Balmer decrement. These values have been rounded, for the precise values we refer to the online tables. The sample column identifies which reference has been used (see Table \ref{metalrefs}).}
\begin{center}
\begin{tabular}{ccccccccccccc} \hline\hline
\scriptsize
name & catID & RA & DEC & $C(\rm{H\beta})$ & $\rm F_{OII}$ & $\rm E_{OII }$& $\rm F_{H\beta}$ & $\rm E_{H\beta}$ & $\rm F_{OIII4959}$ & $\rm E_{OIII4959}$ & $\rm F_{OIII5007}$ & $\rm E_{OIII5007}$ \\ \hline
ESO097-013 & ESO097-013-247 & 213.292890 & -65.343592 & 0.43 & \ldots & \ldots & 1.0 & 0.11& 0.12 & 0.067 & 0.36 & 0.066 \\
NGC0628 & 266 & 24.192 & 15.7499 & 0.19 & 4.9 & 6.5 & 1.0 & 0.35 & 0.11 & 0.15 & 0.34 & 0.47 \\
NGC3381 & 14  & 162.115452 & 34.71817 & 0.22 & 2.41 & 0.18 & 1.0 & 0.041 & 0.124 & 0.026& 0.371& 0.032 \\
NGC4038 & NGC4038-3-62 & 180.470093 & -18.870465 & 0.21 & \ldots & \ldots & 1.0 & 0.02 & 0.154 & 0.004 & 0.429& 0.009 \\
NGC4651 & NGC4651-89 & 190.935774& 16.386419 & 0.46 & \ldots & \ldots & 1.0 & 0.25 & 0.51 & 0.16 & 1.43 & 0.35 \\
NGC5236 & NGC5236-19 & 204.212819 & -29.865541 & 0.32 & \ldots & \ldots & 1.0 & 0.06 & 0.108 & 0.009 & 0.30 & 0.02\\
NGC5457 & 1323-52797-0052 & 210.61172 & 54.28891 & 0.26 & \ldots & \ldots & 1.0 & 0.09 & 0.67 & 0.03 & 2.03 & 0.10 \\
NGC5480 & 50 & 211.589609 & 50.724952& 0.29 & 1.87 & 0.36 & 1.0 & 0.03 & 0.08 & 0.038& 0.24 & 0.04 \\
NGC7320 & 10 & 339.019805 & 33.953993 & 0.062 & 5.5 & 1.1 & 1.0 & 0.2 & 0.44 & 0.08 & 1.4 & 0.2 \\
UGC09299 & 1886 & 217.392430 & -0.016763 & 1.97 & \ldots  & \ldots & 1.0 & \ldots & 0.26 & \ldots & 0.79 & \ldots   \\ \hline
\end{tabular}
\end{center}
\label{app:lines}
\end{table}
\addtocounter{table}{-1}

\begin{table}
\caption{\textit{Continued}}
\begin{center}
\begin{tabular}{ccccccccccc}\hline\hline
\scriptsize
name & catID & $\rm F_{Ha}$ & $\rm E_{Ha}$ & $\rm F_{NII6584}$ & $\rm E_{NII6584}$ & $\rm F_{SII6717}$ & $\rm E_{SII6717}$ & $\rm F_{SII6731}$ & $\rm E_{SII6731}$ & sample \\ \hline
ESO097-013 & ESO097-013-247 & 2.86 & 0.20 & 1.39 & 0.11 & 0.45& 0.035 & 0.338 & 0.028 & MUSE\\
NGC0628 & 266 & 2.86 & 1.24 & 0.86 & 0.53 & 0.50 & 0.33 & 0.43 & 0.31 & Sanchez12\\
NGC3381 & 14 & 2.86 & 0.05 & 0.98 & 0.03 & 0.67 & 0.03 & 0.54 & 0.03 & CALIFA\\
NGC4038  & NGC4038-3-62 & 2.86 & 0.04 & 1.21& 0.02 & 0.646 & 0.008 & 0.456 & 0.006 & MUSE \\
NGC4651 & NGC4651-89 &2.86& 0.53 & 0.73 & 0.14 & 0.29 & 0.05 & 0.22 & 0.04 & MUSE\\
NGC5236 & NGC5236-19 & 2.86 & 0.12 & 1.21 & 0.05 & 0.43 & 0.02 & 0.30 & 0.01 & MUSE\\
NGC5457 & 1323-52797-0052& 2.86 & 0.07& 0.65 & 0.05 & 0.63 & 0.04& 0.48 & 0.04 & SDSS\\
NGC5480 & 50& 2.86 & 0.04 & 1.01 & 0.02& 0.69 & 0.03 & 0.47 & 0.02 & CALIFA \\
NGC7320 & 10 & 2.86 & 0.34 & 0.41 & 0.06 & 0.53 & 0.06 & 0.35 & 0.05 & Rodr14 \\
UGC09299 & 1886 &  2.86 & \ldots & 0.46 & \ldots & 0.78 & \ldots & 0.52 & \ldots & SAMI \\ \hline
\end{tabular}
\end{center}
\end{table}

\begin{table}
\caption{Metallicity measurements using multiple calibrations for individual regions within the galaxy. The same ten randomly selected regions are shown as in Table \ref{metalrefs}. Each of the metallicity columns are labeled by the abbreviation of their calibration (see Section \ref{metalsec}), and are given in $\rm  12+log(O/H)$ together with their asymmetric uncertainties. $r/r_{25}$ and $r/r_{app}$ give the deprojected radius of the region divided by $r_{25}$ and $r_{app}$ (photometry aperture radius) respectively.}
\begin{center}
\scriptsize
\begin{tabular}{ccccccccccccccccc} \hline\hline
name & catID & $C(\rm{H\beta})$ & $\rm PG16S $ & $\rm PG16S_{edown} $ & $\rm PG16S_{eup} $ & $\rm  PG16R  $ & $\rm  PG16R_{edown}  $ & $\rm  PG16R_{eup}  $ & $\rm N2$ & $\rm N2_{edown}$ & $\rm N2_{eup}$ & $\rm O3N2$ & $\rm  O3N2_{edown}$ & $\rm O3N2_{eup}$ & \\ \hline 
ESO097-013 & ESO097-013-247 & 0.43 & 8.64 & 0.04 & 0.03 & \ldots & \ldots & \ldots & 8.85 & 0.06 & 0.06 & 8.77 & 0.03 & 0.03\\
NGC0628$^a$ & 266 & 0.2 & \ldots & \ldots & \ldots & \ldots & \ldots & \ldots & 8.61 & 0.28 & 0.42 & \ldots & \ldots & \ldots \\
NGC3381 & 14 & 0.22 & 8.47 & 0.01 & 0.02 & 8.55 & 0.01 & 0.02 & 8.67 & 0.02 & 0.02 & 8.72 & 0.02 & 0.02\\
NGC4038 & NGC4038-3-62 & 0.21 & 8.559 & 0.006 & 0.006 & \ldots & \ldots & \ldots & 8.77 & 0.01 & 0.01 & 8.728 & 0.005 & 0.005\\
NGC4651 & NGC4651-89 & 0.46 & 8.5 & 0.06 & 0.07 & \ldots & \ldots & \ldots & 8.55 & 0.1 & 0.11 & 8.49 & 0.07 & 0.06\\
NGC5236 & NGC5236-19 & 0.32 & 8.61 & 0.02 & 0.02 & \ldots & \ldots & \ldots & 8.77 & 0.03 & 0.03 & 8.77 & 0.01 & 0.01\\
NGC5457 & 1323-52797-0052 & 0.26 & 8.37 & 0.03 & 0.03 & \ldots & \ldots & \ldots & 8.5 & 0.03 & 0.03 & 8.43 & 0.02 & 0.02\\
NGC5480 & 50 & 0.3 & 8.48 & 0.01 & 0.01 & 8.57 & 0.02 & 0.02 & 8.68 & 0.01 & 0.01 & 8.79 & 0.02 & 0.02\\
NGC7320 & 10 & 0.06 & 8.26 & 0.05 & 0.06 & 8.21 & 0.08 & 0.09 & 8.37 & 0.05 & 0.05 & 8.41 & 0.05 & 0.04\\
UGC09299 & 1886 & 1.97 & 8.21 & \ldots & \ldots & \ldots & \ldots & \ldots & 8.39 & \ldots & \ldots & 8.51 & \ldots & \ldots\\  \hline 
\end{tabular}
\end{center}

\end{table}
\end{landscape}
\newpage
\begin{landscape}
\addtocounter{table}{-1}

\begin{table}
\caption{\textit{Continued}}
\begin{center}
\scriptsize
\begin{tabular}{cccccccccccccccccc}\hline\hline
name  &  $\rm IZI $  &  $\rm IZI_{edown} $ & $ \rm IZI_{eup} $ &  $\rm IZI_{chi2}$  &  $\rm KK04 $  &  $\rm KK04_{edown} $  &  $\rm KK04_{eup} $  &  $\rm T04 $  &  $\rm T04_{edown} $  &  $\rm T04_{eup} $  &  $\rm ONPG16 $  &  $\rm ONPG16_{edown} $  &  $\rm ONPG16_{eup} $  &  $\rm r/r_{25}$  &  $\rm r/r_{app}$  &  sample \\ \hline
ESO097-013  & 8.83 & 0.18 & 0.06 & 0.23 &  \ldots  &  \ldots  &  \ldots  & 9.05 & 0.03 & 0.03 &  \ldots  &  \ldots  &  \ldots  & 0.07 & 0.04 &  MUSE \\
NGC0628  & 8.8 & 0.74 & 0.16 & 0.02 &  \ldots  &  \ldots  &  \ldots  &  \ldots  &  \ldots  &  \ldots  &  \ldots  &  \ldots  &  \ldots & 0.46 & 0.23 &  Sanchez12\\
NGC3381  & 8.64 & 0.21 & 0.15 & 0.03 & 9.03 & 0.02 & 0.02 & 8.99 & 0.02 & 0.02 & -0.88 & 0.03 & 0.03 & 0.73 & 0.35 &  CALIFA\\
NGC4038  & 8.69 & 0.18 & 0.1 & 0.16 &  \ldots  &  \ldots  &  \ldots & 9.002 & 0.005 & 0.005 &  \ldots  &  \ldots  &  \ldots  & 0.06 & 0.04 &  MUSE\\
NGC4651  & 8.59 & 0.23 & 0.15 & 0.13 &  \ldots  &  \ldots  &  \ldots & 8.73 & 0.07 & 0.06 &  \ldots  &  \ldots  &  \ldots  & 0.49 & 0.19 &  MUSE\\
NGC5236  & 8.86 & 0.15 & 0.05 & 0.12 &  \ldots  &  \ldots  &  \ldots  & 9.06 & 0.01 & 0.01 &  \ldots  &  \ldots  &  \ldots  & 0.32 & 0.18 &  MUSE\\
NGC5457  & 8.44 & 0.06 & 0.06 & 2.72 &  \ldots  &  \ldots  &  \ldots & 8.65 & 0.02 & 0.02 &  \ldots  &  \ldots  &  \ldots  & 0.63 & 0.43 &  SDSS\\
NGC5480  & 8.78 & 0.31 & 0.1 & 0.04 & 9.09 & 0.03 & 0.02 & 9.07 & 0.02 & 0.02 & -0.79 & 0.05 & 0.07 & 0.56 & 0.18 &  CALIFA\\
NGC7320  & 8.38 & 0.11 & 0.18 & 0.49 & 8.57 & 0.29 & 0.2 & 8.64 & 0.05 & 0.04 & -1.36 & 0.08 & 0.08 & 0.58 & 0.71 &  Rodr14\\
UGC09299  &  \ldots &  \ldots &  \ldots &  \ldots &  \ldots  &  \ldots  &  \ldots & 8.75 &  \ldots  &  \ldots  &  \ldots  &  \ldots  &  \ldots  & 0.25 & 0.06 &  SAMI\\\hline
\end{tabular}
\end{center}
\label{app:reg}
\end{table}

\begin{table}
\caption{Global metallicity measurements using multiple calibrations. The same ten galaxies are shown as for the randomly selected regions in Table \ref{app:lines}. Each of the metallicity columns are labeled by the abbreviation of their calibration (see Section \ref{metalsec}), and are given in $\rm 12+log(O/H)$ together with their asymmetric uncertainties. The SF, AGN, and Comp columns indicate how many regions in this galaxy are classified as star-forming, AGN or composite respectively on the BPT diagram. }
\begin{center}
\begin{tabular}{cccccccccccccccc} \hline\hline
name  &  SF  &  AGN  &  Comp &  $\rm   PG16S $  &  $\rm   PG16S_{edown} $ &   $\rm  PG16S_{eup} $   &  $\rm  PG16R $  &  $\rm  PG16R_{edown} $  &  $\rm   PG16R_{eup} $  &  $\rm  N2$  &  $\rm N2_{edown}$  &  $\rm N2_{eup}$  &  $\rm O3N2$  &  $\rm O3N2_{edown}$  &  $\rm O3N2_{eup}$ \\ \hline
ESO097-013  & 20 & 60 & 49 & 8.3 & 0.05 &  0.04  &  \ldots  &  \ldots  &  \ldots  & 8.96 & 0.06 & 0.06 & 8.61 & 0.06 & 0.06 \\
NGC0628  & 746 & 0 & 36 & 8.49 & 0.01 &  0.01   & 8.5 & 0.01 & 0.01 & 8.6 & 0.01 & 0.01 & 8.69 & 0.01 & 0.01\\
NGC3381  & 117 & 2 & 1 & 8.43 & 0.01 &  0.01 & 8.54 & 0.02 & 0.02 & 8.62 & 0.02 & 0.02 & 8.69 & 0.02 & 0.01 \\
NGC4038  & 218 & 1 & 196 & 8.51 & 0.01 &  0.01  & 8.61 &  \ldots  &  \ldots  & 8.7 & 0.01 & 0.01 & 8.7 & 0.01 & 0.01\\
NGC4651  & 220 & 1 & 62 & 8.51 & 0.01 &  0.01  & 8.51 & 0.07 & 0.07 & 8.68 & 0.01 & 0.01 & 8.67 & 0.01 & 0.01\\
NGC5236  & 824 & 25 & 633 & 8.56 & 0.01 &  0.01  & 8.58 & 0.03 & 0.04 & 8.73 & 0.01 & 0.01 & 8.84 & 0.01 & 0.01\\
NGC5457  & 300 & 7 & 3 & 8.41 & 0.01 &  0.02   & 8.4 & 0.01 & 0.01 & 8.45 & 0.02 & 0.01 & 8.53 & 0.01 & 0.01\\
NGC5480  & 88 & 7 & 13 & 8.49 & 0.01 &  0.01  & 8.59 & 0.02 & 0.02 & 8.65 & 0.02 & 0.02 & 8.75 & 0.02 & 0.02\\
NGC7320  & 14 & 0 & 0 & 8.25 & 0.02 &  0.02 & 8.23 & 0.03 & 0.03 & 8.38 & 0.02 & 0.02 & 8.44 & 0.02 & 0.02\\
UGC09299  & 771 & 0 & 3 & 8.19 & 0.02 &  0.02  & 8.25 & 0.05 & 0.05 & 8.41 & 0.02 & 0.02 & 8.5 & 0.02 & 0.02\\ \hline
\end{tabular}
\end{center}

\end{table}
\addtocounter{table}{-1}

\begin{table}
\caption{\textit{Continued}}
\begin{center}
\begin{tabular}{ccccccccccccc}\hline\hline
name   &  $\rm  IZI $  &  $\rm  IZI_{edown} $ & $ \rm  IZI_{eup} $   &  $\rm   KK04 $  &  $\rm   KK04_{edown} $  &  $\rm  KK04_{eup} $   &  $\rm   T04 $  &  $\rm   T04_{edown} $  &  $\rm  T04_{eup} $  &  $\rm  ONPG16 $  &  $\rm  ONPG16_{edown} $  &  $\rm  ONPG16_{eup} $ \\\hline 
ESO097-013   & 8.96 & 0.07 &  0.07  &  \ldots  &  \ldots  &  \ldots  & 8.87 & 0.07 & 0.07 &  \ldots  &  \ldots  &  \ldots \\ 
NGC0628   & 8.82 & 0.01 &  0.01 & 8.99 & 0.01 & 0.01 & 8.97 & 0.01 & 0.01 & -0.88 & 0.02 & 0.01  \\
NGC3381   & 8.69 & 0.06 &  0.04  & 9.06 & 0.03 & 0.02 & 8.96 & 0.02 & 0.02 & -0.81 & 0.03 & 0.03 \\
NGC4038   & 8.7 & 0.02 &  0.01    & 8.65 &  \ldots  &  \ldots  & 8.97 & 0.01 & 0.01 & -0.85 &  \ldots  &  \ldots\\    
NGC4651  & 8.69 & 0.01 &  0.01  & 8.97 & 0.08 & 0.08 & 8.94 & 0.01 & 0.01 & -0.91 & 0.08 & 0.08  \\
NGC5236  & 8.9 & 0.01 &  0.01 & 9.13 & 0.02 & 0.02 & 9.13 & 0.01 & 0.01 & -0.63 & 0.04 & 0.03 \\
NGC5457  & 8.62 & 0.04 &  0.04 & 8.87 & 0.01 & 0.01 & 8.77 & 0.01 & 0.01 & -1.01 & 0.01 & 0.01\\ 
NGC5480   & 8.91 & 0.04 &  0.04 & 9.1 & 0.03 & 0.02 & 9.03 & 0.02 & 0.02 & -0.7 & 0.04 & 0.04 \\
NGC7320   & 8.38 & 0.04 &  0.04  & 8.72 & 0.06 & 0.05 & 8.67 & 0.03 & 0.03 & -1.31 & 0.03 & 0.03 \\  
UGC09299  & 8.35 & 0.09 &  0.09 & 8.83 & 0.05 & 0.05 & 8.75 & 0.02 & 0.02 & -1.24 & 0.05 & 0.05 \\ \hline

\hline
\end{tabular}
\end{center}
\label{app:glob}
\end{table}

\end{landscape}

\end{appendix}
\end{document}